\documentclass[12pt, a4paper]{article}
\usepackage[utf8]{inputenc}
\pdfoutput=1

\usepackage[top=1in, bottom=1in, left=0.75in, right=0.75in]{geometry}
\usepackage[T1]{fontenc}
\usepackage{lmodern}
	\DeclareFontFamily{OMX}{lmex}{}
	\DeclareFontShape{OMX}{lmex}{m}{n}{<-> lmex10}{}	
\usepackage{abstract}
\usepackage{amsmath, amssymb, amsfonts, mathtools}	
	\numberwithin{equation}{section}	
\usepackage{graphicx}       
	\graphicspath{ {./figures/} }
\usepackage[caption=false]{subfig} 
\usepackage{caption}   
\usepackage[dvipsnames]{xcolor}         
\usepackage{physics}
\usepackage{tensor}
\usepackage{mathrsfs}
\usepackage{bm}
\usepackage{enumitem}
\usepackage{cite}
\usepackage{bigints}    
\usepackage[bottom]{footmisc}	

\usepackage[subfigure]{tocloft}

\usepackage{hyperref}		
\definecolor{myRed}{rgb}{0.545,0,0}
\definecolor{myDarkBlue}{rgb}{0,0,0.5}
\hypersetup{colorlinks=true, linktocpage, linkcolor=blue, citecolor={myRed}, urlcolor=blue}  

\def \be {\begin{equation}}
\def \ee {\end{equation}}
\def \nn {\nonumber}
\def \del {\partial}

\newcommand \hide[1] {}

\DeclarePairedDelimiterX\inp[2]{\langle}{\rangle}{#1\,\delimsize\vert\,\mathopen{}#2}

\begin{document}

\begin{titlepage}
\vspace*{-10mm}   
\baselineskip 10pt      
\baselineskip 20pt   

\begin{center}
\noindent
{\Large\bf
UV Effects and Short-Lived Hawking Radiation: \\
Alternative Resolution of Information Paradox \\
\par}
\vskip 10mm
\baselineskip 20pt

\renewcommand{\thefootnote}{\fnsymbol{footnote}}

{\large
Pei-Ming~Ho$^{a, b}$\,\footnote[1]{\texttt{\url{pmho@phys.ntu.edu.tw}}},
Hikaru~Kawai$^{a, b}$\,\footnote[2]{\texttt{\url{hikarukawai@phys.ntu.edu.tw}}},
Wei-Hsiang~Shao$^a$\,\footnote[3]{\texttt{\url{whsshao@gmail.com}}}
}

\renewcommand{\thefootnote}{\arabic{footnote}}

\vskip5mm

{\normalsize \it  
$^a$Department of Physics and Center for Theoretical Physics, National Taiwan University, \\
No. 1, Sec. 4, Roosevelt Road, Taipei 106319, Taiwan
\\
\vspace{.2cm}
$^b$Physics Division, National Center for Theoretical Sciences, \\
No. 1, Sec. 4, Roosevelt Road, Taipei 106319, Taiwan
}  

\vskip 15mm

\begin{abstract}
\vspace{-3mm}
\normalsize

This chapter suggests an alternative solution to the black-hole information paradox by proposing that Hawking radiation ceases around the scrambling time due to trans-Planckian effects inherent in string theory. 
We consider two toy models in the literature that incorporate stringy effects. 
The first model utilizes the generalized uncertainty principle, which introduces a minimal length. The second model is inspired by string field theory, where interactions are exponentially suppressed in the UV limit.
Both models indicate an early termination of Hawking radiation around the scrambling time, resulting in negligible evaporated energy and a predominantly classical black hole.

\end{abstract}

\hide{
Invited chapter for the edited book "The Black Hole Information Paradox" (Eds. Ali Akil and Cosimo Bambi, Springer Singapore, expected in 2025).
Invited chapter for the edited book "The Black Hole Information Paradox'' (Eds. Ali Akil and Cosimo Bambi, Springer Singapore, expected in 2025)
}

\end{center}

\end{titlepage}

\newcommand\afterTocSpace{\bigskip\medskip}
\newcommand\afterTocRuleSpace{\bigskip\bigskip}

\hrule
\tableofcontents
\afterTocSpace
\hrule
\afterTocRuleSpace



\newpage
\section{Introduction}

The black-hole information paradox~\cite{Hawking:1976ra,Hawking:1982dj}, arising from Hawking's prediction of black-hole evaporation~\cite{Hawking:1974rv,Hawking:1975vcx}, has been a central challenge in theoretical physics~\cite{Mathur:2009hf}. 
Regarding the fate of the information carried by the collapsing matter (e.g. the wave functions of all particles in the matter), there are three logical possibilities:
\begin{enumerate}
\item 
Information is lost from our universe,\footnote{This includes proposals in which baby universes branch off during the process and preserve the information outside our universe~\cite{Coleman:1988cy, Polchinski:1994zs}.
For example, the black hole might develop a ``neck'' resembling Wheeler's bag of gold~\cite{WheelersBagofGold}.
As the black hole radiates, the neck narrows and eventually pinches off, forming a baby universe.}
and the radiation carries little or no information about the collapsing matter.
\item 
Information is preserved in a small remnant \cite{Chen:2014jwq}, while the radiation carries little or no information.
\item 
Information is preserved in the radiation, which carries all or almost all information about the collapsing matter and no information is lost.
\end{enumerate}
However, a fourth logical possibility has often been overlooked.
\vskip1em
\hskip1.1em
\begin{minipage}{39em}
\hskip-1.3em 4.
{\em
Hawking radiation is turned off at an early stage so that only a tiny fraction of the initial mass is evaporated. 
Almost all information about the collapsing matter remains trapped within the black hole.
}
\end{minipage}
\vskip1em

In this scenario~\cite{Barman:2017vqx, Ho:2022gpg, Akhmedov:2023gqf, Chau:2023zxb, Ho:2023tdq}, the black hole essentially remains classical. 
Given the negligible quantum effects on macroscopic objects in quantum mechanics, this resolution can be considered the most conservative approach to the information paradox.

The prevailing assumption in the literature has been that Hawking radiation persists until the black hole evaporates to a small fraction of its initial mass. The information paradox is commonly associated with the Page time~\cite{Page:1993wv,Page:1993df} when roughly half of the black hole's mass has evaporated. It has been argued that resolving the paradox requires $\mathcal{O}(1)$ corrections to Hawking radiation~\cite{Mathur:2009hf,Almheiri:2012rt}, indicating nonperturbative quantum gravity effects.\footnote{At low energies, quantum gravity effects can normally be expressed as $1/M_p$ corrections, where $M_p$ is the Planck mass. We need $\mathcal{O}(1)$ effects in this sense.}
If such corrections arise before the Page time, they can significantly modify Hawking radiation to achieve the desired information transfer. On the other hand, while such corrections are yet unknown, they may as well turn off Hawking radiation, and resolve the information paradox as the fourth logical possibility outlined above. 

Regardless, as the information paradox is centered around properties of the Hawking radiation, a careful examination of its derivation is essential.
While the motivation for this work stems from the information paradox, our strategy is to focus on a precise description of Hawking radiation, particularly its modifications due to UV physics.

Numerous studies have supported the robustness of Hawking radiation against various hypothetical UV modifications (see, for instance, refs.~\cite{Unruh:1994je, Brout:1995wp, Hambli:1995pp, Corley:1996ar, Corley:1997ef, Corley:1997pr, Himemoto:1999kd, Jacobson:1999ay, Unruh:2004zk, Agullo:2009wt, Coutant:2011in, Kajuri:2018myh, Boos:2019vcz}). These findings provide strong support for the persistence of Hawking radiation beyond the Page time.
Nevertheless, it has been shown that Hawking radiation undergoes $\mathcal{O}(1)$ corrections due to trans-Planckian physics as early as the {\em scrambling time}~\cite{Ho:2020cbf, Ho:2020cvn, Ho:2021sbi, Ho:2022gpg, Akhmedov:2023gqf, Chau:2023zxb, Ho:2023tdq}.\footnote{The possibility that Hawking's derivation may fail after the scrambling time was pointed out in ref.~\cite{Harlow:2014yka}.} For a large black hole of mass $M$, the scrambling time~\cite{Sekino:2008he} ($\sim \mathcal{O}(M\log(M))$) is significantly shorter than the Page time ($\sim \mathcal{O}(M^3)$).
In particular, as we will demonstrate later in this chapter, there are scenarios in which Hawking radiation ceases around the scrambling time, with the radiated energy constituting a negligible fraction of the black hole's mass. This renders Hawking radiation a minor quantum effect that can be effectively ignored.

It is important to note that the series of works~\cite{Ho:2020cbf,Ho:2020cvn,Ho:2021sbi,Ho:2022gpg,Akhmedov:2023gqf,Chau:2023zxb,Ho:2023tdq} reviewed here did not try to deduce the necessary features of quantum gravity
from the assumption that the information paradox has to be resolved. Instead, they focused on identifying overlooked issues in the derivation of Hawking radiation for improvement. The following key issues were uncovered:
\begin{enumerate}
\item
The background geometry is frequently assumed to be static or stationary, ignoring the collapsing matter. This is only justified in the free field theory as in Hawking’s original calculation~\cite{Hawking:1975vcx}.
\item
While renormalizable interactions have been considered~\cite{Leahy:1983vb,Unruh:1983ac,Helfer:2005wz,Frasca:2014gua,Akhmedov:2015xwa},\footnote{Refs.~\cite{Leahy:1983vb,Unruh:1983ac,Frasca:2014gua} showed that the corrections to Hawking radiation can be treated perturbatively, but ref.~\cite{Helfer:2005wz} proposed that ultra-energetic quanta are produced in the gravitaional collapsing region, and ref.~\cite{Akhmedov:2015xwa} reported a large one-loop effect on Hawking radiation.}
non-renormalizable interactions have been largely ignored, except that their potential importance was mentioned in ref.~\cite{Iso:2008sq}.
\item
The claimed robustness of Hawking radiation in the literature typically refers only to the Hawking temperature, not the radiation's magnitude. 
The time dependence of the magnitude of Hawking radiation has been overlooked in most studies, with a few exceptions~\cite{Jacobson:1993hn, Helfer:2003va, Barcelo:2008qe, Barman:2017vqx, Ho:2022gpg, Akhmedov:2023gqf, Chau:2023zxb, Ho:2023tdq}.
\end{enumerate}

When non-renormalizable interactions between the collapsing matter and the radiation field are considered, the time dependence of Hawking radiation can be significantly affected. Recent studies~\cite{Ho:2020cbf,Ho:2020cvn,Ho:2021sbi,Ho:2022gpg} have shown that certain higher-derivative non-renormalizable interactions can lead to exponentially growing contributions to particle creation at large distances, causing the perturbation theory to break down around the scrambling time. This indicates that the low-energy effective theory's prediction of Hawking radiation is unreliable beyond the scrambling time. The origin of this phenomenon lies in the trans-Planckian center-of-mass energy between the collapsing matter and the outgoing Hawking particle after the scrambling time.

As UV physics becomes relevant to Hawking radiation at late times, the information of the collapsing matter can potentially be transmitted into radiation. However, two concrete UV models \cite{Chau:2023zxb,Ho:2023tdq} motivated by string theory and another model \cite{Barman:2017vqx} motivated by loop quantum gravity have shown that Hawking radiation is terminated around the scrambling time. In another model~\cite{Akhmedov:2023gqf} of modified UV dispersion relations, Hawking radiation can be turned off at any time after the scrambling time, depending on the details of the UV physics.

This chapter explores a resolution to the information paradox based on the early termination of Hawking radiation, arguing that trans-Planckian effects inherent in UV-complete theories of quantum gravity lead to its suppression near the scrambling time. Sec.~\ref{sec:Traditional-Model} reviews the traditional model of black-hole evaporation, highlighting assumptions often overlooked in the derivation of Hawking radiation. Sec.~\ref{sec:Information-Problem} addresses prevalent arguments against the relevance of UV physics in resolving the information paradox, alongside the difficulties associated with information transfer through Hawking radiation. In sec.~\ref{sec:UV-physics}, the scattering amplitudes between the collapsing matter and the radiation field due to higher-derivative non-renormalizable interactions are calculated, demonstrating their exponential growth over time and the breakdown of the low-energy effective theory around the scrambling time. Sec.~\ref{sec:Hawking-Radiation-Turned-Off} explores two UV models of the radiation field that lead to the termination of Hawking radiation around the scrambling time. The first model incorporates the generalized uncertainty principle, while the second mimics the exponential suppression of UV interactions in string field theory. Comments on their implications and comparisons with other proposals addressing the information paradox are provided in sec.~\ref{sec:Comments}. Finally, conclusions are presented in sec.~\ref{sec:Conclusion}.

Throughout the chapter, we adopt the following unit system:
\be
c = 1, \qquad 
\hbar = 1, \qquad 
k_B = 1, \qquad 
G_N = \ell_p^2 = M_p^{-2} \, ,
\ee
where $k_B$ is the Boltzmann constant, and $G_N$ is the Newton constant.

\section{Traditional Model}
\label{sec:Traditional-Model}

This section briefly reviews the formation and evaporation of black holes, drawing upon standard references~\cite{Brout:1995rd, Frolov:1998wf}. 
This scenario is called a {\em traditional black hole} in ref.~\cite{Mathur:2009hf}. 
In this chapter, we refer to this framework as the {\em traditional model}. We shall accept all assumptions in the traditional model, except for the properties of Hawking radiation, which will be determined through calculation.

\subsection{Overview}

For simplicity, consider the gravitational collapse of a uniform spherical dust shell with a total mass $M$ and thickness $d \ll a$ (and $d \gg \ell_p$). When the shell enters the horizon at the {\em Schwarzschild radius}
\be
a \equiv 2G_N M \, ,
\label{a}
\ee
the average density $\rho$ of the collapsing shell is
\be
\rho
\simeq 
\frac{1}{8 \pi G_N a d} \propto \frac{1}{a}
\ee 
since the mass is given by $M \simeq \rho V$, where the volume $V \simeq 4\pi a^2 d$. As a result, 
for sufficiently large black holes (i.e., large $a$), the density $\rho \propto 1/a$ becomes very small at the moment of horizon crossing. In this work, we focus on such a large black hole 
formed by the collapse of a dilute dust shell, where $\rho$ is small enough for the spacetime curvature to be small everywhere.\footnote{For example, we can assume that $\rho$ is much smaller than the density of air on Earth, allowing the electromagnetic field to be approximately in the vacuum state.} The traditional model is constructed in such a way that all high-energy events are avoided as much as possible.

We further assume that interactions among the dust particles are negligible so that all dust particles undergo free fall. According to the equivalence principle, particles crossing the horizon will not experience any unusual effects; instead, they will feel like they are floating in empty space. Furthermore, it is assumed in the traditional model that the dust interacts very weakly with the radiation field $\phi$, so that the radiation field can stay in an approximate vacuum state. 

\begin{figure}[t]
\centering
\includegraphics[scale=0.8]{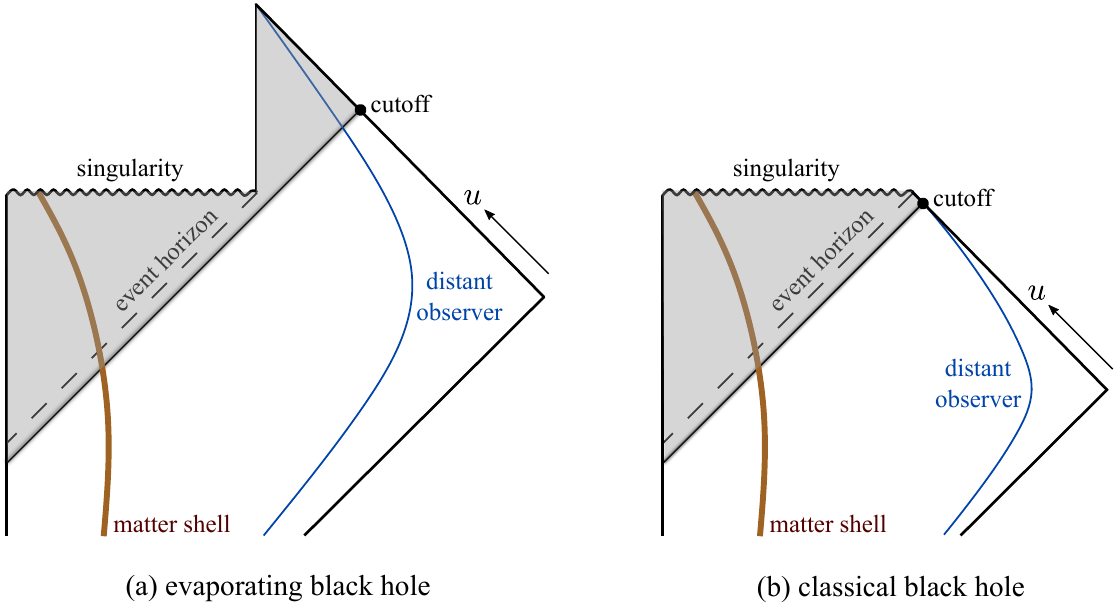}
\caption{\label{fig:penrose0}
Panel~(a) shows the Penrose diagram of a black hole in the traditional model and panel~(b) that of a classical black hole. The shaded areas refer to the regions beyond a certain cutoff retarded time. 
We will only be concerned with Hawking radiation before this cutoff time,
which will later be identified as the scrambling time.
Causally, the region above the cutoff time cannot affect the Hawking radiation below this line. Therefore, the difference between (a) and (b), as well as the event horizon and the singularity, are all irrelevant in our discussion.}
\end{figure}

According to Hawking's calculation~\cite{Hawking:1975vcx} (which will be reviewed in sec.~\ref{sec:HR}), distant observers start to detect outgoing radiation at the same retarded time $u$ when the radius $R(u)$ of the matter shell is close to the horizon (see fig.~\ref{fig:penrose0}), i.e., when 
\be 
R(u) - a \ll a
\, .
\ee 
The emitted radiation exhibits the spectrum of thermal radiation characterized by the Hawking temperature
\be
\label{Hawking-Temperature}
T_H = \frac{1}{4\pi a} \, .
\ee
The energy flux in the radiation is thus $\sim \mathcal{O}(1/a^4)$, and the radiated power is $\sim \mathcal{O}(1/a^2)$. This is an extremely slow process. For a black hole of a solar mass,
it takes $\sim 10^{67}$ years to evaporate.

As Hawking radiation carries energy away from the black hole, the Schwarzschild radius $a(t)$ gradually decreases over time. 
Low-energy effective theories apply to the black hole only if the Schwarzschild radius $a(t)$ is much larger than the Planck length $\ell_p$:
\be
a(t) \equiv 2G_N M(t) \gg \ell_p \, .
\label{a>>l}
\ee
(For a black hole of a solar mass, the Schwarzschild radius is $a_{\odot} \simeq 3 \, km \simeq 2 \times 10^{38} \ell_p$.) We shall always assume the large black hole limit~\eqref{a>>l} below.

%

In the traditional model, Hawking radiation is derived from the low-energy effective theory
in a perturbative expansion in powers of $\ell_p^2/a^2(t)$, or equivalently $1/(a(t) M_p)^2$.
The low-energy effective theory includes the quantum field theory for the radiation field 
propagating on the black-hole background determined by the semi-classical gravity theory.
For a sufficiently large black hole, the curvature near its horizon is even smaller than that around the surface of the Earth. There is consequently no reason why low-energy effective theories such as the Standard Model of particle physics and Einstein's theory of general relativity cannot adequately describe the physics near the black-hole horizon and away from the black hole.

In the following, we introduce the geometry of a Schwarzschild black hole formed by the gravitational collapse of a spherical thin shell in sec.~\ref{sec:BlackHole}, and review the derivation of Hawking radiation in the traditional model in sec.~\ref{sec:HR}.


\subsection{Black-Hole Geometry}
\label{sec:BlackHole}

For explicit calculations, we only focus on spherically symmetric configurations for simplicity,
but we believe that the underlying physics of Hawking radiation can be extended to a generic black-hole formation process.

The Schwarzschild spacetime is the unique static solution to Einstein's equations for a black hole with spherical symmetry. For the dynamic process of black-hole formation, we now discuss how the Schwarzschild geometry should be modified due to the collapsing matter and the back-reaction of Hawking radiation. We will also describe the geometric properties that will be needed for the calculation of Hawking radiation.

\subsubsection{Schwarzschild Solution}

The Schwarzschild solution of Einstein's equations in vacuum describes a static, spherically symmetric classical black hole:
\be
ds^2 = - \left(1 - \frac{a}{r}\right) dt^2 + \frac{dr^2}{\left(1 - a / r \right)} 
+ r^2 \left(d\theta^2 + \sin^2\theta \, d\varphi^2\right) ,
\label{eq:Schwarzschild}
\ee
where $a$ is the Schwarzschild radius \eqref{a}. We refer to $r$ as the areal radius because it directly determines the area of the spatial 2-sphere to be $4\pi r^2$.

The metric \eqref{eq:Schwarzschild} can be rewritten as
\be
ds^2 = - \left(1 - \frac{a}{r}\right) du \, dv
+ r^2 \left(d\theta^2 + \sin^2\theta \, d\varphi^2\right)
\label{Schwarzschild-uv}
\ee
in terms of the Eddington retarded time $u$ and advanced time $v$:
\begin{align}
u &\equiv t - r_{\ast} \, ,
\label{eq:u}
\\ 
v &\equiv t + r_{\ast} \, ,
\label{eq:v}
\end{align}
where $r_{\ast}$ is the tortoise coordinate defined by
\be
r_{\ast} \equiv r - a + a\log\left(r/a - 1\right).
\label{eq:tortoise}
\ee

In terms of the Kruskal light-cone coordinates
\begin{align}
U &\equiv - 2a \, e^{- u / 2a} \, ,
\label{eq:U-u}
\\
V &\equiv 2a \, e^{v / 2a} \, ,
\label{eq:V-v}
\end{align}
the Schwarzschild metric is
\be
ds^2 = - \frac{a}{r} \, e^{- \frac{r - a}{a}} \, dU dV 
+ r^2 \left(d\theta^2 + \sin^2\theta \, d\varphi^2\right) .
\label{eq:Schwarzschild-UV}
\ee
This metric is regular and smooth at the horizon.

The light-cone coordinates $u$ and $v$ align with the Minkowski light-cone coordinates in the asymptotically flat region ($r \rightarrow \infty$), whereas $U$ and $V$ are more naturally employed by freely falling observers. Notably, the retarded coordinates are related by an exponential blue-shift factor
\be
\frac{du}{dU} = e^{u/2a}
\label{eq:blue-shift}
\ee
that goes to infinity at the horizon ($U = 0$ or equivalently $u \rightarrow \infty$).
This exponential relation \eqref{eq:blue-shift} between the retarded times for freely falling observers and distant observers is the single most crucial equation for Hawking radiation.

Since massless outgoing particles move along constant-$u$ (or equivalently, constant-$U$) curves, it is convenient to express their wave functions in terms of the retarded light-cone coordinates $u$ or $U$. Consider the wave function of a Hawking particle detected at a large distance ($r \gg a$) between the retarded times $u_1$ and $u_2$ ($u_2 > u_1$). In terms of the Kruskal coordinate $U$, the wave function lies within a range of $U$ given by 
\begin{align}
\Delta U \equiv U_2 - U_1 = 2a \left(e^{- u_1 / 2a} - e^{- u_2 / 2a} \right)
< 2a \, e^{- u_1 / 2a} \, .
\end{align}

To establish a reference time scale, we define the {\em scrambling time} as
\be
u_{scr} \equiv 2a \log\left(a^2/\ell_p^2\right).
\label{scrambling-time}
\ee
For a large black hole ($a \gg \ell_p$), $u_{scr}$ is much shorter than the Page time $\mathcal{O}(a^3/\ell_p^2)$. 
For instance, a black hole of a solar mass has $u_{scr} \simeq 3 \, ms$
and a Page time $\sim 10^{67}$ years.
Any wave function detected after half the scrambling time (i.e. $u_1 > u_{scr}/2$) lies within two Planck lengths from the horizon ($U = 0$) in terms of $U$:
\be 
\Delta U < |U_1| < 2\ell_p \, .
\ee
The wave function of a Hawking particle detected after the scrambling time $u_{scr}$ is limited to an extremely tiny domain of $\Delta U < 10^{-38} \ell_p$ for a black hole of a solar mass.

\subsubsection{Near-Horizon Geometry}
\label{sec:NHG}

In the near-horizon region defined by
\be
0 < r - a \ll a \, ,
\label{eq:near-horizon}
\ee
the Schwarzschild metric \eqref{eq:Schwarzschild-UV} can be approximated as
\be
ds^2 \simeq - dU dV 
+ r^2(U, V) \left(d\theta^2 + \sin^2\theta \, d\varphi^2\right).
\label{eq:ds2-near-horizon}
\ee
The areal radius $r(U, V)$ is implicitly defined by 
\be
UV = - 4a^2 e^{r_{\ast}(r)/a} \, ,
\ee
where $r_{\ast}(r)$ is the tortoise coordinate \eqref{eq:tortoise}.

It is important to note that this analysis focuses on dynamical black holes formed from gravitational collapse, rather than eternal black holes. The Schwarzschild metric applies only to the region outside the collapsing matter. For simplicity, we consider a thin shell collapsing at the speed of light.
Generalizing this to finite speeds is straightforward, provided that the instantaneous velocity at the moment of horizon crossing is not fine-tuned to approach zero.

Without loss of generality, we can choose the coordinate system such that the shell crosses the horizon at $v = 0$, or equivalently, $V = 2a$. (The Schwarzschild metric~\eqref{eq:Schwarzschild} is invariant under a shift of $t$, which leads to a shift of $v$.)

Inside the collapsing shell, the metric is Minkowskian and has the same form as eq.~\eqref{eq:ds2-near-horizon}, except that the areal radius $r(U, V)$ is now given by
\be
r(U, V) = R(U, V) \equiv \frac{V - U}{2}
\ee
up to an additive constant, 
which can be set to zero by shifting $U$ and $V$ inside the shell.

Across the thin shell at $V = 2a$, the Minkowski light-cone coordinate $U$ inside the shell can be matched with the Kruskal light-cone coordinate $U$ outside the shell in the near-horizon region where $r - a\ll a$. This allows us to use the same symbols $U$ and $V$ throughout, interpreting eq.~\eqref{eq:blue-shift} as the blue-shift factor between the Minkowski patches in the infinite past ($V \rightarrow - \infty$) and the infinite future ($V \rightarrow \infty$) for a massless outgoing wave packet along a fixed $U$ trajectory.

\subsubsection{Energy-Momentum Tensor and Back-Reaction}
\label{sec:Backreaction}

The classical Schwarzschild solution assumes a vanishing energy-momentum tensor outside the collapsing shell. A more accurate description requires the {\em semi-classical Einstein equation}:
\be
G_{\mu\nu} \equiv R_{\mu\nu} - \frac{1}{2} g_{\mu\nu} R 
= 8 \pi G_N \langle T_{\mu\nu} \rangle \, ,
\label{Einstein-eq}
\ee
where $\langle T_{\mu\nu} \rangle$ is the expectation value of the energy-momentum tensor in the low-energy effective theory for the radiation field $\phi$, including the energy flux of the Hawking radiation. The semi-classical Einstein equation \eqref{Einstein-eq} determines
how the background geometry is modified by the quantum energy-momentum tensor.

With regard to the information paradox, we assume that the initial state of the radiation field is the Minkowski vacuum in the infinite past. 
For a matter shell collapsing at the speed of light, a Minkowski vacuum is expected everywhere inside the collapsing shell. 
Hence, $\langle T_{\mu\nu} \rangle = 0$ inside the shell.

The initial state also determines the quantum state outside the shell. In the traditional model, it is well approximated by the Unruh vacuum~\cite{Unruh:1976db}.\footnote{For black holes, two other vacua are often considered~\cite{Frolov:1998wf}. The Boulware vacuum, representing an eternal black hole without radiation, has a diverging energy-momentum tensor at the horizon in the freely falling frame. The Hartle-Hawking vacuum is a time-independent state that assumes a balanced ingoing and outgoing energy flux at the Hawking temperature.} 
(See ref.~\cite{Juarez-Aubry:2018ofz} for a detailed discussion on the validity of this approximation.) 
The Unruh vacuum is defined as the state with zero outgoing energy flux on the shell and zero ingoing energy flux in the infinite past. 
It is characterized by a finite $\langle T_{\mu\nu} \rangle$ for freely falling observers, as required by the equivalence principle.

In the traditional model, the expectation value $\langle T_{\mu\nu} \rangle$ of the Unruh vacuum is assumed to satisfy the bounds
\begin{align}
\langle T_{UU} \rangle &\sim
\left(\frac{du}{dU}\right)^{2} \langle T_{uu} \rangle \sim \mathcal{O}(1/a^4) \, ,
\label{TUU-bound}
\\
\langle T_{UV} \rangle &\sim
\left(\frac{du}{dU}\right)^{-1} \langle T_{uv} \rangle \sim \mathcal{O}(1/a^4) \, ,
\label{TUV-bound}
\\
\langle T_{VV} \rangle &\sim
\langle T_{vv} \rangle \sim \mathcal{O}(1/a^4) \, ,
\label{TVV-bound}
\\
\langle T_{\theta\theta} \rangle &\sim \mathcal{O}(1) \, .
\label{Tthth-bound}
\end{align}
Dictated by the equivalence principle, these bounds have found a lot of supportive evidence~\cite{Davies:1976hi, Davies:1976ei, Christensen:1977jc, Candelas:1980zt, Parentani:1994ij, Ayal:1997ab, Fabbri:2005zn, Fabbri:2005nt, Barcelo:2007yk}. 
As the quantum effect $\langle T_{\mu\nu} \rangle$ is small, one can treat the back-reaction problem perturbatively. A perturbative calculation for a dynamical black hole including the collapsing matter can be found in ref.~\cite{Ho:2018jkm}.


As the back-reaction to the spacetime geometry is very small, one can define an effective time-dependent Schwarzschild radius $a(u)$ such that eq.~\eqref{eq:blue-shift} remains approximately correct. The value of $a(u)$ changes very slowly as
\be
\left|\frac{da(u)}{du}\right| \sim \mathcal{O}\left(\frac{\ell_p^2}{a^2(u)}\right).
\ee
This allows us to treat $a$ as a constant within a time scale much shorter than the Page time:
\be
\Delta u \ll \mathcal{O}\left(\frac{a^3(u)}{\ell_p^2}\right).
\ee

For the Schwarzschild metric~\eqref{Schwarzschild-uv}, eq.~\eqref{TUU-bound} implies
\be
\langle T_{uu} \rangle = 0 \quad \mbox{at} \quad U = 0
\label{TUU=0}
\ee
because
\be
\frac{du}{dU} \sim \frac{1}{r - a}
\ee
in the limit $r \rightarrow a$ on the future horizon. From the perspective of a distant observer,
eq.~\eqref{TUU=0} indicates that the outgoing energy flux $\langle T_{uu} \rangle$ is zero at the future horizon and gradually increases to account for the energy in Hawking radiation at $r - a \gtrsim \mathcal{O}(a)$.

In view of energy conservation, the outgoing energy flux in Hawking radiation is balanced by a negative ingoing energy flux entering the horizon\footnote{For observers staying on top of the trapping horizon, the power of the ingoing energy flux is ${\cal E} \simeq - 1/(2\ell_p^2 a^2)$ \cite{Ho:2019cfw}.} to decrease the black hole's energy.\footnote{Although the negative ingoing energy flux violates the null energy condition, it is accepted as a quantum effect. If there is no negative ingoing energy flux, and we simply extrapolate the energy flux of Hawking radiation from large distances back to the near-horizon region, the energy flux becomes very large around the horizon due to the blue-shift factor~\eqref{eq:blue-shift}. The KMY model~\cite{Kawai:2013mda,Kawai:2014afa,Kawai:2015uya,Ho:2015fja,Ho:2015vga} is based on this energy-momentum tensor. Similar to the firewall proposal~\cite{Almheiri:2012rt, Braunstein:2009my}, the information paradox is alleviated in the KMY model by a Planckian-scale energy-momentum tenor around the horizon.} 
In this sense, the collapsing matter does not lose energy directly to Hawking radiation. Its energy is merely canceled by more and more negative energy inside the black hole. Since the transfer of information is usually associated with the transfer of energy, this suggests that the information in the collapsing matter is not transmitted to the Hawking radiation.

The physical picture behind this energy flow involves a Hawking particle and its partner created in a curved background without directly interacting with the collapsing matter. The Hawking particle carries positive energy away to large distances, while its partner, which carries negative energy, falls into the black hole, effectively reducing its mass.

To illustrate this, we can draw an analogy with the Schwinger effect by substituting the gravitational field and masses with an electric field and charges. Imagine a large positively charged plate that generates a strong electric field, and outgoing positive charges and ingoing negative charges are created in pairs. The background electric field diminishes over time due to the back-reaction resulting from the ingoing negative charges. However, this does not imply that the charges on the plate will gradually disappear, as the ingoing charges are not necessarily the antiparticles of those on the plate. Furthermore, the outgoing charges are not expected to carry any information about the charges on the plate.

\hide{
Determining the exact quantum state outside the shell is very challenging. 
It involves solving simultaneously the coupled equations of the quantum state's time evolution equation and the semi-classical Einstein equation. 
Typically, this is done perturbatively. 
At the zeroth order, the massless scalar's energy-momentum tensor expectation value $\langle T_{\mu\nu} \rangle$ is uniquely determined by the anomaly and the conservation law with suitable boundary conditions \cite{Davies:1976ei}.\footnote{This can be done not only for the Unruh vacuum but also for the Boulware vacuum and the Hartle-Hawking vacuum. The derivation of the energy-momentum tensor in the Schwarzschild background has also been extended to four dimensions~\cite{Christensen:1977jc,Candelas:1980zt} for these vacuum states.} 
Thus, without explicitly writing down the quantum state, $\langle T_{\mu\nu} \rangle$ can be determined for the Schwarzschild black hole after the $s$-wave reduction to two dimensions.
The energy-momentum tensor $\langle T_{\mu\nu} \rangle$ obtained this way is regular at the horizon, and a perturbative calculation can be carried out. A perturbative calculation for a dynamical black hole including the collapsing matter can be found in ref.~\cite{Ho:2018jkm}.
}

\subsubsection{Apparent Horizon}
\label{sec:Apparent-Horizon}

The time-dependent Schwarzschild radius $a(t)$ does not define the event horizon, which is an outgoing light-like surface extrapolated backward in time from the singularity at $r = 0$.
In spacetime configurations with spherical symmetry, the {\em trapping horizon} is defined by the condition
\be
\frac{\partial}{\partial V} \, r(U, V) = 0
\label{trapping}
\ee
on the areal radius $r$. Due to the negative ingoing energy flux, the outer trapping horizon is time-like~\cite{Frolov:1981qr,Roman:1983zza,Hayward:2005gi}
Given the areal radius $r(U, V)$, eq.~\eqref{trapping} allows us to solve for the trapping horizon $r_{trap}(u)$ as a function of $u$.
The Schwarzschild radius $a(u)$ approximates the (outer) trapping horizon with a difference of $\mathcal{O}\left(\ell_p^2/a^2\right)$ \cite{Ho:2019qiu}.

\begin{figure}[t]
\centering
\includegraphics[scale=0.8,bb=0 0 270 270]{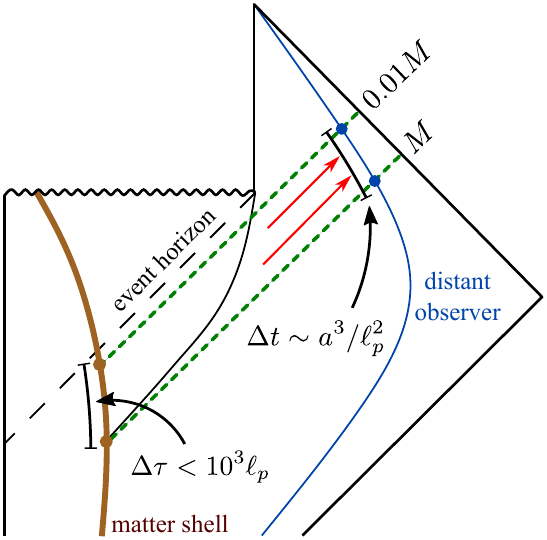}
\caption{\label{fig:app_hor}
In terms of the Eddington retarded time $u$, which agrees with the Schwarzschild time $t$ at a fixed radius $r$, the time for a black hole to evaporate to $1\%$ of its initial mass $M$ is $\mathcal{O}(a^3/\ell_p^2)$.
However, in terms of the proper time of a comoving observer, 
this process occurs over a brief time scale $\sim 1000 \ell_p$~\cite{Ho:2019qiu}.
}
\end{figure}

Interestingly, although the time for a black hole to evaporate to a small fraction $1/n$ of its initial mass (assuming $n \ll (a/\ell_p)^{2/3}$) is very long ($\sim \mathcal{O}(a^3/\ell_p^2)$) for distant observers, from the viewpoint of comoving observers in free fall with the collapsing matter, the proper time available to causally deliver information into the Hawking radiation
is merely $\mathcal{O}\left(n^{3/2} \ell_p\right)$~\cite{Ho:2019qiu}. 
For instance, the time it takes for the black hole to evaporate to $1/100$ of its initial mass is, albeit longer than the Page time for a distant observer, as short a time scale as $\sim 1000 \ell_p$ for a comoving observer. If one expects $99\%$ of the information contained within the collapsing matter to be transmitted to the Hawking radiation, it has to occur within $1000 \ell_p \sim 10^{-40} s$! (See fig.~\ref{fig:app_hor}).

Therefore, if the information of the collapsing matter has to be transferred to Hawking radiation, there must be nonlocal or acausal physics at work.\footnote{
Nonetheless, there have been proposals~\cite{Mukohyama:1998xq, Horowitz:2003he} suggesting that the maximal entanglement between Hawking particles and their partners could provide a causal mechanism for information retrieval.
This approach relies on postselected quantum teleportation, which requires a special final-state boundary condition to be imposed at the black-hole singularity.
}

\subsubsection{Nice Slices}
\label{sec:nice-slices}

\begin{figure}[t]
\centering
\includegraphics[scale=0.8]{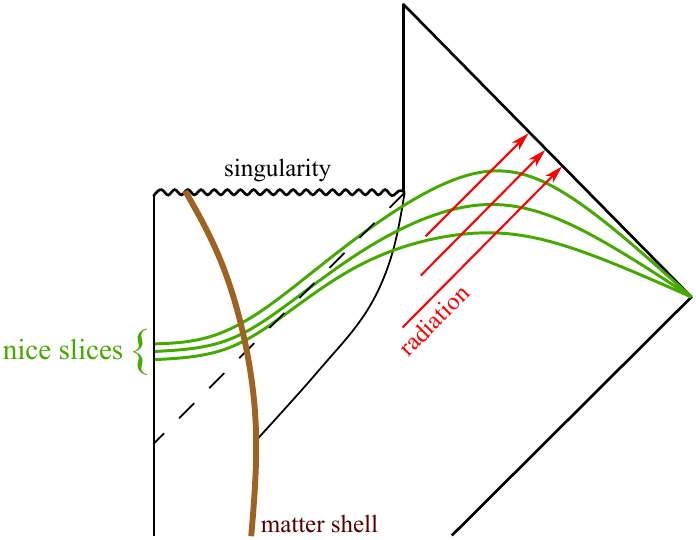}
\caption{\label{fig:nice_slice}
Illustration of nice slices
that intersects the infalling matter and both early and late Hawking radiation.
}
\end{figure}

Away from the singularity, the spacetime geometry of a large black hole is smooth. 
It is possible to construct time slices in the Schwarzschild space (see fig.~\ref{fig:nice_slice}) such that the extrinsic and intrinsic curvatures remain small ($\sim \mathcal{O}(1/a)$) until the black hole becomes microscopic.
These nice slices can extend from the initial time when the radiation field is in the Minkowski vacuum to a late time well past the Page time, so they capture not only the matter falling into the black hole but also the Hawking particles at large distances and their partners inside the black hole~\cite{Mathur:2009hf}.

The adiabatic theorem can be applied to the Hamiltonian evolution on these nice slices, concluding that only low-energy states of energies $\sim \mathcal{O}(1/a)$ can be excited from the Minkowski vacuum.

An example of such nice slices can be found in refs.~\cite{Lowe:1995ac, Polchinski:1995ta}. In the Kruskal coordinates $(U, V)$, the Schwarzschild black hole has an event horizon at $U = 0$ and a curvature singularity at $UV = 4a^2$. 
These slices are constructed by joining two space-like segments smoothly at the surface 
$V = e^{t_{ns}/a} \, U$:
\be
\begin{aligned}
UV &= a^2 
\qquad 
\text{for}
\quad 
V < e^{t_{ns} / a} \, U
\, , 
\\
e^{t_{ns} / 2a} \, U + e^{- t_{ns} / 2a} \, V 
&=
2 a
\qquad 
\text{for}
\quad 
V > e^{t_{ns} / a} \, U
\, .
\label{NS-eq}
\end{aligned}
\ee 

In terms of the coordinates $(u, v)$, eq.~\eqref{NS-eq} implies that 
\be
- e^{(t_{ns} - t)/ 2a} + e^{(t - t_{ns})/ 2a} = e^{- r_{\ast}/2a} \simeq 0
\qquad \mbox{as} \quad r \rightarrow \infty \, ,
\ee
indicating that these slices are asymptotic to surfaces of constant Schwarzschild time, that is, $t = t_{ns}$ in the limit $r \rightarrow \infty$.

Combining the idea of nice slices 
with the equivalence principle and the adiabatic theorem, 
the ``nice-slice'' argument~\cite{Lowe:1995ac, Polchinski:1995ta} 
(which will be summarized in sec.~\ref{sec:Nice-Slice-Argument}) 
establishes that freely falling observers experience an ``uneventful'' horizon, 
with the fields being in an approximate vacuum state.
As such a state cannot carry significant information about the collapsing matter,
Hawking radiation originating from it is also limited in its information content.

\subsection{Derivation of Hawking Radiation}
\label{sec:HR}

The distinction between positive and negative frequency modes in a quantum field theory defines the vacuum state. Due to the blue-shift factor~\eqref{eq:blue-shift}, the Minkowski vacuum in the infinite past (where $U$ is used to define frequencies) differs from the Minkowski vacuum in the asymptotically flat region (where $u$ is used). Hawking radiation is just the interpretation of the Minkowski vacuum in the infinite past by distant observers. The exponential relation~\eqref{eq:U-u} between the Kruskal retarded time $U$ and the Eddington retarded time $u$ implies that the Minkowski vacuum defined inside the collapsing shell (in terms of $U$) appears as an excited state with Hawking particles when viewed from large distances (in terms of $u$).

\subsubsection{Wave Equation}

In principle, Hawking radiation includes all fields. 
However, since the Hawking temperature~\eqref{Hawking-Temperature} is very low for a large black hole, 
the dominant radiation field is the electromagnetic field, 
as it is massless and allows for a free-field approximation.
For simplicity, we consider below a massless scalar field $\phi$
whose Hawking radiation is qualitatively the same as that of the electromagnetic field.

For a static, spherically symmetric background with the line element
\be
ds^2 = - f(r) \, du \, dv + r^2 (d\theta^2 + \sin^2\theta \, d\varphi^2) \, ,
\ee
the Laplace equation
$\nabla^2 \phi 
= 0$
leads to 
\begin{align}
\del_u \del_v \, \psi - \frac{\del_u\del_v \, r}{r} \, \psi
= 0
\label{almost-2D-eq}
\end{align}
for an $s$-wave solution
\be
\label{s-wave}
\phi(u, v) \equiv \frac{\psi(u, v)}{r} \, .
\ee
Eq.~\eqref{almost-2D-eq} would be identical to the 2D wave equation for a massless scalar if the second term $(\del_u\del_v \, r)/r$ can be neglected.

Using the Schwarzschild metric~\eqref{eq:Schwarzschild} where $f(r) = 1 - a/r$, we find 
\be
\frac{\del_u\del_v \, r}{r} = 
- \frac{1}{4}\frac{a}{r^3} \left(1 - \frac{a}{r}\right).
\ee
This term is small both in the near-horizon region ($r - a \ll a$) 
and at large distances ($r \gg a$). 
In the intermediate region, 
it behaves as an effective potential barrier 
that leads to reflection and transmission. 
However, its effect will not be important for an order-of-magnitude estimate of Hawking radiation.
Thus, we can approximate the $s$-wave solution as given by eq.~\eqref{s-wave} with
\be
\psi(u, v) = \psi_{in}(v) + \psi_{out}(u) \, .
\ee
For the purpose of deriving Hawking radiation, 
only the outgoing sector $\psi_{out}(u)$ of the field is needed.

\subsubsection{Uncertainty Relations}
\label{sec:Uncertainty}

Before presenting the derivation of Hawking radiation, 
let us first explain the underlying physics using the uncertainty relation.

Due to the exponential relation~\eqref{eq:U-u}, the conjugate momenta of $u$ and $U$ satisfy the commutation relation~\cite{Ho:2021sbi}
\be
[P_u \, , P_U] = [-i\del_u \, , -ie^{u/2a}\del_u] = - \frac{i}{2a} P_U \, ,
\ee
which leads to the uncertainty relation
\be
\Delta P_u \, \Delta P_U \geq \frac{1}{4a} \langle P_U \rangle \, .
\label{PuPU}
\ee

For a typical Hawking particle, we expect $P_u \sim \mathcal{O}(1/a)$ and $\Delta P_u \ll P_u$. Eq.~\eqref{PuPU} then implies
\begin{align}
\Delta P_U 
\geq \frac{1}{4a \, \Delta P_u} \langle P_U \rangle
\gg \langle P_U \rangle \, .
\label{DPU>PU}
\end{align}
This indicates that the deviation of $P_U$
is significantly larger than its central value $\langle P_U \rangle$, 
suggesting that both positive and negative-$P_U$ modes are present.

Consequently, a Hawking particle composed of purely positive-$P_u$ states 
turns out to be a superposition of both positive and negative-$P_U$ states.
The mixing of positive and negative energy modes for a given outgoing wave with $P_u > 0$ 
implies a non-trivial Bogoliubov transformation 
between the creation and annihilation operator pairs defined in the two coordinate systems,
as will be shown explicitly below.

\subsubsection{Bogoliubov Transformation}
\label{sec:Bogo}

Now we derive the Bogoliubov transformation between the ladder operators $(a_{\Omega}, a^{\dagger}_{\Omega})$ for freely falling observers and those $(b_{\omega}, b^{\dagger}_{\omega})$ for distant observers.
The derivation here has the advantage that it does not rely on the existence of the event horizon.

Assuming a free-field approximation for the propagation of the quantum state, Hawking particles can be detected in the near-horizon region as wave functions with positive frequencies with respect to $u$ (and thus a mixture of positive and negative frequencies with respect to $U$), similar to their detection at large distances. Theoretically, the detection of a quantum state $\psi$ can be equivalently carried out at an earlier time for the state $\Psi$ if they are related via a unitary time evolution.

For a large black hole, the spacetime curvature is small,
and the general $s$-wave solution for the massless scalar field $\phi$ 
in the near-horizon region is approximately
\begin{align}
\phi(x) \simeq 
\int_0^{\infty} \frac{d\omega}{4\pi\sqrt{\omega} \, r}
\left(
b_{\omega} \, e^{- i \omega u} + b^{\dagger}_{\omega} \, e^{i \omega u}
+ \tilde{b}_{\omega} \, e^{- i \omega v} + \tilde{b}^{\dagger}_{\omega} \, e^{i \omega v}
\right),
\label{phi-b}
\end{align}
or equivalently,
\begin{align}
\phi(x) \simeq 
\int_0^{\infty} \frac{d\Omega}{4\pi\sqrt{\Omega} \, r}
\left(
a_{\Omega} \, e^{- i \Omega U} + a^{\dagger}_{\Omega} \, e^{i \Omega U}
+ \tilde{a}_{\Omega} \, e^{- i \Omega V} + \tilde{a}^{\dagger}_{\Omega} \, e^{i \Omega V}
\right),
\label{phi-a}
\end{align}
depending on which coordinate system is used in the Fourier expansion.
We use $\omega$ and $\Omega$ as symbols for the frequencies defined in terms of $u$ and $U$, respectively. In quantum mechanics, they are identified with the conjugate momenta $P_u$ and $P_U$.

Upon quantization, we have the canonical commutation relations
\begin{align}
[b_{\omega} \, , b^{\dagger}_{\omega'}] = \delta (\omega - \omega') \, ,
&\qquad 
[b_{\omega} \, , b_{\omega'}] = 0 = [b_{\omega}^{\dagger} \, , b_{\omega'}^{\dagger}] \, ,
\\
[a_{\Omega} \, , a^{\dagger}_{\Omega'}] = \delta(\Omega - \Omega') \, ,
&\qquad
[a_{\Omega} \, , a_{\Omega'}] = 0 = [a_{\Omega}^{\dagger} \, , a_{\Omega'}^{\dagger}] \, ,
\end{align}
and similarly for the ladder operators associated to the ingoing modes. 
The Unruh vacuum is defined by
\begin{align}
a_{\Omega} | 0 \rangle = 0 
\qquad 
\forall \ \Omega > 0
\, .
\label{Unruh-vacuum}
\end{align}
That is, there are no outgoing particles in the near-horizon region for freely falling observers.

Identifying the field $\phi$ in eqs.~\eqref{phi-b} and~\eqref{phi-a} in the near-horizon region where the areal radius $r$ is approximately a constant, one finds the Bogoliubov transformation
\begin{align}
b_{\omega} = \int_0^{\infty} d\Omega \, 
\bigl( \alpha_{\omega\Omega} \, a_{\Omega} + \beta_{\omega\Omega} \, a^{\dagger}_{\Omega}\bigr) 
\, ,
\qquad
b^{\dagger}_{\omega} = \int_0^{\infty} d\Omega \, 
\bigl( \alpha^{\ast}_{\omega\Omega} \, a^{\dagger}_{\Omega} + \beta^{\ast}_{\omega\Omega} \, a_{\Omega} \bigr) 
\, ,
\label{Bogoliubov}
\end{align}
where the Bogoliubov coefficients are~\cite{Hawking:1975vcx}
\begin{align}
\alpha_{\omega\Omega} 
&\equiv \frac{1}{2\pi} \sqrt{\frac{\omega}{\Omega}} 
\int_{-\infty}^{\infty} du \, e^{i\omega u \, - \, i\Omega U(u)}
\simeq \frac{a}{\pi} \sqrt{\frac{\omega}{\Omega}}
\left( 2a\Omega \right)^{i2a\omega} e^{\pi a\omega} \, \Gamma(-i2a\omega) \, ,
\label{alpha}
\\
\beta_{\omega\Omega} 
&\equiv \frac{1}{2\pi} \sqrt{\frac{\omega}{\Omega}} 
\int_{-\infty}^{\infty} du \, e^{i\omega u \, + \, i\Omega U(u)}
\simeq \frac{a}{\pi} \sqrt{\frac{\omega}{\Omega}}
\left( 2a\Omega \right)^{i2a\omega} e^{- \pi a\omega} \, \Gamma(-i2a\omega) \, ,
\label{beta}
\end{align}
with $U(u)$ given in eq.~\eqref{eq:U-u}.

\subsubsection{Hawking Radiation in Low-Energy Effective Theory}
\label{sec:HR-LEET}

To discuss the time dependence of Hawking radiation, consider the detection of a Hawking particle in a given wave packet
\be
\psi(u) \equiv \int_0^{\infty} \frac{d\omega}{\sqrt{\omega}} \, 
f_{\omega_0}(\omega) \, e^{- i \omega (u - u_0)} \, ,
\label{wavepacket}
\ee
which is centered around the retarded time $u = u_0$. The annihilation operator for this particle state is
\be
b_{\psi} \equiv \int du \, 
\psi^{\ast}(x)\hskip-0.2em\stackrel{\leftrightarrow}{\del_u} \hskip-0.2em (r \phi(x))
= \int_0^{\infty} d\omega \, f^{\ast}_{\omega_0}(\omega) \, e^{- i\omega u_0} b_{\omega} \, ,
\ee
and the corresponding creation operator is
\be
b^{\dag}_{\psi} \equiv
\int_0^{\infty} d\omega \, f_{\omega_0}(\omega) \, e^{i\omega u_0} b_{\omega}^{\dagger} \, .
\label{b-dag}
\ee
The profile function $f_{\omega_0}(\omega)$ is normalized by
\be
\int_0^{\infty} d\omega \, |f_{\omega_0}(\omega)|^2 = 1 \, ,
\ee
so that
\be
[b_{\psi} \, , b^{\dagger}_{\psi}] = 1 \, .
\label{bbd=1}
\ee

To confirm the thermal spectrum of Hawking radiation, we need the profile function $f_{\omega_0}(\omega)$ to have a narrow width $\Delta \omega \ll \omega_0 \sim 1/a$, where $\omega_0$ is the central frequency of the wave packet. 
As a result, we can approximate 
\be
\int_0^{\infty} d\omega \, f_{\omega_0}(\omega) \, h(\omega)
\simeq 
\int_0^{\infty} d\omega \, f_{\omega_0}(\omega) \, h(\omega_0)
\ee
for a generic smooth function $h(\omega)$ with a width $\gg \Delta\omega$.

The number operator for particles in the state $\psi$ is
\be
n_{\psi} \equiv b_{\psi}^{\dagger} b_{\psi} \, .
\ee
Using eqs.~\eqref{Bogoliubov} and~\eqref{b-dag}, 
the expectation value of $n_{\psi}$ in the Unruh vacuum 
can be obtained as
\begin{align}
\langle 0 | n_{\psi} | 0 \rangle
&= 
\int_0^{\infty} d\omega \int_0^{\infty} d\omega' 
f_{\omega_0}(\omega) f^{\ast}_{\omega_0}(\omega')
\int_{0}^{\infty} d\Omega \, 
\beta^{\ast}_{\omega\Omega} \, \beta_{\omega'\Omega}
\nn \\
&\simeq
\int_0^{\infty} d\omega \int_0^{\infty} d\omega' 
f_{\omega_0}(\omega) f^{\ast}_{\omega_0}(\omega') \, 
\frac{a^2}{\pi^2} \, \omega_0 \, 
e^{- 2 \pi a \omega_0} \left|\Gamma(i2a\omega_0)\right|^2
\int_{0}^{\infty} \frac{d\Omega}{\Omega} \, 
(2a\Omega)^{- i2a(\omega - \omega')} 
\nn \\
&\simeq
\frac{1}{e^{4\pi a\omega_0} - 1}
\int_0^{\infty} d\omega \left|f_{\omega_0}(\omega)\right|^2
= 
\frac{1}{e^{4\pi a\omega_0} - 1} \, ,
\label{n-VEV}
\end{align}
which is precisely the Planck distribution at the Hawking temperature~\eqref{Hawking-Temperature}. We have thus completed the derivation of Hawking radiation in the low-energy effective theory.

\subsubsection{UV Cutoff}
\label{sec:cutoff}

We show in this section that the Hawking radiation 
is turned off around the scrambling time $u_{scr}$~\eqref{scrambling-time} 
if there is a UV cutoff at the Planck scale~\cite{Barman:2017vqx, Ho:2022gpg}.

Truncating the lightcone momentum at $\Lambda_{\Omega} = \ell_p^{-1}$ leads to a modification of eq.~\eqref{n-VEV} as
\begin{align}
\langle 0 | n_{\psi} | 0 \rangle
&\simeq
\int_0^{\infty} d\omega \int_0^{\infty} d\omega' 
f_{\omega_0}(\omega) f^{\ast}_{\omega_0}(\omega') \, 
\frac{a^2}{\pi^2} \, \omega_0 \, 
e^{- 2 \pi a \omega_0} \left|\Gamma(i2a\omega_0)\right|^2
\int_{0}^{\ell_p^{-1}} \frac{d\Omega}{\Omega} \, 
(2a\Omega)^{- i2a(\omega - \omega')} 
\nn \\
&\simeq
\frac{1}{2\pi}
\frac{\omega_0}{e^{4 \pi a \omega_0} - 1} 
\int_{-\infty}^{u_{scr}/2} du \, 
\bigl| \psi(u) \bigr|^2 \, .
\label{n-VEV-1}
\end{align}
Note that the scrambling time $u_{scr}$~\eqref{scrambling-time} 
appears in the upper bound of the $u$-integration
as a consequence of applying the change of variable 
$\Omega \mapsto u = 2 a \log \left( a \Omega \right)$ to the $\Omega$-integration.

We observe that the number expectation value of Hawking particles in the state $\psi$
is characterized by a Planck distribution 
at the same Hawking temperature~\eqref{Hawking-Temperature}
but with an amplitude that depends on $u_0$ through the wave function $\psi(u)$~\eqref{wavepacket}.
For a wave function $\psi(u)$ that is localized in a region $|u - u_0| \lesssim \mathcal{O}(a)$ centered around $u = u_0$, the $u$-integral in eq.~\eqref{n-VEV-1} essentially vanishes if $u_{scr}/2$ is much smaller than $u_0$. Therefore, there is a large suppression of the magnitude of Hawking radiation after the scrambling time:\footnote{For a derivation of this outcome in 2D conformal field theory, see appendix B in ref.~\cite{Ho:2022gpg}.}
\be 
\langle 0 | n_{\psi} \ket{0}
\to 0
\quad 
\text{for}
\quad 
u_0 - u_{scr}/2 \gg a
\, .
\ee 

From this result, it becomes clear that Hawking radiation after the scrambling time stems from trans-Planckian modes with respect to the Kruskal retarded time $U$.
Different UV models may in principle predict different behaviors of Hawking radiation at late times.
This is a manifestation of the trans-Planckian problem of Hawking radiation~\cite{tHooft:1984kcu,Jacobson:1991gr}.

In theories with an absolute minimal length scale (such as a lattice theory), Lorentz symmetry only emerges in the low-energy limit, and the minimal length implies a UV cutoff on both the frequency and momentum of quantum modes. 
The derivation here implies that Hawking radiation ceases around the scrambling time in all such theories.

The discussion here can be generalized to a class of theories 
that admit modified canonical commutation relations of the form~\cite{Ho:2022gpg}
\begin{align}
[a_{\Omega} \, , a^{\dagger}_{\Omega'}] &= g(\Omega) \, \delta(\Omega - \Omega') \, ,
\\
[a_{\Omega} \, , a_{\Omega'}] &= [a^{\dagger}_{\Omega} \, , a^{\dagger}_{\Omega'}] = 0 \, ,
\end{align}
where $g(\Omega)$ characterizes the deformation in a given theory.
The model with a UV cutoff $\Omega \leq \Lambda_{\Omega}$
considered above corresponds to the specific case
where $g(\Omega)$ is given by the step function $\Theta(\Lambda_{\Omega} - \Omega)$.

In fact,
as long as one assumes the Unruh vacuum~\eqref{Unruh-vacuum} 
as well as the Bogoliubov transformation~\eqref{Bogoliubov}--\eqref{beta},
it can be shown (see appendix C in ref.~\cite{Ho:2022gpg}) that 
\be
\langle 0 | n_{\psi} | 0 \rangle \simeq 
\frac{1}{2\pi}
\frac{\omega_0}{e^{4 \pi a \omega_0} - 1} \,
\langle 0 | [b_{\psi} \, , b^{\dagger}_{\psi}] | 0 \rangle \, .
\ee
In other words, the standard Hawking radiation persists 
as long as a particle with the wave function 
$\psi(u)$ can be defined, 
meaning there exists a corresponding pair of creation and annihilation operators 
that satisfy the canonical commutation relation~\eqref{bbd=1}. 
Conversely, if Hawking radiation is absent, 
it is simply because no such particles can be defined.

This finding demonstrates the robustness of Hawking radiation:
Hawking particles are not detected only if such particles cannot be defined.
If a black hole ceases to emit Hawking radiation after the scrambling time,
it would imply that a distant observer should also be unable to create 
a particle with frequency $\sim \mathcal{O}(1/a)$ in the same direction.

It is worth noting that although previous studies~\cite{Unruh:1994je,Brout:1995wp,Corley:1996ar,Corley:1997pr,Himemoto:1999kd,Unruh:2004zk} on Hawking radiation using subluminal UV dispersion relations effectively introduced a UV cutoff on the frequency in the freely falling frame, these studies did not report significant deviations from conventional Hawking radiation. 
The key distinction is that the modified dispersion relations considered in these analyses did not exclude the presence of degrees of freedom with trans-Planckian momenta.

\section{Information Paradox}
\label{sec:Information-Problem}

The information paradox has different meanings at different times for different people.
A simple way to phrase the question is this --- {\em ``Is the evaporation of a black hole analogous to burning a piece of coal?''}
In principle, a low-energy effective theory can describe the transfer of information from a piece of coal to radiation with some residual information remaining in the ashes (the remnant) without violating unitarity, provided that the temperature stays well below the theory's cutoff scale but high enough to tear down the structure of the coal and turn it into radiation or ashes.
However, this picture does not apply to the traditional model of black holes.
The key difference is the absence of a high enough temperature to tear down the collapsing matter into radiation and ashes, in contrast to the hot burning process of coal.

\begin{figure}[t]
\centering
\includegraphics[scale=0.6]{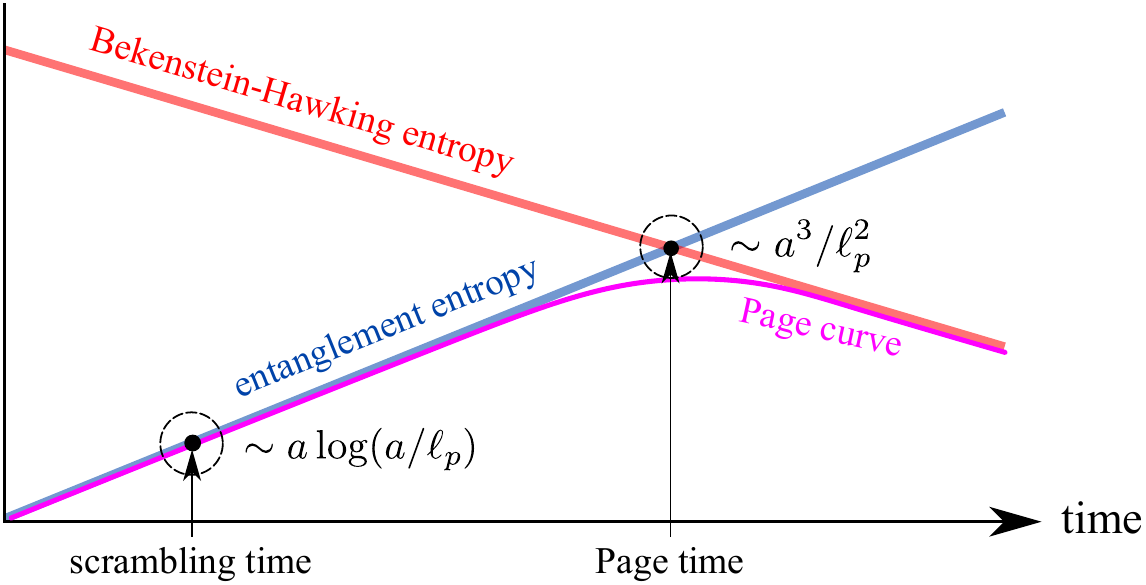}
\caption{\label{fig:page}
Since the Unruh vacuum is a pure state, Hawking particles outside the horizon are entangled with their partners inside the horizon.
Hence, the entanglement entropy of the black hole (blue curve) increases over time.
On the other hand, the Bekenstein-Hawking entropy (red curve) decreases as the black hole evaporates.
There is a conflict at the Page time when the entanglement entropy of the black hole becomes larger than its thermodynamic entropy.
If Hawking radiation is analogous to the radiation from burning a piece of coal, the entropy of the black hole would follow the Page curve \cite{Page:1993df,Page:1993wv}.
}
\end{figure}

But what is wrong with keeping the information inside the black hole?
A notable problem is the contradiction with the Bekenstein-Hawking entropy~\cite{Bekenstein:1973ur,Bekenstein:1974ax}
\be
\mathcal{S}_{\mathrm{BH}} = \frac{\text{horizon area}}{4 G_N} \, ,
\label{BH-entropy}
\ee
which is believed to give the thermodynamic entropy of a black hole.
As the black hole radiates, the entanglement entropy of the black hole increases over time.
On the other hand, the Bekenstein-Hawking entropy (as a bound on the total entropy of the black hole) decreases with time.
There is a conflict at the Page time (see fig.~\ref{fig:page}) when the entanglement entropy exceeds the Bekenstein-Hawking entropy.

The Bekenstein-Hawking entropy and the holographic principle~\cite{tHooft:1993dmi, Susskind:1994vu, Maldacena:1997re, Witten:1998qj} are supported by extensive evidence (e.g. \cite{Strominger:1996sh, Maldacena:1997re, Witten:1998qj, Gubser:1998bc, Dabholkar:2014ema}).
It leads to a sharp conflict with Hawking radiation in the low-energy effective theory after the Page time.
An explanation of the paradox including recent developments can be found in other chapters in this book as well as ref.~\cite{Almheiri:2020cfm}.

\subsection{Is UV Physics Relevant?}
\label{sec:Why-UV}

There would be no immediate paradox if we did not believe in the low-energy effective-theoretic derivation of Hawking radiation.
Here we recall some of the arguments for the folklore that Hawking radiation can be reliably calculated in a low-energy effective theory.
We will also explain the reasons why these arguments are in fact invalid.

\subsubsection{Local Lorentz Symmetry}
\label{sec:LLS}

As the Hawking temperature is given by eq.~\eqref{Hawking-Temperature}, a typical Hawking particle is at a low frequency $\omega \sim \mathcal{O}(1 / a)$ for a distant observer.
But for Hawking particles detected at late times, this frequency can be extremely large from the viewpoint of a freely falling observer due to the infinite blue shift near the horizon.\footnote{Even though Hawking radiation is said to emerge at a place away from the horizon where $r - a \sim \mathcal{O}(a)$, theoretically, 
unitarity allows us to consider the inverse time evolution of a Hawking particle backward in time to a point very close to the horizon.}
Naively, this suggests that a UV theory might be required to describe the origins of Hawking radiation, particularly the radiation quanta that are emitted at late times.

However, frequency is not a local Lorentz invariant. 
Any frequency can be made arbitrarily high or low through a local Lorentz boost. 
This is why we can still use the Standard Model to describe ultra-high-energy cosmic rays ($\sim 10^{6} \; TeV$) when they collide with particles on Earth. 
A UV theory is needed only when Lorentz invariants --- such as the center-of-mass energy of a collision --- reach trans-Planckian scales.
As a result, this so-called ``trans-Planckian problem'' of Hawking radiation is not always taken seriously.

\subsubsection*{Existence of trans-Planckian Lorentz-invariants}
\noindent

Nevertheless, if one considers the scattering process between an outgoing Hawking particle
and a dust particle in the collapsing matter (or the back-reacted geometry due to the quantum energy-momentum tensor --- see sec.~\ref{sec:CBR}), the center-of-mass energy can be extremely large.
When the center-of-mass energy exceeds the Planck energy, the effective theory is expected to break down on predictions about Hawking radiation~\cite{Ho:2020cbf,Ho:2020cvn,Ho:2021sbi,Ho:2022gpg}.

More explicitly, a particle in the infalling matter must have an ingoing momentum 
\be
P_V \gtrsim \mathcal{O}(1/a)
\ee
to fall into the black hole, since it cannot fit inside the horizon if its wavelength is much longer than the Schwarzschild radius.
At the same time, a Hawking particle with the characteristic frequency $\sim \mathcal{O}(1/a)$
has an outgoing momentum
\be
P_U \sim \mathcal{O}\left(\frac{1}{a} \, e^{u/2a}\right)
\ee
in terms of the Kruskal coordinate $U$ due to the blue shift~\eqref{eq:blue-shift}.
Thus, the center-of-mass energy between the infalling matter and a Hawking particle emitted around the retarded time $u_0$ can be estimated as
\be
s^2 \sim P_V P_U 
\sim \frac{1}{a^2} \, e^{u_0/2a} \, .
\label{com-energy}
\ee
This Lorentz-invariant quantity becomes trans-Planckian when
\be
u_0 \geq u_{scr} \, ,
\label{2uscr}
\ee
where the scrambling time $u_{scr}$ is given in eq.~\eqref{scrambling-time}. We will see in sec.~\ref{sec:UV-physics} that this is precisely the condition for low-energy effective theories to break down regarding their predictions of Hawking radiation.

\subsubsection{Nice-Slice Argument}
\label{sec:Nice-Slice-Argument}

The \emph{nice-slice argument}~\cite{Lowe:1995ac, Polchinski:1995ta} is often quoted as the reason why UV physics should be irrelevant to Hawking radiation.
The essence of the argument is that, since all intrinsic and extrinsic curvatures are 
of order $\mathcal{O}(1/a^2)$ or less, 
in the Hamiltonian formalism, the adiabatic theorem ensures that 
only degrees of freedom with low energies $\sim \mathcal{O}(1 / a)$ can be excited from the ground state.
Therefore, assuming that the initial state is the Minkowski vacuum, throughout the spacetime region where nice-slices can be defined, there would be no high-energy excitations.
(See sec.~\ref{sec:nice-slices} for an explicit construction 
of nice slices in the Schwarzschild spacetime.)

The use of low-energy effective theories is thus justified to describe the black-hole geometry.
For instance, if one adds higher-derivative corrections to the Einstein-Hilbert action, 
it will only lead to perturbative corrections to the energy-momentum tensor.
Hence, $\mathcal{O}(1)$-corrections are not expected.

There is thus the widely held belief that low-energy effective theories accurately describe Hawking radiation until the black hole's characteristic energy scale $1/a(t)$ becomes too high for them to apply. 
This suggests that an effective theory valid up to $1 \, TeV$ would suffice to describe the decay of a supermassive black hole of $10^{10}$ solar masses until its Schwarzschild radius shrinks to approximately $a \sim 1 \, TeV^{-1} \sim 10^{-19} \; m$, which is a tiny fraction $\sim 10^{-22}$ of its initial value.


\subsubsection*{Loopholes in nice-slice argument}
\noindent

Consider the time evolution of the Minkowski vacuum in the Minkowski spacetime in an effective theory.
There are nice slices that foliate the full Minkowski space, 
and the effective-theory prediction that the Minkowski vacuum persists under time evolution is reliable.
But for any Planck-scale measurement, say, the detection of particles in a region smaller than the Planck length, the prediction of the effective theory is still unreliable.

Therefore, the question is whether the detection of a Hawking particle is a low-energy event.
If it is a high-energy event, the nice-slice argument does not justify its prediction by a low-energy effective theory.
From the viewpoint of a distant observer, the energy of the Hawking particle is $\sim \mathcal{O}(1/a)$, and the use of effective theories seems to be justified.
On the other hand, the detection of a Hawking particle can be viewed as a ``delayed measurement''\footnote{Here we make the analogy with Wheeler's notion of ``delayed choice''
of the double-slit experiment~\cite{Wheeler:78}.}
of a tiny region of the vacuum state in the near-horizon region.
We can extrapolate the wave packet of a detected Hawking particle backward in time to the near-horizon region, and ask whether the detection of a particle with such a wave packet can be reliably described in the low-energy effective theory in the near-horizon region.

For a Hawking particle detected at a sufficiently late time ($u \geq u_{scr}$), although the frequencies themselves are not Lorentz-invariant, they are associated with a trans-Planckian Lorentz-invariant, as was shown above in eq.~\eqref{com-energy}.
In other words, the nice-slice argument is valid for claiming that the Unruh vacuum in the low-energy effective theory remains approximately the same state in the low-energy effective theory without becoming highly excited.
That said,
we still require UV theories for the derivation of Hawking radiation.

It was proposed in ref.~\cite{Mathur:2000ba} that, even when all intrinsic and extrinsic curvatures are small, quantum gravity effects become important when the hypersurfaces of a foliation are excessively stretched.
The trans-Planckian Lorentz-invariant quantity~\eqref{com-energy} discussed here can be understood as a physical explanation of this criterion for the relevance of quantum gravity effects.

\subsection{Difficulty in Information Transfer}
\label{sec:difficulty}

The derivation of Hawking radiation in sec.~\ref{sec:HR} suggests that there is no information transfer between Hawking radiation and the collapsing matter in the low-energy effective theory.
But even if we consider a complicated UV theory for Hawking radiation, there is a list of reasons why it is still hard for Hawking radiation to carry a substantial portion of the information in the collapsing matter.
We recount some of the well-known reasons here.


\hide{
The standard derivation~\cite{Hawking:1975vcx} of Hawking radiation solely relies on the spacetime geometry, as it was originally conducted in the free field theory.
Different initial states with the same energy-momentum tensor for the collapsing matter necessarily lead to the same Hawking radiation.
At the level of free fields, Hawking radiation can at most carry information about the energy-momentum tensor, without other details of the collapsing matter.

Even in a weakly interacting low-energy effective theory, the interaction between the collapsing matter and the radiation field would only introduce perturbative corrections to the free field result.
This has been verified for renormalizable interactions~\cite{Leahy:1983vb, Unruh:1983ac, Frasca:2014gua}.
However, we will see in sec.~\ref{sec:UV-physics} that non-renormalizable interactions
would in general introduce $\mathcal{O}(1)$ corrections to Hawking radiation.
}

\subsubsection{No-Cloning Theorem}

As long as the equivalence principle remains valid, the infalling matter should be left undisturbed as it falls across the horizon.
Subsequently, 
the no-cloning theorem~\cite{Wootters:1982zz, Dieks:1982dj} does not allow its information to be copied to another quantum system.

The idea of black hole complementarity~\cite{Susskind:1993if} proposed that the cloning of information is allowed if it cannot be verified by a single observer~\cite{Susskind:1993mu}.
However, the compatibility of black hole complementarity with both the equivalence principle and the unitarity of the evaporation process remains an open question~\cite{Almheiri:2012rt,Bousso:2023kdj}.

\subsubsection{Firewall Versus Nonlocality}

Apart from the problem mentioned in sec.~\ref{sec:Apparent-Horizon} that all the information needs to be transmitted within an extremely short period of time~\cite{Ho:2019pjr, Ho:2019qiu},  there is also a problem with how information can be retrieved from the collapsing matter.

Normally, to probe information at a given length scale $L$, we need a wave packet of a shorter wavelength, and thus a momentum larger than $1/L$. Hence, to retrieve the information about the wave functions of the quarks in a nucleus, for example, we need a probe at the energy scale of $GeV$ or larger.
This background of probes at high energies implies a firewall that burns down all ingoing particles at the horizon.
The evaporation is then similar to the burning of a piece of coal,
with both the equivalence principle and the adiabatic principle being violated.

Alternatively, information transfer may occur through 
a nonlocal mechanism~\cite{Giddings:2012gc, Giddings:2013noa}. 
The {\em ER = EPR} proposal \cite{Maldacena:2013xja} 
and the mechanism underlying the entanglement islands~\cite{Penington:2019npb,Almheiri:2019psf,Almheiri:2019hni,Almheiri:2019yqk,Penington:2019kki,Almheiri:2019qdq} fall into this category. 
Naively, these approaches seem to have the advantage of 
circumventing the AMPS firewall~\cite{Almheiri:2012rt}. 
However, it has been argued in ref.~\cite{Marolf:2013dba} that gauge/gravity duality might still imply a firewall for large AdS black holes --- one that cannot be evaded by merely identifying degrees of freedom inside the black hole with those outside.

In any case, if nonlocal physics is important for the discussion of information in Hawking radiation, why should we believe the prediction of Hawking radiation based on a (local) low-energy effective theory in the first place?

\section{Breakdown of Low-Energy Effective Theory}
\label{sec:UV-physics}

In this section,
we examine the validity of the low-energy effective theory in describing the black-hole radiation process~\cite{Ho:2020cbf, Ho:2020cvn, Ho:2021sbi}. 
We shall compute the transition amplitudes of particle creation on top of Hawking radiation due to higher-derivative non-renormalizable interactions during gravitational collapse. 
We show that these amplitudes increase exponentially over time, 
becoming significant when the collapsing matter is roughly 
at the order of a Planck length from the horizon. 
This exponential growth indicates that the 
effective-theory description of Hawking radiation 
breaks down at the scrambling time, 
necessitating Planckian physics 
to understand black-hole evaporation~\cite{Ho:2020cbf, Ho:2020cvn, Ho:2021sbi}.

\subsection{Particle Creation by Higher-Derivative Interactions}

The original derivation of Hawking radiation can be understood as the transition amplitude of particle creation at the free-field level. Later, it was shown that the corrections due to renormalizable interactions only modify Hawking radiation perturbatively~\cite{Leahy:1983vb, Unruh:1983ac, Frasca:2014gua}. However, non-renormalizable interactions have not been considered until more recently in~refs.~\cite{Ho:2020cbf, Ho:2020cvn, Ho:2021sbi}. It turns out that higher-derivative non-renormalizable interactions between the collapsing matter and the radiation field induce particle creation that overlaps with the spectrum of Hawking radiation~\cite{Ho:2020cbf, Ho:2020cvn, Ho:2021sbi}.

The particle creation due to higher-derivative non-renormalizable interactions is suppressed by certain negative powers of the Planck mass, but its amplitude grows exponentially with time for particles with a given frequency $\sim \mathcal{O}(1/a)$. 
After the scrambling time, this effect dominates over ordinary Hawking radiation, and the contributions from interactions containing derivatives of higher orders are even larger~\cite{Ho:2020cbf, Ho:2020cvn, Ho:2021sbi}. 
As a result, the perturbation theory breaks down, and the rate of particle creation can only be reliably derived from the UV theory.

With the particle creation viewed as a scattering process between the radiation field initially in the Minkowski vacuum and the collapsing matter, there is a Lorentz-invariant center-of-mass energy~\eqref{com-energy} that grows exponentially with time, as we explained in sec.~\ref{sec:LLS}.
This is why the contributions of higher-derivative interactions grow exponentially over time and start to dominate at the scrambling time, when the collapsing matter is approximately $\mathcal{O}(\ell_p)$ away from the horizon~\cite{Ho:2020cbf,Ho:2020cvn,Ho:2021sbi}.
The remainder of this section shows this calculation explicitly.

\subsubsection{Higher-Derivative Non-Renormalizable Interactions}

In an effective theory, the action must include all interactions compatible with symmetries (see sec. 12.3 in ref.~\cite{Weinberg:1995mt}). As examples, we consider the following interaction terms~\cite{Ho:2020cbf,Ho:2020cvn,Ho:2021sbi}
\begin{align}
S_n &\equiv \int d^4 x \, \sqrt{-g} \, \frac{\lambda_n}{M_p^{4n-2}} \, 
R^{\mu_1\nu_1} \cdots R^{\mu_n\nu_n} 
\left(\nabla_{\mu_1}\cdots\nabla_{\mu_n}\phi\right)
\left(\nabla_{\nu_1}\cdots\nabla_{\nu_n}\phi\right),
\label{Sn}
\\
S_{mn} &\equiv \int d^4 x \, \sqrt{-g} \, \frac{g_n}{M_p^{2(m+n)}} \, 
g^{\mu_1\nu_1} \cdots g^{\mu_m\nu_m} g^{\lambda_1\rho_1} \cdots g^{\lambda_n\rho_n}
\times
\nn \\
&\qquad
\times
\left(\nabla_{\mu_1}\cdots\nabla_{\mu_m}\psi\right)
\left(\nabla_{\lambda_1}\cdots\nabla_{\lambda_n}\psi\right)
\left(\nabla_{\nu_1}\cdots\nabla_{\nu_m}\phi\right)
\left(\nabla_{\rho_1}\cdots\nabla_{\rho_n}\phi\right),
\label{Smn}
\end{align}
where $\lambda_n$ and $g_n$ are dimensionless coupling constants, $R^{\mu\nu}$ the Ricci curvature, 
$\phi$ the massless scalar field of Hawking radiation, 
and $\psi$ represents the field of the collapsing matter.

We note the following important features of these higher-derivative non-renormalizable interactions:
\begin{enumerate}
\item
In the absence of the collapsing matter, both interaction terms are irrelevant. (Recall that the Ricci tensor vanishes for the Schwarzschild solution.\footnote{At the next order in the $\ell_p^2/a^2$ expansion, the back-reaction of the quantum energy-momentum operator introduces a non-zero Ricci tensor outside the shell and leads to qualitatively the same result through the interaction~\eqref{Sn}. See sec.~\ref{sec:CBR}.})
\item
It is possible to create $\phi$-particles through these interactions.
\item
As non-renormalizable interactions, they are suppressed by negative powers of the Planck mass $M_p$. 
\item
$S_n$ and $S_{mn}$ lead to stronger interactions for higher energy modes and larger $n, m$ (higher-order derivatives).
\end{enumerate}

When a distant observer detects a particle in the radiation field $\phi$ at a given frequency $\omega \sim \mathcal{O}(1/a)$, it is equivalent to detecting a wave packet with a blue-shifted central frequency
\be
\Omega \sim \omega \, e^{u / 2a} 
\ee
with respect to the Kruskal retarded time $U$.
At a later time (larger $u$), wave packets with larger $\Omega$ would have stronger couplings with the background (Ricci tensor or matter field $\psi$) through the interaction terms $S_n$~\eqref{Sn} and $S_{mn}$~\eqref{Smn}.

This expectation can be explicitly verified via a standard $S$-matrix calculation of the transition amplitude from the initial state of the Unruh vacuum to the final state of multiple particles at large distances~\cite{Ho:2020cbf,Ho:2020cvn,Ho:2021sbi}. 
In the ensuing discussion, we shall focus specifically on the effects of the interaction term $S_n$~\eqref{Sn} for simplicity. 

Let us consider the final state obtained by adding two additional particles on top of the original Hawking radiation:
\be
|f\rangle \equiv
b^{\dag}_{\psi_1} b^{\dag}_{\psi_2} | 0 \rangle \, ,
\label{final-state}
\ee
where the creation operator 
corresponding to a wave packet has been defined in eq.~\eqref{b-dag},
and the vacuum $| 0 \rangle$ here refers to the Unruh vacuum~\eqref{Unruh-vacuum}. 
The wave packets $\psi_1$ and $\psi_2$ are assumed to have central frequencies $\omega_1 \sim \omega_2 \sim \mathcal{O}(1/a)$ in the same range as ordinary Hawking particles, and these outgoing wave packets are centered around a given retarded time $u_0$.

In the following, as we calculate the amplitude 
\be
{\cal M}_n \equiv i \, \langle f | S_n | 0 \rangle
\label{Mn}
\ee
of the transition from the Unruh vacuum to the final state $|f\rangle$~\eqref{final-state}, we shall pay extra attention to the time dependence of this particle-creation mechanism.

\subsubsection{Order-of-Magnitude Estimate}
\label{sec:M-estimate}

Let us start with a brief order-of-magnitude estimate of the $S$-matrix~\eqref{Mn}.

In the black-hole background described in sec.~\ref{sec:Traditional-Model}, the operator $S_n$~\eqref{Sn} involves $R_{VV}$, $\nabla_U$, $\phi$, and the spacetime integral. 
The Ricci curvature due to the collapsing matter is
\be
R^{UU} \sim R_{VV} \sim 8\pi G_N T_{VV} \sim \mathcal{O}\left(\frac{1}{a^2}\right) .
\label{RVV-0}
\ee
The covariant derivative $- i \nabla_U$ gives the momentum $P_U$ (or the frequency $\Omega$) conjugate to the Kruskal coordinate $U$. 
For a Hawking particle with frequency $\omega \sim \mathcal{O}(1/a)$ 
detected at $u = u_0$ at large distances, 
each factor of the covariant derivative $\nabla_U$ contributes
\be
\nabla_U \sim 
P_U = \Omega
\simeq 
\omega \, e^{u_0 / 2a} \sim \frac{1}{a} \, e^{u_0 / 2a} \, .
\label{nabla_U}
\ee
The normalization of the field operator $\phi$ for the Hawking particles is $\phi \sim \mathcal{O}(1/a)$. Finally, assuming that the wave packet of the Hawking particle has a width $\Delta u \sim \mathcal{O}(a)$ along the $u$ direction and that the thickness of the collapsing matter is $\Delta v \sim \mathcal{O}(a)$, the spacetime integral $\int dUdV$ contributes a factor of $a^2 \, e^{- u_0/2a}$.

The amplitude of the transition from the Unruh vacuum to the final state~\eqref{final-state} due to the interaction term~\eqref{Sn} is therefore estimated as 
\begin{align}
{\cal M}_n
\sim 
\frac{\lambda_n}{M_p^{4n-2}} \left[a^4 e^{- u_0 / 2a}\right]
\frac{1}{a^{2n}} \left(\frac{1}{a} \, e^{u_0 / 2a}\right)^{2n} \frac{1}{a^2}
\sim \lambda_n \left(\frac{1}{a^2 M_p^2} \, e^{u_0 / 2a}\right)^{2n-1} \, .
\label{Mn-final}
\end{align}

For a coupling $\lambda_n \sim \mathcal{O}(1)$\footnote{In the $\ell_p^2/a^2$ expansion, a small constant like $0.0001$ is still considered $\mathcal{O}(1)$.}, this amplitude~\eqref{Mn-final} is large when
\be
e^{u_0 / 2a} > a^2 M_p^2 \, ,
\label{condition}
\ee
which is equivalent to the condition 
\be
u_0 > u_{scr}
\label{scr-time}
\ee
on the retarded time $u_0$ when the Hawking particles are detected.\footnote{
There is a longstanding observation~\cite{Veneziano:2001ah, Dvali:2007hz, Caron-Huot:2024lbf} 
that in the presence of a large number $N$ of matter fields 
in a $d$-dimensional effective field theory,
the quantum gravity cutoff scale $\Lambda \lesssim N^{- 1 / (d - 2)} M_p$
is parametrically lower than the Planck scale.
This suggests that the breakdown of the 
effective field theory description of Hawking radiation
at the scrambling time $u_{scr}$
is a conservative estimate,
as this breakdown may occur even earlier.
}

Note that eq.~\eqref{scr-time} coincides with the condition~\eqref{2uscr} for an ingoing quantum with $P_V \sim 1/a$ and a Hawking particle with $\omega \sim 1/a$ to have a trans-Planckian center-of-mass energy. In the interaction term $S_n$~\eqref{Sn}, the ingoing quantum is represented by the Ricci tensor $R_{VV}$, while $\nabla_U$ extracts the momentum $P_U$ of the Hawking particle. Thus, the center-of-mass energy squared is reflected in the invariant quantity $R_{VV} \Omega_U^2$ in $S_n$. The calculation above shows that the contribution of the non-renormalizable interaction $S_n$ becomes large precisely when the center-of-mass energy of this scattering process becomes trans-Planckian.

In short, eq.~\eqref{scr-time} indicates that 
after the scrambling time $u_{scr}$, non-renormalizable interactions dominate over renormalizable interactions in terms of their contributions to particle creations in the spectrum of Hawking radiation, and thus the perturbative approximation fails.

\subsubsection{Derivation of the $S$-Matrix}
\label{sec:Detailed-Calculation}

Here we present more details about the calculation of the $S$-matrix~\eqref{Mn} to confirm the order-of-magnitude estimate~\eqref{Mn-final} above. 
The reader may skip this section without missing the physical picture.

The transition amplitude ${\cal M}_n$~\eqref{Mn} 
from $|0\rangle$ to $|f\rangle$~\eqref{final-state} is
\begin{align}
{\cal M}_n &\equiv
i \int d^4 x \, \sqrt{-g} \, \frac{\lambda_n}{M_p^{4n-2}} \, 
R^{\mu_1\nu_1} \cdots R^{\mu_n\nu_n}  \, 
\langle f |
\left(\nabla_{\mu_1}\cdots\nabla_{\mu_n}\phi\right)
\left(\nabla_{\nu_1}\cdots\nabla_{\nu_n}\phi\right)
| 0 \rangle
\, .
\label{Mn-1}
\end{align}
For a collapsing spherical null shell of radius $R_s(U)$, the Ricci tensor vanishes outside the shell in the Schwarzschild spacetime,
and it also vanishes inside the shell in the Minkowski spacetime. 
The non-vanishing component of the Ricci tensor is given by
\be
R_{VV} = \frac{a}{R_s^2(U)} \, \delta_d(V - V_s) \, ,
\ee
where $\delta_d$ is a smooth function that vanishes outside the range $V \in (V_s - d, V_s)$ for a collapsing null shell of thickness $d$ in the $V$ direction, and $V_s$ denotes the location of the outer surface of the collapsing shell. We shall assume that $d \leq a$ so that we do not have to worry about self-collision inside the shell before it enters the horizon. 
Recall from the arguments in sec.~\ref{sec:NHG} that we can also take $V_s = 2a$ without loss of generality.

As other components of the Ricci tensor vanish for a null shell, the first-order tree-level contribution of the interaction term~\eqref{Sn} to the transition from the Unruh vacuum to the final state~\eqref{final-state} can be written as 
\begin{align}
{\cal M}_n &=
i \left(-2 \right)^n 
\frac{\lambda_n}{M_p^{4n-2}} \int d^4 x \, \sqrt{-g} \, \left(R_{VV}\right)^n
\langle f |
\left(\nabla_U^n \phi\right)^2
| 0 \rangle
\, .
\label{Mn-2}
\end{align}
The matrix element $\langle f | \left(\nabla_U^n \phi\right)^2 | 0 \rangle$ above 
is essentially the product of two terms of the form $\langle 0 | b_{\psi_i} \left(\nabla_U^n \phi\right) | 0 \rangle$ for $i = 1, 2$. 
Using the formulas in sec.~\ref{sec:Bogo}, we can derive 
\begin{align}
\langle 0 | b_{\psi} \left(\nabla_U^n \phi\right) | 0 \rangle
\simeq 
\frac{1}{4\pi \left( 2a \right)^n R_s(U(u))} \, F_{\psi}(u-u_0) \, e^{nu / 2a}
\, ,
\label{eq1}
\end{align}
where $U(u)$ \eqref{eq:U-u} is the Kruskal coordinate $U$ expressed as a function of $u$, and
\begin{align}
F_{\psi}(u) &\equiv
\int_0^{\infty} \frac{d\omega}{\sqrt{\omega}} \,
\frac{f^{\ast}_{\omega_0}(\omega) A_n(\omega)}{1 - e^{-4\pi a\omega}} \, 
e^{i \omega u} \, ,
\label{F}
\\
A_n(\omega) &\equiv
(n - 1 + 2ia\omega)(n - 2 + 2ia\omega) \cdots (1 + 2ia\omega) (2ia\omega) \, ,
\label{An}
\end{align}
with $f_{\omega_0}$ being the profile function 
in the definition~\eqref{b-dag} of the wave packet $\psi$.

As we are interested in Hawking radiation, we focus on a small $U$-interval (large $u_0$) when the collapsing shell is just about to enter the horizon, and thus $R_s(U) \simeq a$ is a good approximation. Using eqs.~\eqref{eq1}--\eqref{An}, we can simplify eq.~\eqref{Mn-2} as~\cite{Ho:2021sbi}
\begin{align}
{\cal M}_n &\simeq
\frac{i \lambda_n}{M_p^{4n-2}} \frac{(-1)^n}{2\pi d^{n-1}(2a)^n} \, G_{\psi_1\psi_2} \,
e^{(2n-1)u_0 / 2a} \, ,
\label{Mn-explicit}
\end{align}
where
\begin{align}
G_{\psi_1\psi_2} &\equiv
\int_{-\infty}^{\infty} du \, \frac{e^{(2n-1)u / 2a}}{a^{2n}} \, 
F_{\psi_1}(u) F_{\psi_2}(u)
\label{G}
\end{align}
is a constant depending on the explicit forms of the profile functions characterizing the wave packets $\psi_1$ and $\psi_2$ of the particles in the final state $\ket{f}$~\eqref{final-state}.

Notice that the factor $G_{\psi_1\psi_2}$ is independent of $u_0$, the time at which the particles are detected. 
As a result, 
the transition amplitude ${\cal M}_n$ has the time dependence of 
\be
{\cal M}_n \propto e^{(2n-1)u_0 / 2a} \, .
\label{Mn-exp}
\ee
The proportionality constant is determined by eqs.~\eqref{Mn-explicit} and~\eqref{G} for given profile functions defining the outgoing wave packets $\psi_1$ and $\psi_2$.
Assuming that all characteristic length scales in the problem (e.g. the width of the collapsing shell) is $\sim \mathcal{O}(a)$, dimensional analysis demands the order-of-magnitude of the proportionality constant to be $\sim 1/(aM_p)^{4n-2}$, in agreement with eq.~\eqref{Mn-final} derived above.

We refer to refs.~\cite{Ho:2020cbf,Ho:2020cvn,Ho:2021sbi} for more details about the calculation of the transition amplitudes of particle creation. The calculation for the interaction term \eqref{Smn} can be found in ref.~\cite{Ho:2021sbi}. 
We now proceed to comment on the generalization to 
generic higher-derivative interactions below in sec.~\ref{sec:generic-higher-derivative}.

\subsection{Implications of the $S$-Matrix Calculation}
\label{sec:loophole}

In the above, we have calculated the transition amplitude for the creation of particles on top of Hawking radiation due to a specific higher-derivative non-renormalizable interaction $S_n$~\eqref{Sn}. In the following, we comment on the generalizations of the result above to other interactions and its implications for the information paradox.

\subsubsection{Generic Higher-Derivative Interactions}
\label{sec:generic-higher-derivative}

It is now straightforward to extend the result above to generic higher-derivative interactions. For each derivative $\nabla_U$, we get an exponential factor $e^{u_0/2a}$ as in eq.~\eqref{nabla_U}. There is a suppressing factor of $e^{-u_0/2a}$ from the measure $dU$ in the integration. Therefore, for an interaction with $m$ derivatives, there is an overall exponential factor $e^{(m-1)u_0/2a}$ to enhance the production of Hawking particles at $u = u_0$.

As non-renormalizable interactions are always suppressed by $1/M_p$ to some power $k$, we need 
\be
u_0 > \frac{2k}{(m-1)} \, a \log\left(a/\ell_p\right)
\label{u-formula}
\ee
for the contribution to the amplitude of particle creation to be large. (For $S_n$, $k = 4n - 2$ and $m = 2n$ according to eq.~\eqref{Sn}.) Regardless of the numbers $k$ and $m$, as long as they are finite, the right-hand side is always parametrically the order of magnitude of the scrambling time $\sim u_{scr}$~\eqref{scrambling-time}.
As a result, one can conclude that essentially all higher-derivative interactions contribute to the breakdown of the low-energy effective theory beyond the scrambling time.

\subsubsection{Contribution From Back-Reaction}
\label{sec:CBR}

We mentioned in sec.~\ref{sec:Backreaction} that there is always a negative ingoing energy flux $\langle T_{VV} \rangle$~\eqref{TVV-bound} in the traditional model to balance the outgoing energy flux of Hawking radiation for energy conservation. Through the semi-classical Einstein equation~\eqref{Einstein-eq}, its back-reaction leads to a non-vanishing Ricci tensor 
\be
\langle R_{VV} \rangle \sim \mathcal{O}\left(\ell_p^2/a^4\right).
\ee
Although this is an order of $\ell_p^2/a^2$ smaller than the contribution~\eqref{RVV-0} from the collapsing matter, the higher-derivative interaction term~\eqref{Sn} gets non-trivial contributions from the bulk spacetime outside the collapsing shell. This leads to a larger number $k$ in eq.~\eqref{u-formula}, but for any finite $k$, it is still at the order of magnitude of the scrambling time when these higher-derivative interactions lead to $\mathcal{O}(1)$ corrections to Hawking radiation.


\subsubsection{UV Theories and Information Paradox}

Even if we have complete information about all non-renormalizable interactions, higher-order interaction terms $S_n$~\eqref{Sn} with larger $n$ have exponentially larger contributions to the amplitude~\eqref{Mn-final} of particle creation after the scrambling time. 
Therefore, a perturbative expansion of the UV theory in powers of $1/M_p$ breaks down. 
This is an $\mathcal{O}(1)$ effect that has to be properly dealt with prior to examining the information paradox. 
In this sense, one might be tempted to say that the information paradox is already resolved because the standard description of Hawking radiation using effective field theories is unreliable after the scrambling time.
On the other hand, as we discussed in sec.~\ref{sec:difficulty}, as long as the information is required to be transmitted into Hawking radiation, there are still puzzles even when UV theories are relevant.

\subsubsection{Possible Fates of Hawking Radiation}

Although the perturbative expansion of the amplitude 
displays an exponentially large transition rate of particle creation from higher-order terms, it does not necessarily imply that there is a large enhancement of Hawking radiation in a non-perturbative formulation of the UV theory. For instance, if we introduce an exponential suppression factor $e^{- \ell^2_p \, k^2}$ to all interaction terms in momentum space, the interactions with the background would be suppressed in the UV limit.

Given a self-consistent UV-complete theory that resolves the information paradox, there are two logical possibilities for its non-perturbative (in contrast with $1/M_p$ expansion) prediction of Hawking radiation:
\begin{enumerate}
\item
Some UV mechanism of particle creation dominates over Hawking radiation so that information can be transferred.
\item
Some UV mechanism suppresses Hawking radiation so that the evaporated mass is a negligible portion of the initial black hole mass. There is no need for information transfer.
\end{enumerate}

Naively, the creation of Hawking particles through trans-Planckian interactions between the collapsing matter and the radiation opens the window for information transfer from the collapsing matter to the radiation. However, if higher-derivative interactions are capable of producing trans-Planckian particles, there is in principle a danger of violating the adiabatic principle. In Minkowski space, energy-momentum conservation restricts the production of trans-Planckian particles. But in our expanding universe, we should probably have already seen such an effect. The absence of such events, or the validity of the adiabatic principle reflected in observations, is an indication that trans-Planckian particles are not produced this way.

In this sense, and also in the context of the discussions in sec.~\ref{sec:difficulty}, the information paradox persists even after demonstrating that 
Hawking radiation is modified at the $\mathcal{O}(1)$ level by UV physics. 
It remains problematic if we insist that (almost) all information about the collapsing matter 
is released through Hawking radiation.

Owing to this, the second option mentioned above presents a safer alternative. 
If all higher-derivative interactions align 
to cancel the original Hawking radiation before the Page time, 
the black hole is essentially classical and there would be no paradox. 
The only challenge is determining whether such a UV theory exists. 
We will examine two toy models that exhibit precisely this feature in sec.~\ref{sec:Hawking-Radiation-Turned-Off}.

\hide{
 “We would probably fry…”
(This is what Polchinski said at 1:02:10 of the video:
https://youtu.be/NL9PsYFPv1c?si=Z-70sLKGQToe8zNR
He believed that Hawking radiation does not depend on physics at high scales. See 1:01:45.) 
However, if the UV physics is not to replace the vacuum with an excited state (as the adiabatic principle hints at) but to remove the UV mode, there would be no conflict between Hawking radiation and cosmology.
}

\subsubsection{Firewall?}

Assuming that some UV mechanism dominates Hawking radiation with a large production rate of particles at large distances, does it imply a firewall around the horizon due to the large blue shift? If we trace the particle created by the interaction $S_n$~\eqref{Sn} backward to the near-horizon region, their wave packets are composed of extremely high-frequency modes in terms of the Kruskal coordinates. 
Therefore, naively speaking, 
the particle creation that we discussed in secs.~\ref{sec:M-estimate}--\ref{sec:Detailed-Calculation} implies a firewall around the horizon~\cite{Ho:2020cbf,Ho:2020cvn}. However, the situation could be more subtle than that~\cite{Ho:2021sbi}.

As we mentioned in sec.~\ref{sec:Uncertainty}, 
there is a large uncertainty $\Delta P_U > P_U$~\eqref{DPU>PU} in the momentum $P_U$ conjugate to the Kruskal coordinate $U$ for a Hawking particle with a well-defined momentum $P_u \sim \mathcal{O}(1/a)$. With a large uncertainty $\Delta P_U$, the requirement of energy conservation is relaxed (see sec. 4 of ref.~\cite{Ho:2021sbi}). This is why the Unruh vacuum can be turned into a multi-particle state with a large amplitude.

On the other hand, a state with a very large $\Delta P_U$ is not a typical particle state. To illustrate the idea, let us consider the following state
\be
|\Psi\rangle \equiv \frac{1}{\sqrt{\Lambda}} \int_0^{\Lambda} d\Omega \, |\Omega\rangle \, ,
\label{constant-spectrum}
\ee
where $\Lambda$ is a large UV cutoff of this superposition of energy eigenstates. This state has a large $\Delta P_U \sim \Lambda$ with a large central value $P_U = \Lambda/2$. Naively, it should be identified with a high-energy particle state. However, the probability of detecting such a particle at any given energy $\Omega_0$ over a width $\Delta\Omega_0$ is
\be
\frac{\Delta\Omega_0}{\Lambda} \, ,
\ee
which goes to $0$ as $\Lambda \rightarrow \infty$. 
For a state with the upper bound $\Lambda = M_p$, the probability of finding a particle with energy between $0$ and $1 \; TeV$ is as small as $10^{-16}$.

Nevertheless, the total energy flux of such a non-particle state becomes very large due to the large blue-shift factor after the scrambling time. This means that the adiabatic principle will be violated, and there is a firewall. To avoid the violation of the adiabatic principle, the contributions of the higher-derivative non-renormalizable interactions have to sum up with a large cancellation.

\subsubsection{Scrambling Time Versus Page Time}

The ``scrambling time'' was first named to refer to the lower bound on the order of magnitude of the minimal time it takes a quantum system to scramble (thermalize) information, and black holes are said to saturate the bound as ``fast scramblers''~\cite{Sekino:2008he}. 
For large black holes $(a \gg \ell_p)$, the scrambling time is of order $\mathcal{O}(a\log(a/\ell_p))$, which is orders of magnitude shorter than the Page time $\sim \mathcal{O}(a^3/\ell_p^2)$. 

While the information paradox becomes significant around Page time, new physics related to black-hole evaporation has typically been proposed to emerge at or near Page time. 
For example, loop corrections to correlation functions in a black-hole background exhibit secular growth~\cite{Akhmedov:2015xwa, Burgess:2018sou}, signaling the breakdown of perturbation theory around the Page time. 
Similarly, the firewall scenario~\cite{Almheiri:2012rt} is concerned with the Page time.

In contrast to these proposals, we propose below that Hawking radiation ceases around the scrambling time. 
The mass lost to Hawking radiation within the scrambling time is negligible, scaling as $\mathcal{O}(\log(a/\ell_p)/a)$, which is a tiny fraction $\sim \mathcal{O}(\ell_p^2\log(a/\ell_p)/a^2)$ of the black hole's total mass.

\section{UV Models of Hawking Radiation}
\label{sec:Hawking-Radiation-Turned-Off}

In this section, we investigate Hawking radiation in two UV-modified quantum field theories 
of massless fields that have been studied in the literature. 
The first model is based on the generalized uncertainty principle (sec.~\ref{sec:GUP}), and the second model is rooted in the universal UV suppression characteristic of string field theories (sec.~\ref{sec:StringyModel}). 
We demonstrate that in both models, Hawking radiation ceases around the scrambling time.

These models exhibit nonlocality, though restricted to trans-Planckian modes. While nonlocality in the IR is strongly constrained by observations, nonlocality in the UV appears to be a prevalent feature across quantum gravity theories. Both models incorporate a ``minimal length'' characteristic, as widely anticipated in quantum gravity theories. Furthermore, the nonlocalities in these models share the common feature of a larger spatial extension at higher momenta inspired by string theory.

\subsection{Nonlocality in String Theory}
\label{sec:QG}

To motivate the UV-modified models for Hawking radiation, we first discuss the nonlocality inherent in string theory. 

In the study of quantum gravity, there is the general expectation that spacetime as we know it breaks down at scales shorter than the Planck length. This implies that excitations with wavelengths shorter than the Planck length or string length\footnote{We will not distinguish the Planck length from the string length in this chapter for simplicity in discussion as they are of the same order in the $\ell_p^2/a^2$ expansion.} should be eliminated in quantum gravity, leading to some form of nonlocality. Indeed, nonlocality is a common feature in quantum gravity theories, while locality is expected only at low energies.

For example, studies of string scattering amplitudes in the high-energy limit showed that strings cannot probe distances shorter than the string length~\cite{Amati:1988tn, Fabbrichesi:1989ps, Konishi:1989wk, Guida:1990st}. To explore smaller distances, one is tempted to increase the momentum of the probe string, but this also leads to higher energy and a longer string extension.


As a manifestation of nonlocality in quantum gravity and string theory, the generalized uncertainty principle (GUP)~\cite{Amati:1988tn, Fabbrichesi:1989ps, Veneziano:1989fc, Konishi:1989wk, Guida:1990st} is proposed as
\be
\Delta x \, \Delta p \geq \frac{1}{2}\left(1 + \ell_s^2 \, \Delta p^2\right) ,
\label{GUP}
\ee
where $\ell_s = \sqrt{\alpha'}$ denotes the string length.\footnote{
Strictly speaking, there are two types of uncertainty at play here. 
The first term $1 / 2$ on the right-hand side of eq.~\eqref{GUP} arises from the standard uncertainty relation due to the wave nature of particles. The second term originates from the spatial extension characteristic of strings. 
When strings are used to probe spacetime geometry, both types of uncertainty contribute to the overall uncertainty in the spacetime geometry.
}
It is an extension of the Heisenberg uncertainty principle that effectively characterizes the nonlocality inherent in string theory. 
It also accounts for the presence of a minimal length scale $\Delta x \geq \ell_s$ as motivated by a variety of quantum gravity considerations~\cite{Maggiore:1993rv, Garay:1994en, Scardigli:1999jh, Adler:1999bu, Capozziello:1999wx, Scardigli:2003kr,Ali:2009zq,Hossenfelder:2012jw}. 

Another proposal is the space-time uncertainty relation (STUR)~\cite{Yoneya:1987gb, Yoneya:1989ai, Yoneya:1997gs, Yoneya:2000bt}
\be
\Delta x \, \Delta t \geq \ell_s^2
\, .
\label{STUR}
\ee
Roughly speaking, as the uncertainty in energy or momentum increases, the uncertainty in the size of the string grows as well.

Both relations \eqref{GUP} and \eqref{STUR} have been argued to arise from the stringy nature of interactions in string theory.
They are actually interconnected.
Combining the GUP~\eqref{GUP} with the usual uncertainty relation $\Delta E \, \Delta t \geq 1$ under the assumption
\be
\Delta E \sim \Delta p \, ,
\label{DE=Dp}
\ee
we derive
\begin{align}
\Delta x 
\gtrsim 
\ell_s^2 \, \Delta p 
\sim
\ell_s^2 \, \Delta E 
\geq 
\frac{\ell_s^2}{\Delta t}
\, ,
\end{align}
which is simply the spacetime uncertainty relation \eqref{STUR}.

For an outgoing massless field fluctuation with $E = p$ and satisfying eq.~\eqref{DE=Dp}, both the GUP and the STUR imply
\be
\Delta x \gtrsim \ell_s^2 \, \Delta E
\, .
\label{Dx>DE}
\ee
For an order-of-magnitude estimate, when $\Delta E \sim E$ or $\Delta E \gtrsim E$ (as is the case for Hawking particles in the near-horizon region shown in eq.~\eqref{DPU>PU}), we find that $\Delta x > \ell_s^2 E$. 
This uncertainty $\ell_s^2 E$ exceeds the wavelength $\sim 1/E$ only when the energy  of the particle is trans-Planckian: $E > \ell_s^{-1}$. Thus, even though the spatial uncertainty $\Delta x$ can be macroscopic, it remains strictly confined to trans-Planckian degrees of freedom.

Eq.~\eqref{Dx>DE} implies that probing a structure of length scale $L$ requires not only selecting a probe with a wavelength shorter than $L$ 
but also ensuring that the probe's energy does not exceed $ L / \ell_s^2$. 
This condition sets a minimum length scale $\Delta x \sim \ell_s$ for defining spacetime geometry, thus limiting the smallest resolvable scale to $\ell_s$.

As shown in sec.~\ref{sec:Uncertainty}, in the freely falling frame we have $\Delta E \simeq \Delta P_U \gg \langle P_U \rangle$, and thus eq.~\eqref{Dx>DE} leads to 
\be
\Delta x 
\gg
\ell_s^2 \, \langle P_U \rangle
\sim 
\ell_s^2 \, \langle P_u \rangle \, e^{u/2a}
\sim 
\frac{\ell_s^2}{a} \, e^{u/2a}
\label{Dx}
\ee
for a Hawking particle with $\langle P_u \rangle \sim \mathcal{O}(1/a)$. 
According to eq.~\eqref{Dx}, the uncertainty $\Delta x$ grows exponentially for Hawking particles at late times.
Specifically, $\Delta x$ becomes much larger than the Schwarzschild radius $a$ after the scrambling time $u \geq 2a \log(a^2/\ell_s^2)$.

\hide{
In string theory, scattering amplitudes are exponentially suppressed in the high-energy limit. 
Corresponding to this UV suppression feature, in string field theories, every interaction vertex with external momenta $\{k_i\}_{i=1}^{n}$ includes a factor proportional to~\cite{Witten:1985cc, Kostelecky:1989nt, Pius:2016jsl, deLacroix:2017lif, DeLacroix:2018arq}
\be
e^{- c \ell_s^2 \sum_{i=1}^{n} k_i^2} \, ,
\label{expk2}
\ee
where $c$ is a positive numerical constant of order $1$, apart from multiplicative factors of polynomials of the particle momenta. This exponential suppression factor eliminates UV divergences in loop momentum integrations. It also leads to the spacetime uncertainty relation in terms of the light-cone coordinates \cite{Ho:2023tdq} (see sec.~\ref{sec:StringyModel}).
}

The potential significance of nonlocal effects in string theory for the black-hole information problem has also been suggested in refs.~\cite{Lowe:1995ac, Lowe:1995pu, Giddings:2007bw, Giddings:2012gc, Giddings:2013noa, Dodelson:2015toa, Dodelson:2015uoa}.

\subsection{Hawking Radiation Under GUP}
\label{sec:GUP}

In this section, we examine how the GUP influences late-time Hawking radiation~\cite{Chau:2023zxb}.

As discussed in sec.~\ref{sec:LLS}, the presence of the collapsing matter and the quantum energy-momentum tensor breaks local Lorentz symmetry in the near-horizon region. Although this breaking is weak with a small curvature $\sim \mathcal{O}(1/a^2)$, it is crucial to consider its effects on Hawking particles after the scrambling time as we have demonstrated in sec.~\ref{sec:UV-physics}.

Rather than examining the effects of trans-Planckian interactions between the collapsing matter and the outgoing radiation field in a fundamental UV theory, it is simpler to focus on UV modifications to the wave equation of the outgoing field in a preferred reference frame.

To illustrate this, we first implement the GUP in the Vaidya coordinates, showing that Hawking radiation ceases at the scrambling time. 
As we will discuss in sec.~\ref{sec:FFF}, 
selecting other freely falling frames yields the same conclusion~\cite{Chau:2023zxb}. 
Thus, the termination of Hawking radiation at the scrambling time is a robust prediction of the GUP, independent of the choice of the local frame.

\subsubsection{Low-Energy Effective Description}

\begin{figure}[t]
\centering
\includegraphics[width=0.35\textwidth]{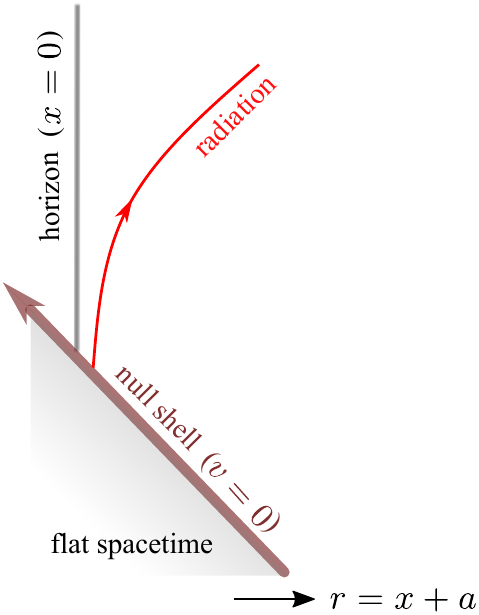}
\caption{\label{fig:vaidya}
A thin null shell with a flat interior collapses to form a black hole.
The calculation of Hawking radiation involves tracing an outgoing wave packet backward in time and decomposing it in terms of the Minkowski modes inside the shell.
}
\end{figure}

We consider a black hole formed from the collapse of a light-like infalling matter shell. 
The background geometry was reviewed in sec.~\ref{sec:BlackHole}.
The Schwarzschild metric outside the shell and the Minkowski space inside the shell can be combined in the ingoing Vaidya metric as
\be 
\label{vaidya}
ds^2 = -\left( 1 - \Theta(v) \, \frac{a}{x + a} \right) dv^2 + 2 dv \, dx \, ,
\ee  
where $v \equiv t + r_*$ \eqref{eq:v} is the light-cone time, and $x \equiv r - a$ represents the radial coordinate from the horizon.
To streamline our analysis, we focus on the $s$-wave sector of the radiation field and omit the angular part of the metric.
Without loss of generality, we assume that the trajectory of the null shell coincides with the time slice $v = 0$ (see fig.~\ref{fig:vaidya}).
Outside the shell ($v > 0$), the metric describes a Schwarzschild geometry, with the event horizon located at $x = 0$.
In the region inside the shell ($v < 0$), the spacetime reduces to flat Minkowski space.

Adopting the line element~\eqref{vaidya} in the near-horizon region where $x \equiv r - a \ll a$, 
the free-field action $S_0$ of the radiation field in the low-energy effective theory is
\be 
\label{gup_S0_x}
S_0[\phi] = 
- \frac{1}{2} \int dv \int dx 
\left[ \del_x \, \phi(v, x) \right]
\biggl\{ 
2 \del_v + \left[ 1 + \Theta(v) \left( \frac{x}{a} - 1 \right) \right] \del_x 
\biggr\} \, 
\phi(v, x) \, .
\ee 
Via the Fourier transform
\be
\phi(v, x) = 
\int_{-\infty}^{\infty} \frac{dp}{\sqrt{2\pi}} \, \Phi (v, p) \, e^{i p x} 
\, ,
\ee
the Klein-Gordon inner product on a constant-$v$ hypersurface of this background can be expressed as 
\be 
\label{inp0}
\langle \phi_1 \, , \phi_2 \rangle 
\equiv 
- i \int_{-\infty}^{\infty} dx \, 
\phi_1^{\ast}(v, x) \, \overset{\leftrightarrow}{\del_x} \, \phi_2 (v, x)
=
2 \int_{-\infty}^{\infty} dp \, 
\Phi_1^{\ast} (v, p) \, p \, \Phi_2 (v, p)
\, .
\ee 
In terms of the Fourier modes, the action can be brought to the form
\begin{align}
S_0 [\Phi]
&= 
\int dv 
\int_{-\infty}^{\infty} dp \,
\Phi^{\ast}(v, p) \,
p 
\left[
i \del_v - \frac{p}{2} 
- \Theta(v) 
\left( \frac{\hat{x}}{2a} \, p - \frac{p}{2} \right) 
\right] 
\Phi (v, p)
\nn \\
&=
\frac{1}{2}
\bigintssss dv 
\left\langle \Phi(v, p)
\, , 
\left[
i \del_v - \frac{p}{2} 
- \Theta(v) 
\left( \frac{\hat{x}}{2a} \, p - \frac{p}{2} \right) 
\right] 
\Phi (v, p) 
\right\rangle \, ,
\label{gup_S0_p}
\end{align}
where $\hat{x} \equiv i \del_p$. 

\subsubsection{Implementing GUP}

Following the proposal of ref.~\cite{Brout:1998ei}, we incorporate the GUP~\eqref{GUP} into the radial coordinate $x$ of the background~\eqref{vaidya} by deforming the commutation relation between $x$ and the radial momentum $p$ as~\cite{Kempf:1994su}
\be 
\label{GUP-comm}
\left[ 
\hat{x} \, , 
\hat{p} 
\right] 
= 
i \left( 1 + \ell_s^2 \, \hat{p}^2 \right) 
,
\ee 
which implies the desired uncertainty relation 
\be 
\Delta x \, \Delta p 
\geq 
\frac{1}{2} 
\left[ 1 + \ell_s^2 \left( \Delta p \right)^2 + \ell_s^2 \, \langle p \rangle^2 \right] 
\geq 
\frac{1}{2} \left[ 1 + \ell_s^2 \left( \Delta p \right)^2 \right]
.
\ee 
This modification can be realized in momentum space 
via the representation
\be 
\label{gup_rep}
\hat{x} \equiv i \left( 1 + \ell_s^2 \, p^2 \right) \del_p 
\, , 
\qquad \hat{p} = p 
\, .
\ee 

To incorporate the GUP into this system, we take the action \eqref{gup_S0_p} and modify the definition of $\hat{x}$ and the inner product. We utilize the deformed momentum-space representation of $\hat{x}$~\eqref{gup_rep} while modifying the inner product~\eqref{inp0} accordingly as 
\be 
\label{gup_inp}
\langle \Phi_1 \, , \Phi_2 \rangle_{\mathrm{GUP}}
\equiv 
2 \int_{-\infty}^{\infty} 
\frac{dp}{1 + \ell_s^2 \, p^2} \, 
\Phi_1^{\ast} (v, p) \, p \, \Phi_2 (v, p)
\, .
\ee 
This ensures the Hermiticity of $\hat{x}$~\cite{Kempf:1994su} and thus preserves the reality of the GUP-modified action
\be 
\label{S_gup}
S_{\mathrm{GUP}} [\Phi]
= 
\frac{1}{2}
\bigintssss dv 
\left\langle \Phi(v, p)
\, , 
\left[
i \del_v - \frac{p}{2} 
- \Theta(v) 
\left( \frac{\hat{x}}{2a} \, p - \frac{p}{2} \right) 
\right] 
\Phi (v, p) 
\right\rangle_{\mathrm{GUP}} 
\ee 
for the radiation field.
As $\phi$ is by assumption a real scalar,
we shall assume the reality condition $\Phi^{\ast}(v, p) = \Phi (v, -p)$. 

In the presence of a minimal uncertainty $\Delta x \geq \ell_s$, the conventional description of the field in terms of the radial coordinate $x$ is no longer available. We view eq.~\eqref{S_gup} as the starting point for this investigation. 
\hide{
More explicitly, the action~\eqref{S_gup} takes the form 
\be 
\label{S_gup2}
S_{\mathrm{GUP}} [\Phi]
=
\int dv \int_0^{\infty} 
\frac{dp}{1 + \ell_s^2 \, p^2} \, 
\Phi^{\ast}(v, p) \,
p 
\left\{ 
i \del_v - \frac{p}{2} 
- 
\Theta(v) 
\left[ \frac{i (1 + \ell_s^2 \, p^2)}{2a} \, \del_p \, p - \frac{p}{2} \right]
\right\} 
\Phi (v, p) 
\, ,
\ee 
}

Although this realization of the GUP~\eqref{GUP} violates Lorentz symmetry, this is justified for the following reasons. 
First, as it was demonstrated in sec.~\ref{sec:UV-physics}, the local Lorentz symmetry in the derivation of Hawking radiation is lost due to the presence of the collapsing matter and the back-reaction of the quantum energy-momentum tensor. 
There is thus a preferred reference frame.

Second, there are numerous examples in which the Hawking radiation of the traditional model is only mildly modified by Lorentz-violating UV physics implemented in a freely falling frame (see, e.g.~\cite{Unruh:1994je,Brout:1995wp,Corley:1996ar,Corley:1997ef,Corley:1997pr,Himemoto:1999kd,Jacobson:1999ay,Unruh:2004zk,Coutant:2011in}). 
Furthermore, even though the realization of the GUP is mathematically inequivalent for each freely falling frame, we will see below that the conclusion of turning off Hawking radiation around the scrambling time is independent of the choice of the frame~\cite{Chau:2023zxb}, including the Vaidya coordinates as a limit (see sec.~\ref{sec:FFF}).

We note that there have been attempts~\cite{Hossenfelder:2006cw, Shibusa:2007ju, Kober:2010sj, Kober:2011dn, Husain:2012im, Faizal:2014dua, Todorinov:2018arx, Bosso:2020fos} in the literature to integrate the GUP into quantum field theories in a Lorentz-covariant manner, which could provide the necessary tools to extend our analysis in a more rigorous framework. It will be interesting to extend the calculation here to a Lorentz-covariant version. Let us also mention that, in sec.~\ref{sec:StringyModel}, the spacetime uncertainty relation that will be responsible for turning off Hawking radiation is Lorentz-covariant.

\hide{
In this formalism, the corrections introduced by the GUP are encoded in the wave equation governing the radiation field, leading to significant alterations in the behavior of propagating modes probing the black hole background at trans-Planckian momenta.
}

Variation of the action~\eqref{S_gup} yields the field equation 
\be 
\label{gup_eom}
\left\{
i \del_v - \frac{p}{2} 
- \Theta(v) 
\left[ 
\frac{i \left( 1 + \ell_s^2 \, p^2 \right)}{2a} \, 
\del_p \, p 
- 
\frac{p}{2} 
\right] 
\right\} 
\Phi (v, p) = 0 
\ee 
in the outgoing sector, with the continuity of $\Phi (v, p)$ required across the null shell at $v = 0$. 

The field equation \eqref{gup_eom} has to be supplemented with a suitable boundary condition. According to the definition of the inner product~\eqref{gup_inp}, the time derivative of the norm is
\be 
\del_v 
\left\langle \Psi \, , \Psi \right\rangle_{\mathrm{GUP}}
=
\frac{\Theta(v)}{2a} 
\left[
\bigl|
p \, \Psi (v, p) 
\bigr|^2
\right]^{\infty}_{p \, = \, -\infty}
\ee 
for a wave $\Psi$ obeying the outgoing field equation~\eqref{gup_eom}. 
Therefore, the norm of $\Psi$ is conserved under time evolution outside the matter shell only if the boundary condition
\be
\lim_{p \to \pm \infty} p \, \Psi (v, p) = 0
\label{BC-GUP}
\ee
is satisfied. Hence, unitarity demands that the Hilbert space of the quantized theory is only composed of wave functions satisfying this boundary condition.

According to the GUP commutation relation \eqref{GUP-comm}, it is natural to define the conjugate momentum of $x$ as
\be
K(p) \equiv \ell_s^{-1} \tan^{-1} \left( \ell_s \, p \right),
\label{K(p)}
\ee
so that the usual Heisenberg algebra $\comm*{\hat{x}}{\hat{K}(p)} = i$ is satisfied. This explains the measure of the inner product \eqref{gup_inp}:
\be
dK = \frac{dp}{1+\ell_s^2 p^2}.
\ee

The existence of a minimal length $\Delta x \geq \ell_s$ is now reflected in the UV cutoff of the conjugate momentum:
\be
\abs{K} \leq \pi / 2 \ell_s.
\label{K-bound}
\ee
In terms of this new momentum $K$, the free field described by eq.~\eqref{S_gup} exhibits a superluminal dispersion relation $\omega(K) = \tan\left( \ell_s K \right) / 2 \ell_s$.

While the final outcome of Hawking radiation is independent of the choice of momentum variables, we shall adopt the interpretation where the field is dispersionless ($\omega(p) = p / 2$), and take $p$ instead of $K$ as the momentum. 

\subsubsection{Hawking Particle's Wave Packet}
\label{sec:HPWP}

In the Schwarzschild spacetime outside the shell ($v > 0$), the monochromatic Hawking modes $\Phi_{\omega}(v, p)$ with frequency $\omega$ (defined with respect to the Eddington time $v$) can be solved from the wave equation~\eqref{gup_eom} as 
\be 
\label{gup_mode}
\Phi_{\omega}(v, p) 
= 
\mathcal{N}_{\omega} \, 
\frac{e^{-i \omega v}}{p - i 0^+} 
\exp\left\{-2 i a \omega \log\left[ \frac{ a \left( p - i 0^+ \right)}{\sqrt{1 + \ell_s^2 \, p^2}} \right]\right\} ,
\ee 
where $\mathcal{N}_{\omega}$ is a normalization factor that will be determined later.
The $i 0^+$-prescription for analytic continuation across the pole (and branch point) at $p = 0$ 
ensures that the expression~\eqref{gup_mode} is both analytic and bounded in the lower half of the complex $p$-plane. This guarantees that $\Phi_{\omega}$ describes an outgoing wave 
with no support behind the horizon ($x < 0$), thereby respecting the causal behavior of the wave equation~\cite{Damour:1976jd}.

As will become clear soon, the difference $\exp\left[ - 2 \pi a \omega \, \Theta(-p) \right]$ resulting from the discontinuity across the branch cut of the logarithm in eq.~\eqref{gup_mode} gives rise to the Boltzmann factor $e^{- 4 \pi a \omega}$ characterizing the Hawking temperature.

As stated in sec.~\ref{sec:HR-LEET}, to capture the time dependence of Hawking radiation, rather than working with spacetime-filling waves, we construct a localized wave packet $\Psi_{\omega_0 , u_0}$ as a superposition of the positive-frequency modes~\eqref{gup_mode}:
\be 
\label{gup_packet}
\Psi_{\omega_0 , u_0} (v, p)
= 
\int_0^{\infty} d \omega \,
f_{\omega_0} (\omega) \, 
e^{i \omega u_0} \, 
\Phi_{\omega}(v, p) \, .
\ee 
Its profile $f_{\omega_0} (\omega)$ in the frequency domain is supposed to be compactly supported around $\omega = \omega_0$ with a narrow width $\ll \omega_0$.
Under this assumption, the Hawking wave packet~\eqref{gup_packet} can be written more explicitly as 
\begin{align}
\Psi_{\omega_0 , u_0} (v, p)
&= 
\int_0^{\infty} d \omega \,
f_{\omega_0} (\omega) \, 
\mathcal{N}_{\omega} \, 
\frac{e^{- i \omega (v - u_0)}}{p - i 0^+} \, 
\exp \, 
\biggl\{ 
-2 i a \omega 
\log 
\biggl[ \frac{a \left( p - i 0^+ \right)}{\sqrt{1 + \ell_s^2 \, p^2}} \biggr]
\biggr\}
\nn \\
&\simeq 
\mathcal{N}_{\omega_0} \, 
\frac{\exp\left[ - 2 \pi a \omega_0 \, \Theta(-p) \right]}{p - i 0^+} 
\int_0^{\infty} d \omega \,
f_{\omega_0} (\omega) \, 
\exp 
\left\{ 
- i \omega 
\bigl[
v - u_0 
+ 
2 a 
\log \abs{a P(p)} \, 
\bigr] 
\right\} 
,
\label{gup_packet2}
\end{align}
where the correction from the GUP is encoded in the \emph{effective momentum}
\be 
\label{gup_P(p)}
P(p) 
\equiv 
\frac{p}{\sqrt{1 + \ell_s^2 \, p^2}} 
\, .
\ee 
In deriving eq.~\eqref{gup_packet2}, we took advantage of the compactness of $f_{\omega_0} (\omega)$ to approximate the slowly varying term $\mathcal{N}_{\omega} \exp\left[ - 2 \pi a \omega \, \Theta(-p) \right]$ by a constant evaluated at $\omega_0$.

Defining the following quantities:
\begin{align}
F(p) 
&
\equiv 
\log \bigl| a P(p) \bigr|
\, , 
\label{eq:F}
\\
\tilde{f}_{\omega_0} (v)
&\equiv 
\int_0^{\infty} 
\frac{d \omega}{\sqrt{2\pi}} \,
f_{\omega_0} (\omega) \, 
e^{- i \omega v}
\, ,
\label{gup_rho}
\end{align}
we further simplify eq.~\eqref{gup_packet2} as 
\be 
\label{gup_packet3}
\Psi_{\omega_0 , u_0} (v, p)
\simeq 
\sqrt{2 \pi} \, 
\mathcal{N}_{\omega_0} \, 
\frac{\exp\left[ - 2 \pi a \omega_0 \, \Theta(-p) \right]}{p - i 0^+} \, 
\tilde{f}_{\omega_0} \bigl( v - u_0 + 2 a F(p) \bigr) 
\, .
\ee 
From this expression, it is clear that the momentum-space wave packet $\Psi_{\omega_0 , u_0} (v, p)$ for a Hawking particle is centered along the characteristic trajectory 
\be 
\label{gup_chrcurve}
v + 2a F(p) = \text{constant}
\, .
\ee 

At sufficiently late times (large $v$), eqs.~\eqref{gup_P(p)}, \eqref{eq:F}, and \eqref{gup_chrcurve} imply that 
the central momentum $\bar{p}$ of the wave packet is cis-Planckian, i.e., $\bar{p} \ll \ell_s^{-1}$. In this regime, it is reasonable to approximate $F(p) \simeq \log |a p|$.
As a result, the wave packet~\eqref{gup_packet3} eventually evolves in accordance with the low-energy theory. It is then natural to normalize the wave packet~\eqref{gup_packet3} by matching it to the ordinary Hawking wave packet in the unmodified theory at late times.

In the unmodified theory, the wave packet is a superposition of the outgoing plane waves $e^{- i \omega u} / \sqrt{4 \pi \omega} = e^{- i \omega (v - 2 r_*)} / \sqrt{4 \pi \omega}$. Tracing these modes backward in time to the near-horizon region where $0 < x \ll a$, the form of each individual mode function in momentum space is given by 
\be 
\int_{-\infty}^{\infty}
\frac{dx}{\sqrt{2 \pi}} \, e^{-i p x} 
\left[ 
\Theta(x) \, 
\frac{e^{-i \omega v}}{\sqrt{4 \pi \omega}} 
\left( \frac{x}{a} \right)^{2 i a \omega}
\right]
=
\frac{a}{\pi} 
\sqrt{\frac{\omega}{2}} \, e^{\pi a \omega} \, \Gamma(2 i a \omega)
\left\{ 
\frac{e^{-i \omega v}}{p - i 0^+}
\left[ a (p - i 0^+) \right]^{-2 i a \omega}
\right\}
.
\ee 
The counterpart of~\eqref{gup_packet3} in the low-energy theory is 
\be 
\Psi_{\omega_0 , u_0}^{(0)} (v, p)
\simeq 
a \sqrt{\frac{\omega_0}{\pi}} \, 
e^{\pi a \omega_0} \, \Gamma(2 i a \omega_0) \, 
\frac{\exp\left[ - 2 \pi a \omega_0 \, \Theta(-p) \right]}{p - i 0^+} \, 
\tilde{f}_{\omega_0} \bigl( v - u_0 + 2 a \log|a p| \bigr) 
\, .
\ee 

By demanding that the GUP-modified wave packet~\eqref{gup_packet3} matches its unmodified counterpart ${\displaystyle\Psi_{\omega_0 , u_0}^{(0)} (v, p)}$ during the late-time evolution within the IR regime:
\be 
\Psi_{\omega_0 , u_0} (v \gg u_0 \, , p)
\simeq 
\Psi_{\omega_0 , u_0}^{(0)} (v \gg u_0 \, , p)
\, ,
\ee 
one can uniquely fix the $p$-independent overall constant $\mathcal{N}_{\omega}$ in the mode functions~\eqref{gup_mode} as
\be 
\label{gup_normalization}
\mathcal{N}_{\omega} 
= 
\frac{a}{\pi}
\sqrt{\frac{\omega}{2}} \, 
e^{\pi a \omega} \, 
\Gamma(2i a \omega)
\, .
\ee 
As the correspondence with an ordinary Hawking particle is established at late times, the phase factor $e^{i \omega u_0}$ that was included in the definition~\eqref{gup_packet} of the wave packet now takes on a clear interpretation as localizing the wave packet around the Eddington retarded time $u = u_0$.

\begin{figure}[t]
\centering
\includegraphics[scale=0.5]{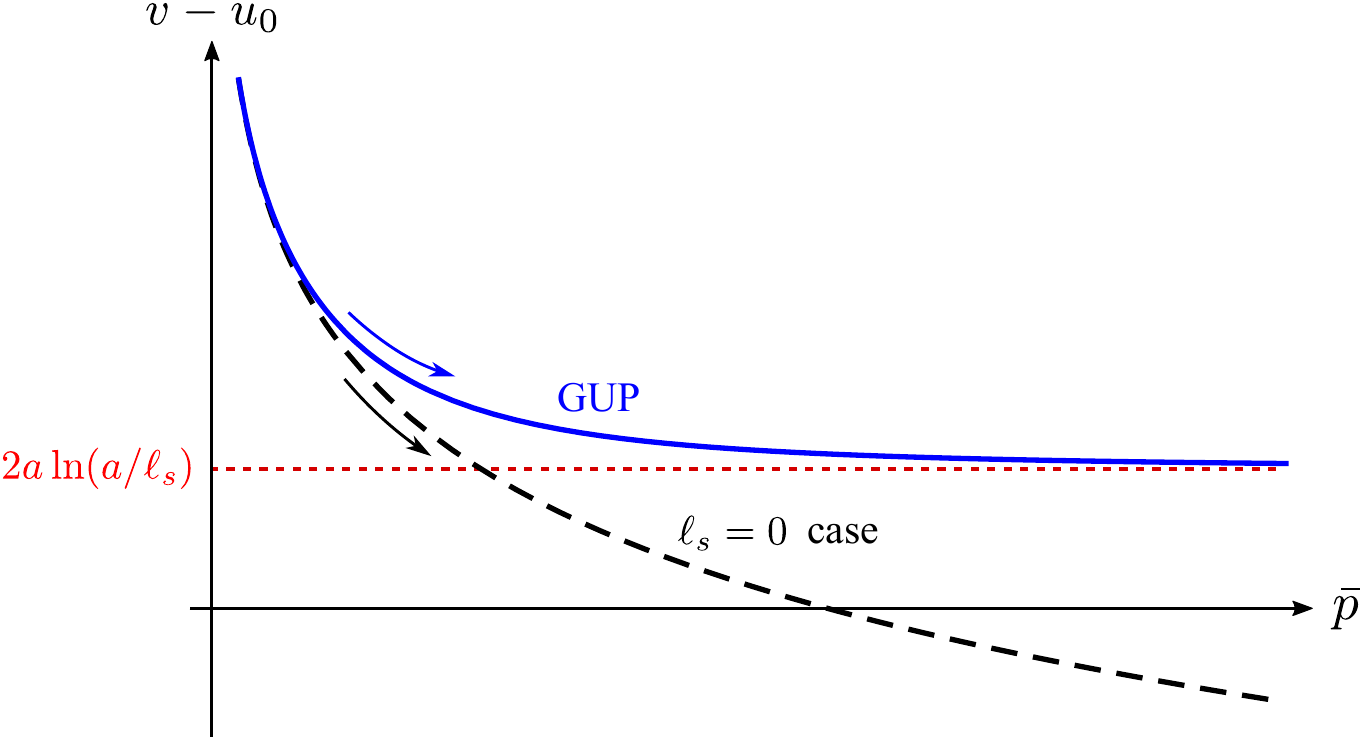}
\caption{\label{fig:gup_p_central}
A graph illustrating the positive-$p$ branch of the characteristic trajectory $v - u_0 = - 2 a F(p)$~\eqref{gup_chrcurve} (solid blue curve), alongside its counterpart in the low-energy theory (dashed black curve). Arrows indicate the flow direction toward the past (decreasing $v$). With the introduction of the GUP, the momentum reaches infinity in a finite time marked by the dotted red line.
}
\end{figure}

One can solve for the central momentum $\bar{p}$ of a Hawking wave packet from eq.~\eqref{gup_chrcurve} as
\be 
\label{gup_centralp}
\bar{p}(v)
=
\begin{dcases}
\frac{1}{a} \, 
\frac{e^{(u_0 - v) / 2a}}{\sqrt{1 - e^{(u_0 - v) / a} \, \ell_s^2 / a^2}}
& \text{for} \ \bar{p} > 0 
\\
- \frac{1}{a} \, 
\frac{e^{(u_0 - v) / 2a}}{\sqrt{1 - e^{(u_0 - v) / a} \, \ell_s^2 / a^2}}
& \text{for} \ \bar{p} < 0 
\end{dcases}
\, .
\ee 
As time $v$ decreases, the magnitude of the central momentum $\bar{p}(v)$ grows sharply 
due to the rapid decrease of its denominator.
An illustration of the characteristic evolution~\eqref{gup_chrcurve} is presented in fig.~\ref{fig:gup_p_central}. 
During the backward evolution of a wave packet, the blue shift of its central momentum $\bar{p} > 0$ occurs far more rapidly than the exponential growth seen in the low-energy effective theory. 

Based on these results, a Hawking particle that reaches the asymptotic region at around the scrambling time $u_0 \sim 2 a \log(a / \ell_s)$ would have originated from the collapsing shell at $v = 0$ with a central momentum $\abs{\bar{p}} \gg 1 / \ell_s$ greatly exceeding the Planck scale. 
As it turns out, eq.~\eqref{gup_centralp} indicates that $\abs{\bar{p}(v)}$ is in fact infinite when $v = u_0 - 2 a \log(a / \ell_s)$. Namely, the central momentum would already be blue-shifted to reach the boundary $p = \pm \infty$ within a finite duration of backward propagation. This peculiar feature of the GUP system has profound implications to be discussed in more detail later in this section.

\subsubsection{Wave Function Inside the Shell}

Now that the mode configuration outside the collapsing shell is set up, we proceed to quantize the field inside the shell ($v < 0$). In the flat spacetime inside the shell, the field equation~\eqref{gup_eom} simply reduces to 
\be
\left(i \del_v - \frac{p}{2} \right) \Phi(v \leq 0 \, , p) = 0 \, ,
\ee
which suggests the mode expansion
\be 
\label{gup_decom}
\Phi(v \leq 0 \, , p) 
= 
e^{-i p v / 2} \sqrt{\frac{1 + \ell_s^2 \, p^2}{2|p|}} \, 
\bigl[ a_p \, \Theta(p) + a_{-p}^{\dagger} \, \Theta(-p) \bigr] .
\ee  

Upon canonical quantization, where the field $\Phi$ and its conjugate momentum 
\be 
\Pi (v , p) 
\equiv 
\frac{\delta S_{\mathrm{GUP}}}{\delta [ \del_v \Phi(v, p) ]}
= 
\frac{2 i p}{1 + \ell_s^2 \, p^2} \, 
\Phi^{\ast}(v , p)
\ee 
are promoted to operators satisfying the equal-time commutation relation 
\be
\comm*{\Phi(v, p)}{\Pi(v, p')} = i \, \delta(p - p') \, ,
\ee 
the creation and annihilation operators obey
\be 
\bigl[ a_p \, , a_{p'}^{\dagger} \bigr]
=
\delta(p - p')
\, ,
\qquad 
\left[ a_p \, , a_{p'} \right]
=
0
\, , 
\qquad 
\bigl[ a_p^{\dagger} \, , a_{p'}^{\dagger} \bigr]
=
0
\, .
\ee 
The quantum state of the field inside the shell is assumed to be the Minkowski vacuum $\ket{0}$ defined by 
\be
a_p \ket{0} = 0 \qquad \forall \ p > 0 \, .
\label{gup_vacuum}
\ee

\subsubsection{Derivation of Hawking Radiation Under GUP}

In analogy with the approach in sec.~\ref{sec:HR-LEET}, we can associate an annihilation operator $b_{\Psi}$ to the Hawking particle described by the wave packet $\Psi_{\omega_0 , u_0}$~\eqref{gup_packet3}. 
This operator $b_{\Psi}$ is defined as the inner product between $\Psi_{\omega_0, u_0}$ and the field operator $\Phi$:
\be 
\label{gup_b}
b_{\Psi} 
\equiv
\bigl\langle 
\Psi_{\omega_0, u_0} \, , 
\Phi 
\bigr\rangle_{\mathrm{GUP}} \, .
\ee 
The number expectation value $\ev{b_{\Psi}^{\dagger} b_{\Psi}}{0}$ of such Hawking particles in the Minkowski vacuum $\ket{0}$~\eqref{gup_vacuum} is determined by how the operator $b_{\Psi}$ decomposes into the creation and annihilation operators $( a_p \, , a_p^{\dagger} )$ inside the shell.
Hence, the calculation of Hawking radiation amounts to tracing the wave packet $\Psi_{\omega_0, u_0} (v, p)$ back in time to the collapsing shell at $v = 0$, where the field $\Phi(v = 0 \, , p)$ can be matched with the mode expansion~\eqref{gup_decom} in Minkowski space.

According to the form~\eqref{gup_inp} of the inner product under the GUP, eq.~\eqref{gup_b} reads 
\be 
b_{\Psi} 
=
\int_0^{\infty} dp \, 
\sqrt{\frac{2 p}{1 + \ell_s^2 \, p^2}} 
\left[
\Psi_{\omega_0, u_0}^* (0, p) \, a_p 
- 
\Psi_{\omega_0, u_0}^* (0, -p) \, a_p^{\dagger} 
\right]
,
\ee 
which is essentially a decomposition of the wave packet $\Psi_{\omega_0, u_0} (0 , p)$ into 
positive and negative momentum components. It follows that the vacuum expectation value of the number of Hawking particles is given by 
\be 
\label{gup_Nb}
\ev{b_{\Psi}^{\dagger} b_{\Psi}}{0} 
= 
\int_0^{\infty} dp \, 
\frac{2 p}{1 + \ell_s^2 \, p^2} \, 
\abs\big{\Psi_{\omega_0, u_0} (0, -p)}^2
=
\bigl\langle 
\Psi_{\omega_0, u_0} (0, p) 
\, ,
\Theta (- p) \, \Psi_{\omega_0, u_0} (0, p) 
\bigr\rangle_{\mathrm{GUP}}
\, ,
\ee 
which is determined by the norm of the negative-momentum components in the wave packet $\Psi_{\omega_0, u_0} (0, p)$ as anticipated.

Plugging the expression~\eqref{gup_packet2} for the wave packet into eq.~\eqref{gup_Nb}, we find 
\begin{align}
\ev{b_{\Psi}^{\dagger} b_{\Psi}}{0} 
&\simeq 
\abs{\mathcal{N}_{\omega_0}}^2 \, 
e^{- 4 \pi a \omega_0}
\int_0^{\infty} d \omega 
\int_0^{\infty} d \omega' \, 
f_{\omega_0}^{\ast}(\omega) \, 
f_{\omega_0}(\omega') \, 
e^{- i (\omega - \omega') u_0}
\int_0^{\infty} dp \, 
\frac{2 \, e^{2 i a (\omega - \omega') F(p)}}
{p \left( 1 + \ell_s^2 \, p^2 \right)} 
\nn \\
&= 
2 \, \abs{\mathcal{N}_{\omega_0}}^2 \, 
e^{- 4 \pi a \omega_0}
\int_0^{\infty} d \omega 
\int_0^{\infty} d \omega' \, 
f_{\omega_0}^{\ast}(\omega) \, 
f_{\omega_0}(\omega') \, 
e^{- i (\omega - \omega') u_0}
\int_{-\infty}^{\log(a / \ell_s)} dF \, 
e^{2 i a (\omega - \omega') F}
\, ,
\end{align}
from which we observe that the effect of the GUP on Hawking radiation appears in the form of a finite upper bound $\log(a / \ell_s)$ on the $F$-integration. Subsequently, through a change of variable 
\be 
u \equiv 2 a F - u_0
\, ,
\ee 
we arrive at~\cite{Chau:2023zxb}
\begin{align}
\ev{b_{\Psi}^{\dagger} b_{\Psi}}{0} 
&\simeq 
\frac{a}{\pi} \, 
\frac{e^{- 4 \pi a \omega_0}}{1 - e^{- 4 \pi a \omega_0}}
\int_0^{\infty} d \omega 
\int_0^{\infty} d \omega' \, 
f_{\omega_0}^{\ast}(\omega) \, 
f_{\omega_0}(\omega')
\int_{-\infty}^{2 a \log(a / \ell_s) - u_0}
\frac{du}{2 a} \, 
e^{i (\omega - \omega') u}
\nn \\
&=
\frac{1}{e^{4 \pi a \omega_0} - 1}
\int_{-\infty}^{2 a \log(a / \ell_s) - u_0} 
du \, 
\biggl| 
\int_0^{\infty} 
\frac{d \omega'}{\sqrt{2 \pi}} \,
f_{\omega_0} (\omega') \, 
e^{- i \omega' u}
\biggr|^2
\nn \\
&= 
\frac{1}{e^{4 \pi a \omega_0} - 1}
\int_{-\infty}^{2 a \log(a / \ell_s) - u_0} du \,
\left|\tilde{f}_{\omega_0} (u) \right|^2
\, ,
\label{gup_Nb2}
\end{align}
where we made use of the definition~\eqref{gup_rho}, as well as the fact that 
\be 
\abs{\mathcal{N}_{\omega}}^2
=
\frac{a}{2 \pi} \, 
\frac{1}{1 - e^{- 4 \pi a \omega}}
\ee 
based on eq.~\eqref{gup_normalization}. 
Note that the Hawking temperature remains the same while the amplitude changes over time.

We thus arrive at a result similar to the scenario involving a non-covariant UV cutoff described in sec.~\ref{sec:cutoff}. Specifically, by employing the effective momentum $P(p) \equiv p / \sqrt{1 + \ell_s^2 \, p^2}$~\eqref{gup_P(p)} and redefining the field as 
\be 
\nn 
\chi (v, P) 
\equiv 
p(P) \, \Phi \bigl( v, p(P) \bigr)
\, ,
\ee 
where $p(P)$ is the inverse function of $P(p)$,
the action~\eqref{S_gup} can be expressed as 
\be 
\nn 
S_{\mathrm{GUP}}[\chi]
=
\int dv 
\int_0^{\ell_s^{-1}} \frac{dP}{P} \, 
\chi^{\ast} (v, P)
\left( 
2 i \partial_v 
- 
\frac{i}{a} \, \Theta(v) P \, \partial_P 
\right) 
\chi (v, P) 
\, ,
\ee 
which indeed differs from the low-energy action $S_0[\chi]$~\eqref{gup_S0_p} merely by a UV cutoff $\ell_s^{-1}$.
Given that the profile function $\tilde{f}_{\omega_0}(u)$ defined in eq.~\eqref{gup_rho} is centered around the retarded time $u = 0$ with a width of $\mathcal{O}(a)$, the probability~\eqref{gup_Nb2} of detecting Hawking particles centered around the retarded time $u_0$ diminishes with increasing $u_0$. Eventually, this probability approaches zero when
$u_0 \gtrsim 2 a \log(a / \ell_s)$, signifying that Hawking radiation is turned off around the scrambling time.

\subsubsection{Freely Falling Frames}
\label{sec:FFF}

In the above,
we have considered Hawking radiation with the GUP realized in the Vaidya coordinates. The same calculation can be carried out in a generic freely falling frame \cite{Chau:2023zxb}.

Consider a generic freely falling observer traveling radially inward along a timelike geodesic in the Schwarzschild spacetime.
We can parametrize the initial speed of the geodesic at the infinite past ($t \rightarrow - \infty, r \rightarrow \infty$) as 
\be 
\lim_{t\rightarrow-\infty}\left(\frac{dr}{d t}\right)^2 = 1 - \gamma \, , 
\qquad
\text{where} \ \gamma \in (0, 1] \, .
\ee 
By introducing the proper time coordinate
\be 
\tau (t , r) = \gamma^{-1/2} \left[ t + h_{\gamma}(r) \right] \, ,
\ee
where 
\begin{align}
h_{\gamma}(r) &\equiv \int^r dr' \, \frac{\sqrt{1 - \gamma (1 - a / r')}}{1 - a / r'}
\nn \\
\begin{split}
&= \sqrt{1 - \gamma ( 1 - a/r )} \, r 
+ \frac{(2 - \gamma)}{2 \sqrt{1 - \gamma}} \, a \log \left\lvert \frac{\sqrt{1 - \gamma} + \sqrt{1 - \gamma(1 - a / r)}}{\sqrt{1 - \gamma} - \sqrt{1 - \gamma(1 - a / r)}} \right\rvert
\\
&\quad - a \log \left\lvert \frac{1 + \sqrt{1 - \gamma(1 - a / r)}}{1 - \sqrt{1 - \gamma(1 - a / r)}} \right\rvert
,
\end{split}
\end{align}
we can define a class of freely falling frames $(\tau, r)$, in which the Schwarzschild line element can be expressed as
\be 
d s^2 = - d \tau^2 + \gamma \, \bigl( d r - v_{\gamma}(r) \, d \tau \bigr)^2 \, ,
\label{metric-FFF}
\ee 
where 
\be 
v_{\gamma}(r) = - \gamma^{-1/2} \sqrt{1 - \gamma \left( 1 - \frac{a}{r} \right)} \, .
\ee 
The Eddington-Finkelstein line element that we worked with previously is the limiting case with $\gamma \to 0$.

Following similar calculation steps as in the previous section 
(with the detailed derivation provided in ref.~\cite{Chau:2023zxb}),
it can be shown that the result~\eqref{gup_Nb2} of Hawking radiation with the GUP implemented in the Vaidya coordinates is slightly modified in a generic freely falling frame as
\begin{align}
\ev{b_{\Psi}^{\dagger} b_{\Psi}}{0} 
&\simeq
\frac{1}{e^{4 \pi a \gamma^{-1/2} \omega_0} - 1}
\int_{-\infty}^{2 a \log(a / \ell_s) - u_0} 
\left|\tilde{f}_{\omega_0} (u)\right|^2 \, 
du
\, ,
\label{bb8}
\end{align}
where there is a factor $\gamma^{-1/2}$ in front of the central frequency $\omega_0$ defined with respect to the free-fall proper time $\tau$.
The conclusion remains the same that Hawking radiation is turned off around the scrambling time $u \sim u_{scr}$~\eqref{scrambling-time}.
Notice that the proper time interval $\Delta \tau$ is related to the Schwarzschild time interval $\Delta t$ via $\Delta \tau = {}\gamma^{-1/2} \Delta t$ at a fixed radius.
Hence, by combining the Lorentz boost factor $\gamma^{-1/2}$ with the frequency $\omega_0$ into the Schwarzschild frequency (defined with respect to $t$) in the Planck distribution factor in eq.~\eqref{bb8}, we find that the same Hawking temperature is displayed across all freely falling frames.

Since the results of the UV modification by the GUP to Hawking radiation are insensitive to the free parameter $\gamma$ labeling different freely falling frames, we can infer that this UV effect is robust and insensitive to the details of the gravitational collapse.

\subsubsection{Comments on Hawking Radiation Under GUP}

We showed above that Hawking radiation is terminated around the scrambling time by the GUP implemented in any freely falling frame. In this section, we discuss the mathematics and physics behind this result.


The termination of Hawking radiation can be attributed to the UV cutoff in the Hilbert space, analogous to what was shown in sec.~\ref{sec:cutoff}.
While the momentum $p$ within the Hilbert space may, in theory, take on arbitrarily large values, the conjugate variable $K$~\eqref{K(p)} is limited to $|K| \leq \pi/(2\ell_s)$.
States with $|K| > \pi/(2\ell_s)$ in the near-horizon region, which are necessary for the evolution into Hawking particles post-scrambling time, do not exist within the Hilbert space.
This phenomenon is related to the ``non-conservation of the norm'' of a Hawking wave packet, 
as highlighted in refs.~\cite{Brout:1998ei, Chau:2023zxb}. 

If we start with a state with momentum constrained within $|K| \leq \pi/(2\ell_s)$ on an initial time slice in the past, it will not evolve into a Hawking particle beyond the scrambling time. 
Conversely, if we consider a Hawking particle that emerges after the scrambling time at a large distance and we allow it to propagate backward in time, it will violate the boundary condition~\eqref{BC-GUP} as demonstrated in sec.~\ref{sec:HPWP}. 
This violation leads to a decrease in the norm during backward evolution until the wave packet falls out of the Hilbert space in the near-horizon region.

The question now is whether it is sensible to say that, at large distances, there are low-energy particle states that cannot be created from the vacuum when these states originate from a small region close to the horizon of a black hole.
If this is not acceptable, there is a violation of unitarity.
It was suggested in ref.~\cite{Brout:1998ei} that 
the late-time Hawking modes originate from massive string states contained within a ``reservoir'' of size $\mathcal{O}(\ell_s)$ surrounding the horizon. The origin of the non-conservation of the norm is that these states are excluded from the Hilbert space.\footnote{However, ref.~\cite{Brout:1998ei} missed the fact that Hawking radiation stops around the scrambling time because the magnitude of Hawking radiation was not derived.}

Another possibility is that a proper description of these missing degrees of freedom will become accessible once we achieve a deeper understanding of spacetime.
For instance, this includes how to characterize the Schwarzschild background for a trans-Planckian mode with $\Delta x \gg a$.


In any case, recall that the motivation behind the GUP in quantum gravity is to establish a minimal length scale, below which no degrees of freedom exist.
A key observation from the GUP~\eqref{GUP} is that the uncertainty $\Delta x$ in position increases with momentum. Under the influence of the GUP, the width $\Delta p$ of a Hawking wave packet
$\Psi_{\omega_0, u_0}(v, p)$~\eqref{gup_packet3} in momentum space scales with its central momentum $\bar{p}$ as 
\be 
\Delta p 
\sim 
\left( \frac{d F}{d p} \right)^{-1} 
\bigg\vert_{p \, = \, \bar{p}}
=
\bar{p} 
\left( 1 + \ell_s^2 \, \bar{p}^2 \right)
.
\ee 
This implies that at early times in the past, when the Hawking particle state has a highly blue-shifted momentum $\bar{p} \gg a / \ell_s^2$, its wave function not only exhibits a much broader spectrum $\Delta p \gg \bar{p}$ compared to the low-energy theory, but also an enormous spread in position since 
\be 
\Delta x 
\, \gtrsim \, 
\ell_s^2 \left( \Delta p \right)
\, \gg \, 
\ell_s^2 \, \bar{p}
\, \gg \, a
\, .
\ee 
This indicates that the particle state is unlikely to be confined within the near-horizon region 
and that the uncertainty $\Delta x$ in its location far exceeds the size of the black hole itself.~\footnote{This feature may seem incompatible 
with the assumption of the near-horizon approximation $x \ll a$ that we started with in eq.~\eqref{gup_S0_x}. That said, this issue was addressed in the appendix of ref.~\cite{Chau:2023zxb}, where a result similar to~\eqref{gup_Nb2} was obtained without relying on the assumption of small $x$.}

Notice that in the freely falling frames defined in eq.~\eqref{metric-FFF}, $x = r - a$ measures the proper radial distance from the horizon (up to an $\mathcal{O}(1)$ factor $\gamma$) on a constant proper time slice.
Therefore, the relation $\Delta x \gg a$ represents a genuinely large distance in length scales.

For this reason, it was proposed in ref.~\cite{Chau:2023zxb} that highly trans-Planckian momentum states with $p \gg a / \ell_s^2$ should be interpreted as \emph{stringy} states that span macroscopic distances far beyond the characteristic length scale of the black hole. 
Such states are expected to be ignorant of the geometry of the near-horizon region, or anything that has a characteristic scale $\lesssim \mathcal{O}(a)$. 
Effectively they propagate in a smooth background of curvature $\ll \mathcal{O}(1/a^2)$.
The horizon is invisible to these extremely high-energy modes.
With their initial conditions being the Minkowski vacuum in the infinite past, the adiabatic theorem ensures that they remain approximately in the vacuum at a scale $\gg \mathcal{O}(a)$. 
This means that their quantum states shall agree with the asymptotic vacuum defined for distant observers, rather than the Unruh vacuum in the near-horizon region. 
Consequently, they do not contribute to Hawking radiation, leading to the termination of Hawking radiation after the scrambling time.

The fact that the vacuum state of a given propagating mode can only be defined in a sufficiently large space is not new:
The vacuum state of a fluctuation mode of wavelength $L$ can only be defined in a region much larger than $L$.
Furthermore, this mode is insensitive to spacetime structures at a distance scale much shorter than $L$.
When its wavelength $L$ significantly exceeds the Schwarzschild radius $a$ of a black hole, the black hole can effectively be ignored in the metric.
This is why, with the galaxies and black holes smeared out, the Friedmann-Robertson-Walker (FRW) metric is suitable for describing the fluctuations at the scale of the universe.

In addition to the wave nature of a probe, there is also a stringy nature for trans-Planckian modes described above.
The abovementioned idea of background smearing that applies to waves can be applied analogously to high-energy stringy probes.

\hide{
In any case, we argue that when the stringy modes' vacua need to be defined over a large region including both the near-horizon region and a distant region, they cannot contribute to Hawking radiation, which is an effect of having different vacua in the two regions.
Therefore, even though we are missing stringy modes in the calculation, we claim that they do not contribute to Hawking radiation.
}

\subsection{Hawking Radiation in String Field Theory}
\label{sec:StringyModel}

In this section, we explore the nonlocal structure of string field theory (SFT), aiming to provide a more direct justification from string theory regarding the shutdown of Hawking radiation after the scrambling time.

\subsubsection{Nonlocality in String Field Theory}

A common feature of string field theories~\cite{Kaku:1974zz, Witten:1985cc, Zwiebach:1992ie, deLacroix:2017lif} is the exponential suppression of UV interactions. 
In $d$-dimensional Minkowski space $x^{\mu} = (t, x^i)$, each component field in the theory can be described by an action of the schematic form
\be
\label{SFT_action}
S_{\mathrm{SFT}}[\phi]
= 
\bigintssss d^d x
\left[
\frac{1}{2} \, 
\phi \, \Box \, \phi
- 
V( \tilde{\phi} \, , \del \tilde{\phi} )
\right]
,
\ee
where $\Box \equiv - \del_0^2 + \del_i^2$.
The field $\phi$ appears in the interaction vertices $V( \tilde{\phi} \, , \del \tilde{\phi} )$ only in the nonlocal form
\be
\tilde{\phi}
\equiv
e^{\ell_s^2 \, \Box / 2} \, \phi \, ,
\label{eq:tildephi}
\ee
while $V( \tilde{\phi} \, , \del \tilde{\phi} )$ itself contains at most derivative couplings of finite order in $\tilde{\phi}$.
Actions of the form~\eqref{SFT_action} have been employed as toy models to study various aspects of string field theories, such as causality, UV finiteness, and unitarity of the perturbation theory~\cite{Tomboulis:2015gfa, Pius:2016jsl, Carone:2016eyp, Briscese:2018oyx, Chin:2018puw, Pius:2018crk, Buoninfante:2018mre, DeLacroix:2018arq, Buoninfante:2018gce, Koshelev:2021orf, Erbin:2021hkf, Buoninfante:2022krn} . 

It is crucial to note that all interaction vertices are dressed with the exponential operator $e^{\ell_s^2 \, \Box / 2}$, so any measurement of field observables via coupling to an Unruh-DeWitt detector~\cite{Unruh:1976db, DeWitt:1980hx} would be determined by the Wightman correlator $\ev{\hat{\tilde{\phi}} (x_1) \cdots \hat{\tilde{\phi}} (x_n)}{0}$ for the UV-suppressed field $\tilde{\phi}$ instead of $\phi$~\cite{Ho:2023tdq}.
It is therefore useful to rewrite the action in terms of $\tilde{\phi}$ as~\cite{Siegel:2003vt}
\be
S_{\mathrm{SFT}}[\tilde{\phi}]
= 
\bigintssss d^d x
\left[
\frac{1}{2} \,
\tilde{\phi} \, \Box \, e^{- \ell_s^2 \, \Box} \, \tilde{\phi}
- 
V( \tilde{\phi} \, , \del \tilde{\phi} )
\right] 
,
\ee
which we will refer to as the \emph{stringy model} for convenience.

The interaction terms are now local, and the nonlocality resulting from the operator $e^{- \ell_s^2 \, \Box}$ can be studied \emph{non-perturbatively} with respect to the string length parameter $\ell_s$ by focusing on the free-field action:
\be 
\label{SFT_stringy}
S_{\mathrm{SFT}}^{(0)} [\tilde{\phi}]
= 
\frac{1}{2} 
\int d^d x \, 
\tilde{\phi} \, \Box \, e^{- \ell_s^2 \, \Box} \, \tilde{\phi}
\, .
\ee 

Previous efforts~\cite{Kajuri:2018myh, Kajuri:2017jmy} to address the stringy nonlocality have argued that it does not affect Hawking radiation, since transcendental functions of $\Box$ do not change the initial value problem for the original field equation $\Box \phi = 0$~\cite{Barci:1995ad, Barnaby:2007ve}.
However, we provide below a recap of ref.~\cite{Ho:2023tdq} on how a proper non-perturbative treatment of the nonlocal operator $e^{- \ell_s^2 \, \Box}$ in the stringy model~\eqref{SFT_stringy} indicates significant modifications to the amplitude of late-time Hawking radiation.

The behavior of late-time Hawking radiation is encoded in the Wightman function $\langle 0 | \hat{\tilde{\phi}} (x) \, \hat{\tilde{\phi}} (x') | 0 \rangle$ of the field operator $\hat{\tilde{\phi}} (x)$ at arbitrarily short distances. 
%
However, the operator $e^{-\ell_s^2 \, \Box}$ in the action~\eqref{SFT_stringy} introduces exponential UV divergences along time-like directions, which is apparent in the two-point correlation function
\be 
\label{SFT_2pt}
\ev{\tilde{\phi}(x) \, \Tilde{\phi}(x')}
=
\int \frac{d^d p}{(2 \pi)^d} \, 
\frac{- i e^{- \ell_s^2 \, p^2}}{p^2 - i \epsilon} \, 
e^{i p \cdot (x - x')} 
\ee 
from the path integral formalism. Consequently, some form of analytic continuation, such as Euclideanizing the Minkowski spacetime~\cite{Pius:2016jsl, Briscese:2018oyx, Chin:2018puw, Pius:2018crk, Buoninfante:2022krn}, is inevitable to regularize the divergence and extract meaningful results from the stringy model.

In the following, we shall focus on the case of $d = 2$ for simplicity.
Recently, a Hamiltonian approach for the stringy model in real time was developed based on consistency with its path-integral correlation functions~\cite{Ho:2023tdq,Hamiltonian-SFT}. The idea is to work with the analytically continued string length parameter:
\be 
\label{ellE}
\ell_E^2 = - i \ell_s^2 > 0
\, ,
\ee 
and formulate the theory in the light-cone frame
\be 
x^{\pm} \equiv t \pm x
\, .
\ee 
This is based on the observation that for modes with fixed retarded frequency $p_-$ with respect to the coordinate $x^-$, the effect of the nonlocal operator $\exp\left( - i \ell_E^2 \, \Box \right) = \exp\left( 4 i \ell_E^2 \, \del_{x^+} \del_{x^-} \right)$ simplifies to a finite shift $\exp\left( 4 \ell_E^2 \, p_- \, \del_{x^+} \right)$ in the light-cone time $x^+$.\footnote{
Since it is the combination $\ell_s^2 \, \del_{x^+}$ that appears in the exponent, this approach can alternatively be carried out with a real string length parameter $\ell_s$
while adopting the Euclidean time $x_E^+ = i x^+ \in \mathbb{R}$ \cite{Hamiltonian-SFT}.}

In the continued space of $\ell_E^2 > 0$, the two-point function~\eqref{SFT_2pt} can be obtained as
\begin{align}
&\ev{\tilde{\phi}(x^+ , x^-) \, \tilde{\phi}(y^+ , y^-)}
\nn \\
&
=
\int_{-\infty}^{\infty} dp_-
\int_{-\infty}^{\infty} \frac{dp_+}{2 \pi^2} \, 
\frac{-i \, e^{4 i \ell_E^2 p_- \, p_+}}{-4 p_- \, p_+ - i \epsilon} \, 
e^{-i p_- (x^- - \, y^-)} \, e^{-i p_+ (x^+ - \, y^+)}
\nn \\
&\hskip0em
=
\int_{-\infty}^{\infty}
\frac{dp_-}{4 \pi p_-} \, e^{-i p_- (x^- - \, y^-)}
\int_{-\infty}^{\infty} \frac{dp_+}{2 \pi} \, 
\frac{i}{p_+ + i \epsilon / p_-} \, 
e^{- i p_+ (x^+ - \, y^+ - 4 \ell_E^2 p_-)}
\nn \\
&\hskip0em
=
\int_0^{\infty} 
\frac{dp_-}{4 \pi p_-} \, 
\bigl[
e^{-i p_- (x^- - \, y^-)} \, 
\Theta(x^+ - y^+ - 4 \ell_E^2 p_-)
+
e^{i p_- (x^- - \, y^-)} \, 
\Theta(y^+ - x^+ - 4 \ell_E^2 p_-)
\bigr]
\, .
\end{align}
The goal is to construct the field operator $\hat{\tilde{\phi}} (x^+ , x^-)$ such that the correlation function can be reproduced through the correspondence
\begin{align}
&\ev{\tilde{\phi}(x^+ , x^-) \, \tilde{\phi}(y^+ , y^-)}
=
\ev*{\mathcal{T} \, 
\bigl\{ \hat{\tilde{\phi}} (x^+ , x^-) \, \hat{\tilde{\phi}} (y^+ , y^-) \bigr\}}{0}
\nn \\
&\hskip0em
\equiv 
\Theta(x^+ - y^+) 
\ev*{
\hat{\tilde{\phi}} (x^+ , x^-) \, \hat{\tilde{\phi}} (y^+ , y^-)}{0}
+
\Theta(y^+ - x^+) 
\ev*{
\hat{\tilde{\phi}} (y^+ , y^-) \, \hat{\tilde{\phi}} (x^+ , x^-)}{0}
\label{SFT_corres}
\end{align}
between the path integral formalism and the vacuum expectation value of time-ordered operators. 
(We have added hats on operators in the Hamiltonian formalism.)
By decomposing the outgoing sector of the field operator $\hat{\tilde{\phi}} (x^+ , x^-)$ into Fourier modes $e^{-i p_- x^-}$, we write
\be 
\label{SFT_phiop}
\hat{\tilde{\phi}} (x^+ , x^-)
= 
\int_0^{\infty} 
\frac{dp_-}{\sqrt{4\pi p_-}} 
\left[
\hat{a}_{p_-}(x^+) \, e^{- i p_- x^-} 
+ 
\hat{a}^{\dagger}_{p_-}(x^+) \, e^{i p_- x^-} 
\right] 
, 
\ee 
and define the vacuum state $\ket{0}$ by 
\be 
\hat{a}_{p_-} (x^+) \ket{0}
=
0
\qquad 
\forall \ 
p_- > 0
\, .
\ee 
We can then deduce from the correspondence~\eqref{SFT_corres} a self-consistent algebra for the creation and annihilation operators~\cite{Ho:2023tdq}:
\begin{align}
\bigl[ \hat{a}_{p_-}(x^+) \, , \hat{a}^{\dagger}_{p_-'}(y^+) \bigr] 
&= 
\Theta \left(\abs{x^+ - y^+} - 4 \ell_E^2 p_-\right) \, 
\delta(p_- - p_-') 
\, ,
\label{SFT_comm} \\
\bigl[ \hat{a}_{p_-}(x^+) \, , \hat{a}_{p'_-}(y^+) \bigr] 
&=
\bigl[ \hat{a}_{p_-}^{\dagger} (x^+) \, , \hat{a}^{\dagger}_{p_-'}(y^+) \bigr] 
=
0
\, .
\end{align}
Interestingly, the nonlocality of the stringy model is translated to a $p_-$-dependent minimal $x^+$ separation in the commutation relation of the creation-annihilation operators.

These unequal time commutation relations allow us to construct an operator formalism where 
the outgoing Wightman function is given by 
\be 
\langle 0 | 
\hat{\tilde{\phi}} (x^+ , x^-) \, 
\hat{\tilde{\phi}} (y^+ , y^-) 
| 0 \rangle
= 
\int_{0}^{|x^+ - \, y^+| / 4 \ell_E^2} 
\frac{dp_-}{2\pi} \, 
\frac{e^{- i p_- (x^- - \, y^-)}}{2 p_-}
\, .
\ee 
This reveals that the light-cone formulation of the stringy model~\eqref{SFT_stringy} exhibits nonlocality along light-cone directions, with a UV cutoff 
\be 
\label{SFT_cutoff}
p_-
\, \leq \, 
\frac{\abs{x^+ - \, y^+}}{4 \ell_E^2}
\ee 
on the retarded frequency, or equivalently, a minimal uncertainty in the advanced light-cone direction:
\be
\label{Dx+>Dp-}
\Delta x^+ \, \geq \,4 \ell_E^2 p_- \, ,
\ee
where $\Delta x^+ \equiv \abs{x^+ - \, y^+}$.

As noticed in ref.~\cite{Ho:2023tdq}, this reflects a \emph{UV/IR connection} in the stringy model, where the correlator between two spacetime points is nonvanishing only when their separation in the light-cone direction $x^+$ (or $x^-$) exceeds the nonlocality length scale 
$4 \ell_E^2 \, p_-$ (or $4 \ell_E^2 \, p_+$), which becomes macroscopic for large $p_-$ (or $p_+$). Approaching the UV limit $p_- \rightarrow \infty$ demands the IR limit $\abs{x^+ - y^+} \rightarrow \infty$.
Moreover, due to the standard uncertainty principle $\Delta x^- \Delta p_- \geq 1$ between conjugate variables, the upper bound~\eqref{SFT_cutoff} also implies 
\be 
\Delta x^-
\, \geq \,  
\frac{1}{\Delta p_-}
\, \gtrsim \,  
\frac{1}{p_-}
\, \geq \,  
\frac{4 \ell_E^2}{\Delta x^+}
\, .
\ee 
Thus, we find that a \emph{Lorentz-invariant} space-time uncertainty relation~\cite{Ho:2023tdq}
\be
\Delta x^+ \Delta x^- \gtrsim 4\ell_E^2
\label{Dx+Dx-}
\ee
is naturally incorporated into the operator formalism for the stringy model~\eqref{SFT_stringy} 
under a ``Wick rotation'' of $\ell_s^2$.
\hide{
Analogous to how we derived eq.~\eqref{Dx>DE} from eq.~\eqref{STUR}, we can also derive 
\be
\Delta x^+ > \ell_E^2 \Delta p_- 
\label{Dx+>Dp-}
\ee
from eq.~\eqref{Dx+Dx-}.
}
\hide{
The UV/IR condition~\eqref{SFT_cutoff} also imposes a constraint on quantum statesl~\cite{Ho:2023tdq,Hamiltonian-SFT}. For instance, a generic one-particle $p_-$-eigenstate
\be
\ket{1_{p_-}}
\equiv 
\int dx^+ \, 
\psi_{p_-}(x^+) \, 
\hat{a}_{p_-}^{\dagger}(x^+) 
\ket{0}
\ee
has a norm proportional to
\begin{align}
\inp{1_{p_-}}{1_{p_-}}
\propto 
\int dx^+ \int dy^+ \, 
\psi_{p_-}^{\ast}(x^+) \, \psi_{p_-}(y^+) \, 
\Theta ( \abs{x^+ - y^+} - 4 \ell_E^2 p_-) 
\, ,
\end{align}
which vanishes if $\psi_{p_-}(x^+)$ has finite support less than an interval $4 \ell^2_E p_-$.
In other words, a quantum mode with retarded frequency $p_-$ in the stringy model must be defined over a range of $\Delta x^+ > 4 \ell^2_E p_-$.
}

Due to the nontrivial $x^+$-dependence of the ladder operators $\hat{a}_{p_-}(x^+)$ and $\hat{a}_{p_-}^{\dagger}(x^+)$ required by the correspondence~\eqref{SFT_corres} with the path integral result, the field operator $\hat{\tilde{\phi}}(x^+ , x^-)$ constructed in eq.~\eqref{SFT_phiop} does not satisfy the equation of motion, i.e. $e^{-i \ell_E^2 \Box} \, \Box \, \hat{\tilde{\phi}} \neq 0$. 
This is expected in higher-derivative theories of infinite order. 
The equation of motion should instead be imposed as a constraint defining the physical Hilbert space~\cite{Llosa94, Woodard:2000bt, Gomis:2000gy, Heredia:2021wja, Heredia:2022mls, Hamiltonian-SFT}. 
This feature marks a major difference from the perturbative treatments of nonlocal theories~\cite{Barci:1995ad, Cheng:2001du, Kajuri:2017jmy}. 

The $x^+$-dependence of these operators appears to introduce an infinite number of extra degrees of freedom into the state space of the stringy model --- a common feature of theories with infinite time derivatives that typically leads to instabilities or violations of unitarity.
However, as shown in ref.~\cite{Hamiltonian-SFT}, for the particular model under consideration, the physical-state constraint ends up eliminating the negative-norm states and, apart from the decoupled zero-norm states, reducing the physical degrees of freedom to match those of standard local quantum field theory. 
(See ref.~\cite{Hamiltonian-SFT}  for more details.)

For our purpose, we only need a free field theory (including a background field interaction).
While our calculation of Hawking radiation only relies on the Wightman function, it has been shown that its Hamiltonian formalism can be constructed consistently~\cite{Hamiltonian-SFT}.
The progress is based on a light-cone formulation with a light-cone time coordinate and the analytic continuation of the string tension.
This is a remarkable result, as the Hamiltonian formalisms for nonlocal theories are typically pathological~\cite{Ostrogradsky:1850fid, Eliezer:1989cr, Woodard:2015zca}. 
After all, when there is nonlocality in the time direction, it is not clear what the meaning of a Hamiltonian formalism is.
It remains to be seen whether a complete, consistent Hamiltonian formalism including all component fields in SFT can be achieved by the same approach.



\subsubsection{Derivation of Hawking Radiation in Stringy Model}

We now apply the newly developed operator formalism for the stringy model to the radiation field $\tilde{\phi}$ in a Schwarzschild background. We shall focus on the near-horizon region again, where the Kruskal coordinates $U$, $V$ are identified with the light-cone coordinates $x^-$, $x^+$ above, and the frequency $\Omega$ with $p_-$.

Following the procedure outlined in sec.~\ref{sec:HR-LEET}, a Hawking particle with central frequency $\omega = \omega_0$ localized around the Eddington retarded coordinate $u = u_0$ can be described by the wave packet
\be 
\label{SFT_packet}
\psi (u)
=
\int_0^{\infty} 
\frac{d \omega}{{\sqrt{\omega}}} \, 
f_{\omega_0} (\omega) \, 
e^{-i \omega (u - u_0)}
\, .
\ee 
Let $\hat{b}_{\omega} (v)$ and $\hat{b}^{\dagger}_{\omega} (v)$ denote the ladder operators associated with an outgoing Hawking mode $e^{-i \omega u}$ with positive frequency $\omega > 0$ near asymptotic infinity.
The annihilation operator $\hat{b}_{\psi}$ corresponding to the wave packet~\eqref{SFT_packet} is defined as 
\be 
\hat{b}_{\psi} (v)
\equiv 
\int_0^{\infty} d\omega \, 
f_{\omega_0}^{\ast} (\omega) \, 
e^{-i \omega u_0} \, 
\hat{b}_{\omega} (v)
\, ,
\ee 
where the only deviation from the standard case is the dependence of the operators on the advanced Eddington time $v$ due to the nonlocality in the stringy model.

In the freely falling frame near the horizon, the outgoing sector of the radiation field $\hat{\tilde{\phi}}$ has the Minkowski mode expansion
\be 
\hat{\tilde{\phi}} (U, V)
= 
\int_0^{\infty} 
\frac{d\Omega}{\sqrt{4\pi\Omega}} 
\left[
\hat{a}_{\Omega}(V) \, e^{- i \Omega U} 
+ 
\hat{a}^{\dagger}_{\Omega}(V) \, e^{i \Omega U} 
\right] 
.
\ee 
The quantity of interest is the expectation value of the number operator $\hat{n}_{\psi} (v, v') = \hat{b}_{\psi}^{\dagger} (v) \, \hat{b}_{\psi} (v')$ in the free-fall vacuum state $\ket{0}$ defined by 
\be 
\hat{a}_{\Omega}(V) \ket{0} = 0
\qquad 
\forall \ \Omega > 0 
\, .
\ee 
Although the ladder operators acquire time dependence in the stringy model, they are still related by the same Bogoliubov transformation \eqref{Bogoliubov} in the conventional theory:
\be 
\label{SFT_bb}
\langle 0 | \, 
\hat{b}_{\omega}^{\dagger} \bigl( v(V) \bigr) \, 
\hat{b}_{\omega^{\prime}} \bigl( v'(V') \bigr) 
\ket{0}
=
\int_{0}^{\infty} d \Omega 
\int_{0}^{\infty} d \Omega' \, 
\beta_{\omega \Omega}^{\ast} \, 
\beta_{\omega'\Omega'} \, 
\langle 0 | \, 
[ \hat{a}_{\Omega}(V) \, , \hat{a}_{\Omega'}^{\dagger}(V') ]
\ket{0}
\, ,
\ee 
since the Bogoliubov coefficients $\beta_{\omega \Omega}$ are completely determined by the exponential relation $U(u) = - 2a e^{-u / 2a}$ between the retarded coordinates.

Inserting the free-field commutator~\eqref{SFT_comm}
\be 
[\hat{a}_{\Omega}(V) \, , \hat{a}_{\Omega'}^{\dagger}(V')]
= 
\delta(\Omega - \Omega') \, 
\Theta( \abs{V - V'} - 4 \ell_E^2 \Omega )
\ee 
into eq.~\eqref{SFT_bb} leads to the number expectation value~\cite{Ho:2023tdq}
\begin{align}
&
\langle 0 | \, 
\hat{n}_{\psi} \bigl( v(V), v'(V') \bigr)
\ket{0}
\nn \\
&= 
\int_0^{\infty} d \omega 
\int_0^{\infty} d \omega' \, 
f_{\omega_0}^{\ast}(\omega) \, 
f_{\omega_0} (\omega') \, 
e^{i \left( \omega - \omega' \right) u_0} \, 
\langle 0 | \, 
\hat{b}_{\omega}^{\dagger} \bigl( v(V) \bigr) \, 
\hat{b}_{\omega^{\prime}} \bigl( v'(V') \bigr) 
\ket{0}
\nn \\
&\simeq 
\frac{a}{\pi} \frac{1}{e^{4 \pi a \omega_0} - 1}
\int_0^{\infty} d \omega 
\int_0^{\infty} d \omega' \, 
f_{\omega_0}^{\ast} (\omega) \, 
f_{\omega_0} (\omega') \, 
e^{i \left( \omega - \omega' \right) u_0} 
\int_0^{\abs{V - V'} / 4 \ell_E^2} 
\frac{d \Omega}{\Omega} 
\left( 2 a \Omega \right)^{-2 i a \left( \omega - \omega' \right)}
\, .
\label{SFT_bb_psi}
\end{align}
By performing a change of variable $\Omega \mapsto u(\Omega) = 2 a \log \left( 2 a \Omega \right)$, the $\Omega$-integral above can be evaluated as 
\be 
\label{SFT_dOm_du}
\int_0^{\abs{V - V'} / 4 \ell_E^2} 
\frac{d \Omega}{\Omega} 
\left( 2 a \Omega \right)^{-2 i a \left( \omega - \omega' \right)}
=
\int_{-\infty}^{u_{\Lambda} (\abs{V - V'} \, , \, \ell_E^2)}
\frac{du}{2 a} \,
e^{-i \left( \omega - \omega' \right) u}
\, ,
\ee 
where 
\be 
\label{SFT_uLam}
u_{\Lambda} (\abs{V - V'} \, , \ell_E^2)
\equiv 
2 a
\log 
\left( \frac{a \abs{V - V'}}{2 \ell_E^2} \right)
.
\ee 
Ultimately, the physical observables of the original theory are recovered through an analytic continuation $\ell_E^2 \to -i \ell^2$ back from the complexified string length.
Under this continuation, $u_{\Lambda} (\abs{V - V'} \, , \ell_E^2)$ as a function of $\ell_E^2$ gets turned into 
\be 
u_{\Lambda} (\abs{V - V'} \, , \ell_E^2)
\quad \to \quad 
u_{\Lambda} (\abs{V - V'} \, , \ell_s^2)
+
i \pi a
\, .
\ee 
The additional term $i \pi a$ introduces an extra factor $e^{\pi a \left( \omega - \omega' \right)}$ in the integrand of eq.~\eqref{SFT_dOm_du}, which contributes merely a factor of $\mathcal{O}(1)$ to the final outcome~\eqref{SFT_bb_psi} since the compact profile $f_{\omega_0}$ restricts both $\omega$ and $\omega'$ to values near $\omega_0$.
Thus, the expectation value becomes 
\be 
\label{SFT_bb_psi2}
\langle 0 | \, 
\hat{n}_{\psi} (V, V')
\ket{0}
\simeq 
\frac{1}{2 \pi} \, 
\frac{\omega_0}{e^{4 \pi a \omega_0} - 1}
\int_{-\infty}^{u_{\Lambda} (\abs{V - V'} \, , \, \ell_s^2)} 
\abs{\psi (u)}^2 \, 
du 
\, ,
\ee 
which again showcases an amplitude that decays in $u_0$, this time due to the UV cutoff $\Omega \leq \abs{V - V'} / 4 \ell_E^2$.
The temperature remains the same as the Hawking temperature~\eqref{Hawking-Temperature}.

However, we emphasize that the physical interpretation here differs significantly from a simple truncation of the effective field theory by imposing a UV cutoff on the retarded frequency $\Omega$, as it was done in sec.~\ref{sec:cutoff}. 
Throughout the discussion, we have not excluded the presence of trans-Planckian modes.
Instead, as we will argue, these modes simply do not contribute to Hawking process.

In a physical setting, a stationary observer would detect radiation over a finite duration $V_{\mathrm{d}}$ of light-cone time and measure the expectation value
\be 
N_{\psi}(V_{\mathrm{d}})
\equiv 
\frac{1}{V_{\mathrm{d}}^2}
\int dV \, s(V) 
\int dV' \, s(V') \, 
\langle 0 | \, 
\hat{n}_{\psi} (V, V')
\ket{0}
\, ,
\ee 
where the switching function $s(V)$~\cite{Takagi:1986kn} associated with a particle detector
is normalized as 
\be 
\int s(V) \, dV = V_{\mathrm{d}} \, .
\ee 
Since the wave packet $\psi(u)$~\eqref{SFT_packet} of a Hawking particle is localized around the retarded time $u = u_0$ with a width of $\mathcal{O}(a)$, eq.~\eqref{SFT_bb_psi2} indicates that the mean number $N_{\psi}(V_{\mathrm{d}})$ of Hawking particles measured would vanish for sufficiently large $u_0$ when 
\be 
u_0 
\, \gg \, 
u_{\Lambda} (V_{\mathrm{d}} \, , \ell_s^2)
+
\mathcal{O}(a)
\, ,
\ee
where $u_{\Lambda}$ is defined in eq.~\eqref{SFT_uLam}.
In particular, the expression~\eqref{SFT_uLam} for $u_{\Lambda} (V_{\mathrm{d}} \, , \ell_s^2)$ implies that a Hawking particle emitted on the order of the scrambling time scale with
\be 
u_0 
= 
2 \left( n + 1 \right) a 
\log \left( \frac{a}{\ell_s} \right)
\label{u0}
\ee 
requires a detection range
\be 
V_{\mathrm{d}}
\, \gg \, 
\frac{\ell_s^2}{a} \, e^{u_0 / 2 a} 
=
\left( \frac{a}{\ell_s} \right)^{n - 1} 
a 
\label{Vd>>}
\ee 
that vastly exceeds the size $a$ of the black hole as long as $n > 1$.

\begin{figure}[t]
\centering
\includegraphics[width=0.4\linewidth]{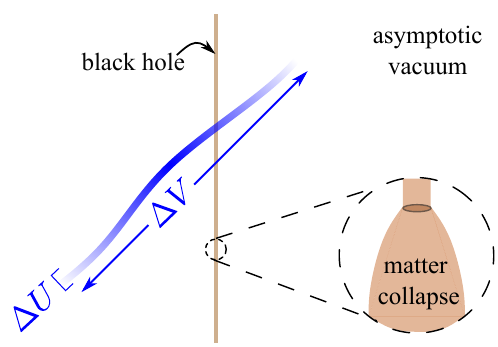}
\caption{
A schematic illustration of a high-frequency outgoing wave packet with a narrow width $\Delta U$. The wave packet behaves like an extended string spanning a range $\Delta V > 4 \ell_s^2 / \Delta U$. The vertical line represents the worldvolume of matter that eventually collapses into a black hole with a size considerably smaller than $\Delta V$. Since the wave packet cannot be localized near the horizon, it does not contribute to Hawking radiation.
}
\label{fig:stringy}
\end{figure}

The result~\eqref{Vd>>} is simply a reflection of the UV/IR constraint~\eqref{SFT_cutoff} for a large $p_- = \Omega$.
A Hawking particle with central frequency $\omega_0\sim 1/a$ at retarded time $u_0$~\eqref{u0} involves modes with trans-Planckian free-fall frequencies 
\be 
\Omega 
\sim 
\frac{1}{a} \, e^{u_0 / 2a}
=
\left( \frac{a}{\ell_s} \right)^{n - 1}
\frac{a}{\ell_s^2} \, .
\ee
Rather than being confined within a Planckian distance $\sim 1/\Omega$ around the horizon,
these trans-Planckian modes should be viewed as extremely long stringy states that span an interval (see eq.~\eqref{SFT_cutoff} and identify $|x^+ - y^+|$ with $\Delta V$)
\be 
\Delta V
\, > \, 
4 \ell_s^2 \Omega 
\sim 
\left( \frac{a}{\ell_s} \right)^{n - 1} 
a \, ,
\label{eq:DV>4l2}
\ee 
which far exceeds the size of the black hole for $n > 1$, as illustrated in fig.~\ref{fig:stringy}.
Eq.~\eqref{Vd>>} is reproduced as the requirement $V_d \gg \Delta V$.
That is, the detector has to be switched on for a duration $V_d$ much longer than the uncertainty $\Delta V$.

The black-hole evaporation process can be viewed as a collection of scattering processes. 
The initial states of the scattering include the ingoing collapsing matter and the Minkowski vacua of outgoing modes.
The final states include a black hole and outgoing Hawking particles.
In the low-energy approximation, the Schwarzschild geometry accounts for the effect of the exchanges of virtual gravitons.\footnote{Recall that the Schwarzschild metric can be reconstructed from a perturbative expansion of scattering amplitudes in Minkowski space in low-energy effective theories~\cite{Bern:2019nnu, Bern:2019crd, Jakobsen:2020ksu, Mougiakakos:2020laz, Driesse:2024xad, Damgaard:2024fqj}.}
This low-energy effective description fails for Hawking particles after the scrambling time, as we saw in sec.~\ref{sec:UV-physics}.
In the high-energy limit, the UV/IR relation~\eqref{SFT_cutoff} obtained in the stringy model suggests that the underlying geometry is smeared over at a scale $\Delta V \gg a \left( a/\ell_s \right)^{n-1}$~\eqref{eq:DV>4l2}.
From the viewpoint of Hawking particles after the scrambling time, the black hole is smeared by a length scale much larger than $a$.

For example, consider a solar-mass black hole whose Schwarzschild radius is $3 \; km$ and assume that $\ell_s = \ell_p$.
For $n = 2$ or equivalently $u_0 \sim 5 \; ms$, the required detection period $V_{\mathrm{d}}$ has to be much larger than $10^{15}$ times the size of the observable universe!

Similar to the situation with the GUP, the black hole background would have had little to no impact on the evolution of these high-frequency modes.\footnote{This statement is supported by demonstrating that the corrections to the correlation function $\ev{\tilde{\phi}(U, V) \, \tilde{\phi}(U' , V')}$ due to interactions between high-energy modes and a time-dependent background with certain compact profiles are highly suppressed~\cite{Hamiltonian-SFT}.}
These modes remain in the asymptotic vacuum state and do not contribute to particle creation in the first place. As a result, we conclude that Hawking radiation no longer exists after the scrambling time $\mathcal{O} \bigl( a \log (a / \ell_s) \bigr)$.

\section{Comments}
\label{sec:Comments}

We examined two UV models studied in the literature for Hawking radiation but arrived at the surprising conclusion that Hawking radiation is turned off as early as the scrambling time.
Why has this simple resolution of the information paradox not been proposed earlier? 
How does this proposal contradict or align with the existing literature on black-hole evaporation? 
We comment on related questions in this section.

\subsection{Comments on Theoretical Issues}

Before commenting on related works in sec.~\ref{sec:RelatedWorks}, we first discuss the theoretical issues involved in our proposal.

\subsubsection{Robustness of Hawking Radiation}

While the magnitude of Hawking radiation is sensitive to UV physics, the Hawking temperature ($T_H = 1 / 4 \pi a$ as given in eq.~\eqref{Hawking-Temperature}) remains robust against Planck-scale effects, with perturbative corrections of $\mathcal{O}(\ell_p^2/a^2)$~\cite{Chau:2023zxb}.

The mathematical reason is that the Boltzman factor $e^{- 4 \pi a \omega}$, which appears from the ratio $\abs{\beta_{\omega \Omega} / \alpha_{\omega \Omega}}^2$ of the Bogoliubov coefficients~\eqref{alpha}--\eqref{beta}, is an infrared feature essentially determined by the analytic continuation of the factor $(2 a \Omega - i \epsilon)^{2 i a \omega}$ across $\Omega = 0$ from positive to negative values of $\Omega$. 
The $i \epsilon$-prescription that one should adopt in the Fourier transform of an outgoing Hawking mode $e^{- i \omega u}$ is fixed by the causality and locality of the mode propagation. 

Most claims in the literature regarding the robustness of Hawking radiation refer specifically to the robustness of the Hawking temperature. In particular, the insensitivity of the Hawking temperature and its associated thermodynamic properties to the GUP has been noted repeatedly in past studies~\cite{Amelino-Camelia:2005zpp, Majumder:2012rtc, Miao:2014jea, Wang:2014cza, Chen:2014xgj, Bargueno:2015tea, Mu:2015qta, Sakalli:2016mnk, Gecim:2017zid, Lambiase:2017adh, Kanazawa:2019llj, Buoninfante:2019fwr, Anacleto:2021nhm, Anacleto:2023ali}. However, discussions often overlook details about the magnitude of radiation. The time dependence of the magnitude of Hawking radiation has not been thoroughly examined as the temperature has been.

Since the Hawking temperature remains the same, the usual derivation of the Bekenstein-Hawking entropy \eqref{BH-entropy} from $dS = dM/T_H$ remains valid. 

As a dynamic process, a time-dependent magnitude of radiation from the collapsing matter is somewhat expected. This is analogous to the thermal Larmor radiation emitted by a moving point charge, which has an exceedingly small amplitude if the final speed is sufficiently low \cite{Ievlev:2023inj}.

\subsubsection{Local Lorentz Symmetry and Nice Slices}

In sec.~\ref{sec:Why-UV}, we explained that local Lorentz symmetry and the nice-slice argument have been utilized to claim that UV physics is irrelevant to Hawking radiation, and so Hawking radiation is expected to persist until the black hole becomes microscopic; we also clarified there that these are misconceptions.

In sec.~\ref{sec:UV-physics}, we demonstrated explicitly how the low-energy effective theory fails in $S$-matrix calculations concerning Hawking radiation when higher-derivative interactions are considered. The nice-slice argument merely ensures that a low-energy effective theory can describe the time evolution of the background geometry defined in the low-energy limit; however, it says nothing about the trans-Planckian scattering involving the background and the Hawking particle. The nice-slice argument cannot ensure that Hawking radiation persists adiabatically.

We have also shown in sec.~\ref{sec:StringyModel} that a Lorentz-invariant uncertainty relation \eqref{Dx+Dx-} at the Planck scale inhibits Hawking radiation around the scrambling time.
Just like long wavelength modes perceive a smoother background, eq.~\eqref{Dx+>Dp-} implies that trans-Planckian modes do, too.
Instead of saying that local Lorentz symmetry is broken, we would rather say that the spacetime geometry is probe-dependent, and the notion of local Lorentz symmetry is different for different probes.

\subsection{Comments on Related Works}
\label{sec:RelatedWorks}

The information paradox has attracted a lot of attention over the past 50 years. 
Numerous works have been devoted to this topic. 
Here we comment on the connections with other works in the literature.

\subsubsection{AdS/CFT Duality and Entanglement Entropy}

The AdS/CFT duality \cite{Maldacena:1997re, Witten:1998qj} serves as a concrete illustration
of the holographic principle of quantum gravity \cite{tHooft:1993dmi, Susskind:1994vu}, having collected extensive evidence over time. This duality offers profound insights into the physics underlying the Bekenstein-Hawking entropy \eqref{BH-entropy} and sheds light on the information paradox. On the other hand, while it provides a strong argument for unitarity (no information loss), it does not explicitly reveal the underlying physical mechanism.

The AdS/CFT duality motivated a formula for the generalized fine-grained entropy~\cite{Ryu:2006bv, Hubeny:2007xt, Faulkner:2013ana, Engelhardt:2014gca}. Recently, with the help of ``entanglement islands''~\cite{Penington:2019npb, Almheiri:2019psf, Almheiri:2019hni, Almheiri:2019yqk, Penington:2019kki, Almheiri:2019qdq}, the generalized entropy formula reproduced the Page curve for an evaporating black hole.
However, despite these advancements, the island proposal still faces various criticisms~\cite{Geng:2020qvw, Geng:2021hlu, Geng:2023zhq, Landgren:2024ccz} and lacks a clear explanation of the physical mechanism for information retrieval~\cite{Guo:2021blh} or the ultimate fate of the black hole. 

Notably, the generalized entropy formula applies to any spacetime geometry, whether it describes a classical black hole or one undergoing continuous evaporation. 
Regarding the scenario proposed in this chapter, where Hawking radiation terminates near the scrambling time, the entanglement island is irrelevant, as its effects would only become significant if Hawking radiation persists beyond the Page time.


It is important to note that nonlocality is a crucial aspect of the discussion surrounding entanglement islands, as well as many other proposed resolutions to the information paradox. Indeed, the essence of the AdS/CFT duality --- the holographic principle~\cite{tHooft:1993dmi, Susskind:1994vu} --- may stem from a type of nonlocality that reveals a UV/IR connection. 

In our analysis, we have explored specific forms of nonlocality (GUP and STUR) inspired by string theory studies. 
However, it remains uncertain whether these formulations accurately reflect the nonlocality inherent in string theory. 
It will be interesting to see whether they correspond to the same type of nonlocality that contributes to the holographic nature of string theory.

\subsubsection{Fuzzball}

The fuzzball proposal \cite{Mathur:2005zp, Skenderis:2008qn} claims that horizons exist only in coarse-grained spacetime geometries.
(See, for example, refs.~\cite{Bena:2022ldq, Bena:2022rna} for recent reviews.) 
What we traditionally understand as a black hole is actually an ensemble of horizonless microstate configurations that share the same mass, charge, and angular momentum as a classical black hole.

Since fuzzballs lack horizons, the exponential relation \eqref{eq:U-u} between $U$ and $u$ is altered near the Schwarzschild radius.
Consequently, the conventional derivation of Hawking radiation does not apply.
This aspect of the fuzzball paradigm is similar to our proposal in the sense that Hawking radiation occurs only for a brief period.
On the other hand, fuzzballs exhibit enhanced radiation \cite{Chowdhury:2007jx} due to classical instabilities in spacetime geometry. This situation is analogous to the burning of a piece of coal. The fuzzball's radiation has a fundamentally different origin compared to traditional Hawking radiation. 

\hide{
The formation of a fuzzy horizon may seem to violate the equivalence principle. But it was argued that, even though the probability for a collapsing matter shell to tunnel into a horizonless fuzzball configuration is extremely small, the plethora of such states $\sim e^{S_{BH}}$ \cite{Kraus:2015zda} compensates this small number to make the probability of $\mathcal{O}(1)$.
}


Both the fuzzball proposal and the approach presented in this chapter are motivated by string theory, though each emphasizes different aspects of it.

In general, whether a collapsing matter shell tunnels into a fuzzball~\cite{Mathur:2008kg,Bena:2015dpt} may depend on its initial state.
This suggests that one might be able to classify black holes based on their ultimate fates, with possibly multiple scenarios for their final state configurations.

\subsubsection{Firewall}


It was argued that, assuming that information is transmitted from matter to Hawking radiation,
there must be a firewall before Page time~\cite{Almheiri:2012rt}.
As the Unruh vacuum is a pure state, Hawking particles outside the apparent horizon are maximally entangled with their partners inside the black hole, hence the quantum state inside the apparent horizon can no longer be viewed as a vacuum state.
At a later time, when the partner emerges from the shrinking apparent horizon, its identity is already determined by its maximally entangled Hawking particle already detected in the early radiation.
As such, the quantum state at the apparent horizon can no longer be viewed as a pure state.
It must be viewed as a particle at the apparent horizon even for freely falling observers, and its frequency is trans-Planckian due to the large blue-shift factor.\footnote{The firewall can also be argued from the AdS/CFT duality \cite{Marolf:2013dba} without using the entanglement argument.}

This proposal argues that the firewall is a necessity for information to come out in Hawking radiation after the Page time.
In our proposal, Hawking radiation stops well before the Page time and information stays inside the black hole, so there is no need for a firewall or the violation of the equivalence principle.

\subsubsection{Massive Remnants}

Our proposal bears resemblance to the concept of ``\emph{massive remnants}''~\cite{Giddings:1992hh}, which suggests that Hawking radiation terminates at a time well beyond the Page time, leaving behind a macroscopic remnant characterized by an entropy density limited by the Planck density. 
This type of remnant was proposed to retain only a small fraction $\sim (\ell_p M)^{-1/3}$ of its initial mass (but still macroscopic).
It does not resolve the conflict between the Bekenstein-Hawking entropy and the entanglement entropy that arises around the halfway point of evaporation (see sec.~\ref{sec:Information-Problem}).

A similar idea is the memory burden effect~\cite{Dvali:2020wft}, which suggests that the further decay of a black hole is suppressed beyond the Page time.
Ultimately, these proposals require a mechanism to stop Hawking radiation well before the black hole becomes microscopic.
In out approach, this early termination is accomplished through the nonlocal nature of strings.

\subsubsection{Analogue Gravity}

The results from analogue gravity have contributed to the belief in the adiabaticity of Hawking radiation.
The original concept emerged from Unruh's sonic model \cite{Unruh:1980cg} of a black hole, 
which is characterized by an inhomogeneous fluid flow and a sonic horizon. 
In this analogy, the spontaneous emission of sound waves from the acoustic horizon can be likened to Hawking radiation \cite{Garay:1999sk}. 
The breakdown of the continuous description of the fluid, caused by atomic effects at submolecular scales, allows analogue black holes to be used as a platform to investigate Hawking radiation and its connections to short-distance physics.

Nevertheless, in much of the analogue gravity literature~\cite{Barcelo:2005fc}, the introduction of new UV physics is primarily limited to modifications of the dispersion relation in the UV regime. 
Although Lorentz symmetry may be broken, such modifications are relatively mild compared to a UV cutoff or nonlocality. 
Consequently, it is plausible that many studies report only a slightly modified version of Hawking radiation.

An exception is recently reported in ref.~\cite{Akhmedov:2023gqf}, which presents a UV dispersion relation that alters the propagation of the Hawking modes, making them sensitive to the boundary conditions at the black hole singularity. 
Under specific boundary conditions, this modification also leads to the termination of Hawking radiation.

\subsubsection{Eternal Black Holes}

The information paradox is initially posed in the context of a dynamical black hole, 
as discussed throughout this work. 
However, an analogous issue arises 
for an eternal black hole in $AdS$ space~\cite{Maldacena:2001kr}. 
In $AdS_4$, if Hawking radiation does not carry information
and proceeds as the low-energy effective theory predicts, 
a bulk correlation function should decay exponentially over time. 
The paradox lies in the fact that boundary correlators in the CFT on $S^3$, 
which are linked to the bulk correlators via the extrapolate dictionary, 
can only decay to a minimum value $e^{-c \, \mathcal{S}_{\mathrm{BH}}}$, 
where $c$ is a constant of order one and $\mathcal{S}_{\mathrm{BH}}$
is the Bekenstein-Hawking entropy~\eqref{BH-entropy}.

Our proposal offers a potential resolution to this discrepancy 
by suggesting that the effects of Hawking radiation may effectively be disregarded. 
In this scenario, nearly all information remains within the black hole (which only has a horizon in the low-energy effective theory), 
which possesses a degeneracy of approximately $e^{c \, \mathcal{S}_{\mathrm{BH}}}$. 
Consequently, it is natural for the correlation function 
to asymptotically approach $e^{-c \, \mathcal{S}_{\mathrm{BH}}}$ as $t \rightarrow \infty$.

\subsubsection{Cosmology}

A challenge analogous to the trans-Planckian problem in Hawking radiation 
also arises in cosmological inflation, 
where low-energy effective theories are utilized to derive primordial cosmological density perturbations despite the involvement of trans-Planckian modes~\cite{Martin:2000xs,Brandenberger:2012aj}.

A UV cutoff that halts Hawking radiation at the scrambling time is translated to a UV cutoff in the static patch of de Sitter space that stops the Gibbons-Hawking radiation~\cite{Blamart:2023ixr}.
A bound ($t \sim H^{-1} \log(H^{-1}M_p)$) analogous to the scrambling time has been established as the time scale of the validity of effective field theory in de Sitter space~\cite{Blamart:2023ixr}, agreeing with the predictions from the Trans-Planckian Censorship Conjecture (TCC)~\cite{Bedroya:2019snp,Bedroya:2019tba}.
Turning off Hawking radiation at the scrambling time is thus closely related to the TCC.

The TCC suggests that the universe avoids revealing details about trans-Planckian modes to the macroscopic world.
This idea aligns with the intuition of a minimal uncertainty in the spacetime geometry.
In ref.~\cite{Brandenberger:2024vgt}, the spacetime uncertainty relation is examined in the context of the TCC.

In the early universe, primordial black holes formed during the Big Bang. 
Smaller black holes, typically assumed to evaporate completely, radiate at higher temperatures.
If Hawking radiation stops in the early stages and these black holes survive, their implications for cosmology could be significant as they may act as dark matter candidates.
Further exploration of this new mass window for primordial black holes inspired by string theory considerations would be intriguing.
Similar studies motivated by the memory burden effect that suppresses black-hole decay near the halfway point of evaporation have been considered in refs.~\cite{Alexandre:2024nuo, Thoss:2024hsr, Dvali:2024hsb}.

\section{Conclusion}
\label{sec:Conclusion}

In the string-motivated models discussed in sec.~\ref{sec:Hawking-Radiation-Turned-Off},
the most important ingredient that turns off Hawking radiation is the nonlocality exhibiting a UV/IR connection: there is an IR uncertainty in the UV limit.
Although the nonlocality is restricted to the UV regime, it has an IR effect. 
It is reminiscent of the UV/IR connections observed in T-duality, noncommutative field theories, AdS/CFT duality, etc. in string theory.
This UV/IR connection is the reason why quantum modes for would-be Hawking particles after the scrambling time have spatial uncertainties too large to fit inside the near-horizon region, causing Hawking radiation to cease.

\begin{figure}[t]
\centering
\includegraphics[scale=0.8]{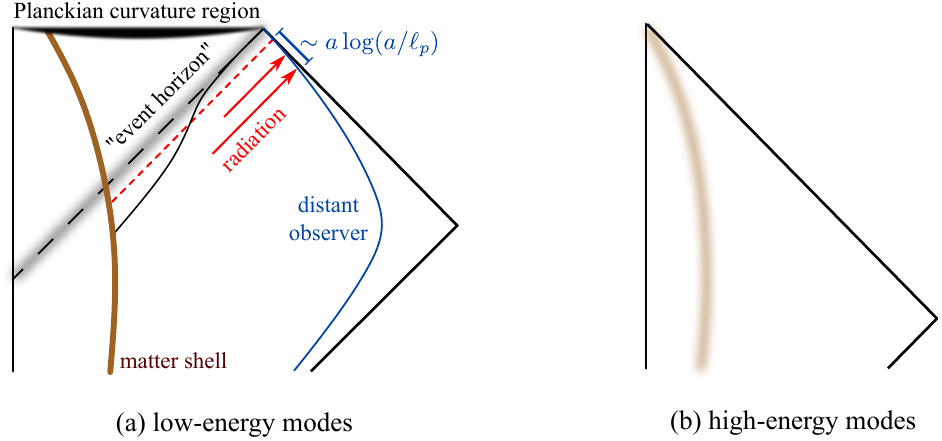}
\caption{\label{fig:penrose}
Panel~(a) shows the Penrose diagram of a black hole with Hawking radiation terminated at an early stage marked by the dashed red line.
The background is defined in the low-energy limit. 
While the spacetime metric may not be well-defined for trans-Planckian modes, panel~(b) schematically illustrates the Penrose diagram of a highly smeared black-hole geometry, representing the background on which the vacuum states for extremely high-energy modes with $\Delta x \gg a$ are defined.
}
\end{figure}

From the perspective of a scale $\gg \mathcal{O}(a)$, Hawking radiation can be viewed as a scattering process between the radiation field and the collapsing matter through gravitational interactions. 
In the UV models discussed in secs.~\ref{sec:GUP} and \ref{sec:StringyModel}, the larger nonlocality associated with higher-frequency modes is expected to result in a smeared geometry.
When this nonlocality for trans-Planckian modes surpasses the size of the black hole, the black hole is smeared out as a smooth background (see fig.~\ref{fig:penrose}), leading to a cessation of Hawking radiation.

From the perspective of a UV theory that may radically differ from our conventional understanding of spacetime, geometric notions such as the Schwarzschild metric only emerge as effective descriptions in the low-energy limit. 
They are low-energy effective concepts only good for low-energy probes.
In string field theories, the nonlocality in spacetime originates from exponentially suppressed interactions in the UV limit.
This feature of a weaker gravity in the high-energy limit is shared by other theories, such as asymptotically safe gravity~\cite{Bonanno:2000ep,Reuter:2019byg,Bonanno:2020bil,Saueressig:2023irs}. 
Our idea is also reminiscent of gravity's rainbow~\cite{Magueijo:2002xx}, that is, the spacetime geometry depends on the energy scale of the propagating modes, but we are free from the dangers of violating Lorentz symmetry in flat space~\cite{Hossenfelder:2010tm}.
In our proposal, there is no correction to the Minkowski spacetime.\footnote{Hence, we also do not expect significant modifications to the Unruh effect in Minkowski spacetime, which is consistent with the findings of ref.~\cite{Hata:1995di}.}

This article proposes an alternative resolution to the black hole information paradox. 
Instead of focusing on the details of Hawking radiation at late times to look for the desired information transmission from the collapsing matter, we argue that Hawking radiation is effectively suppressed before the Page time, stopping around the scrambling time due to the dominance of trans-Planckian effects.
This suppression is arguably a consequence of the fundamental nonlocality present in UV-complete theories of quantum gravity such as string theory.

As black holes can approximately be treated as classical objects in this scenario, all paradoxes associated with Hawking radiation are alleviated. 
This perspective offers a conceptually straightforward resolution to the information paradox.

We support this claim with the following discussions:
\begin{enumerate}
\item
Analyzing the breakdown of the low-energy effective theory: We demonstrate that higher-derivative non-renormalizable interactions, always present in any UV completion of general relativity, lead to exponentially growing contributions to particle creation around the scrambling time, invalidating perturbative calculations of Hawking radiation.
(See sec.~\ref{sec:UV-physics}.)
\item
Modeling UV physics using GUP and SFT: Two concrete UV models, one incorporating the generalized uncertainty principle (GUP) and the other based on string field theory (SFT), are analyzed to show that trans-Planckian modes effectively do not ``see'' the black hole geometry and thus do not contribute to late-time Hawking radiation.
The suppression of Hawking radiation is attributed to the inherent nonlocality in these models.
(See sec.~\ref{sec:Hawking-Radiation-Turned-Off}.)
\item
Avoiding firewalls and other exotic physics: The challenges assocaited with transmitting information from the collapsing matter to the radiation, such as the firewall or violation of the no-cloning theorem (see sec.~\ref{sec:Information-Problem}), are avoided.
\hide{
\item
Connecting to other ideas: We comment on the compatibility and relevance of our proposal with existing proposals for resolving the information paradox, such as the firewall, the fuzzball, the AdS/CFT duality, etc., highlighting similarities and differences.
(See sec.~\ref{sec:Comments}.)
}
\end{enumerate}

A UV suppression and a corresponding uncertainty relation are important in our theory.
In lower (two or three) dimensions, a UV-finite theory does not always need a UV suppression factor like $e^{-\ell^2_s \, \Box}$, and a spacetime uncertainty relation may not be present.
Therefore, the problem of Hawking radiation in lower dimensions~\cite{Callan:1992rs,Giddings:1992ff} can be fundamentally different from higher dimensions.

The early cessation of Hawking radiation, driven by trans-Planckian nonlocality, offers a conceptually simple resolution to the information paradox, suggesting that black-hole evaporation is a far less dramatic process than previously assumed and that most information remains within the black hole.
We need a nonlocal effect to exclude trans-Planckian modes in a small region near the horizon, but we do not need a nonlocal mechanism for information transfer.
In addition to offering the simplest resolution to the information paradox, this work also sheds light on our understanding of the nonlocal, UV/IR feature of quantum gravity.

\section*{Acknowledgement}

This work is an invited chapter for the edited book ``The Black Hole Information Paradox'' (Eds. Ali Akil and Cosimo Bambi, Springer Singapore, expected in 2025).
We thank Emil Akhmedov, Robert Brandenberger, Ronny Chau, Chong-Sun Chu, Nick Dorey, Shinji Mukohyama, Toshifumi Noumi, Nobuyoshi Ohta, Domenico Orlando, Susanne Reffert, Yuki Yokokura, and Cheng-Tsung Wang for valuable discussions. 
This work is supported in part by the Ministry of Science and Technology, R.O.C. (MOST 110-2112-M-002 -016-MY3), and by the National Taiwan University. 
H.K. thanks Prof. Shin-Nan Yang and his family for their kind support through the Chin-Yu chair professorship.
H.K. is partially supported by the Japan Society of Promotion of Science (JSPS), Grants-in-Aid for Scientific Research (KAKENHI) Grants No.\ 20K03970 and 18H03708, by the Ministry of Science and Technology, R.O.C. (MOST 111-2811-M-002-016), and by the National Taiwan University.

\small

\bibliographystyle{myJHEP}
\bibliography{bibliography}

\end{document}